\documentclass[a4paper]{aa}
\usepackage{graphicx}
\usepackage{natbib}
\usepackage{longtable,lscape}
\usepackage{slashbox}

\usepackage{txfonts}  

\def\atlas9{{\sc ATLAS9}}

\begin{document}

\title{Long-term variability of the optical spectra of NGC 4151:  II.
Evolution  of the broad H$\alpha$ and H$\beta$ emission-line  profiles}

\author{A.I. Shapovalova \inst{1} \and L.\v C. Popovi\'c\inst{2,3} \and A.N.Burenkov\inst{1} \and
V.H. Chavushyan\inst{4}  \and D.Ili\'c \inst{3,5} \and
A. Kova\v cevi\'c\inst{3,5} \and \\
 N.G. Bochkarev\inst{6}  \and J. Le\'on-Tavares\inst{4,7}  }

\titlerunning{Variability of NGC 4151}
\authorrunning{A.I. Shapovalova et al.}
\offprints{A. I. Shapovalova, \\ \email{ashap@sao.ru}\\ Tables 2,4
and Figures 1,2,5,7,9,13,15 are available electronically only}

\institute{Special Astrophysical Observatory of the Russian AS,
Nizhnij Arkhyz, Karachaevo-Cherkesia 369167, Russia
\and
Astronomical Observatory, Volgina 7, 11160 Belgrade 74, Serbia
\and
Isaac Newton Institute, Yugoslav branch
\and
Instituto Nacional de Astrof\'{\i}sica, \'{O}ptica y
Electr\'onica, Apartado Postal 51, CP 72000, Puebla, Pue. M\'exico
\and
 Department of Astronomy, Faculty of Mathematics, University
of Belgrade, Studentski trg 16, 11000 Belgrade, Serbia
\and Sternberg Astronomical
 Institute, Moscow, Russia
 \and Mets\"ahovi Radio Observatory, Helsinki University of
Technology TKK, Mets\"ahovintie 114, FIN-02540 Kylm\"al\"a, Finland
}
\date{Received  / Accepted }

\abstract
{}
{Results of the long-term (11 years, from 1996 to 2006) H$\alpha$
and H$\beta$ line variations of the active galactic nucleus of NGC
4151 are presented.}
{High quality spectra (S/N$>50$ and R$\approx$8 \ \AA) of H$\alpha$
and H$\beta$ were investigated. We analyzed line profile variations
during monitoring period. Comparing the line profiles of H$\alpha$
and H$\beta$, we studied different details (bumps, absorbtion
features) in the line profiles. The variations of the different
H$\alpha$ and H$\beta$ line profile segments have been investigated.
Also, we analyzed the Balmer decrement for whole line and for line
segments. }
{We found that the line profiles were strongly changing during  the
monitoring period, showing blue and red asymmetries. This indicates
a complex BLR geometry of NGC 4151 with, at least, three
kinematically distinct regions: one that contributes to the blue
line wing, one to the line core and one to the red line wing. Such
variation can be caused by an accelerating outflow starting very
close to the black hole, where the red part may come from the region
 { closer to the black hole than the blue part, which is coming} from the
region having the highest outflow velocities.}
{Taking into account the fact that the BLR of NGC 4151 has a complex
geometry (probably affected by an outflow) and that a portion of the
broad line emission seems to have not a pure photoionization origin,
one can ask the question whether the study of the BLR by
reverberation mapping may be valid in the case of this galaxy.}

\keywords{galaxies: active - galaxies: individual: NGC 4151}

\maketitle

\section{Introduction}

In spite of many papers  devoted to the physical properties (physics
and geometry, see e.g. Sulentic et al. 2000) of the broad line
region (BLR) in active galactic nuclei (AGN), the true nature of the
BLR is not well known. The broad emission lines, their shapes and
intensities can give us many information about the BLR geometry and
physics. In the first instance, the change in the line profiles and
intensities could be used for investigating the BLR nature. It is
often assumed that variations of line profiles on long time scales
are caused by dynamic evolution of the BLR gas, and on short time
scales by reverberation effects (Sergeev et al. 2001). In a number
of papers it was shown that individual segments in the line profiles
change independently on both long and short time scales (e.g.,
Wanders and Peterson 1996; Kollatschny and Dietrich 1997; Newman et
al. 1997; Sergeev et al. 1999). Moreover, the broad line shapes may
reveal some information about the kinematics and structure of the
BLR (see Popovi\'c et al. 2004).

One of the most famous and best studied Seyfert galaxies is NGC 4151
(see e.g. Ulrich 2000; Sergeev et al. 2001; Lyuty 2005, Shapovalova
et al. 2008 - Paper I, and reference therein). This galaxy, and its
nucleus, has been studied extensively at all wavelengths.   The
reverberation investigation indicates a small BLR size in the center
of NGC 4151 (see e.g. Peterson and Cota 1988: $6\pm4$ l.d.; Clavel
et al. 1990: $4\pm3$ l.d.; Maoz et al. 1991: $9\pm2$ l.d.; Bentz et
al. 2006: $6.6_{+1.1}^{-0.8}$ l.d.).

Spectra of NGC 4151 show a P Cygni Balmer and He I absorption with
an outflow velocity from -500 ${\rm km \ s^{-1}}$ to -2000 $\ {\rm
km \ s^{-1}}$, changing with the nuclear flux (Anderson and Kraft
1969; Anderson 1974; Sergeev et al. 2001; Hutchings et al. 2002).
This material is moving outward along the line of sight, and may be
located anywhere outside $\sim$15 light-days (Ulrich and Horne
1996).  An outflow is also seen in higher velocity emission-line
clouds near the nucleus (Hutchings et al. 1999), such as multiple
shifted absorption  lines in C IV and other UV resonance lines
(Weymann et al. 1997; Crenshaw et al. 2000), while warm absorbers
are detected in X-ray data (e.g. Schurch and Warwick 2002).

Some authors assumed that a variable absorption is responsible, at
least partially, for the observed continuum variability of AGN
(Collin-Souffrin et al.1996; Boller et al. 1997; Brandt et al. 1999;
Abrassart and Czerny 2000; Risaliti, Elvius, Nicastro 2002). Czerny
et al. (2003) considered that most of variations are intrinsic to
the source, though a variable absorption cannot be quite excluded.

The nucleus of NGC 4151 emits also in the radio range. The radio
image reveals a 0.2 pc two-sided base to the well-known arc-second
radio jet (Ulvestad et al. 2005). The apparent speeds of the jet
components relative to the radio AGN are less than 0.050c and less
than 0.028c at  nuclear distances of 0.16 and 6.8 pc, respectively.
These are the lowest speed limit yet found in a Seyfert galaxy and
indicates non-relativistic jet motions, possibly due to thermal
plasma, on a scale only an order of magnitude larger than the BLR
(Ulvestad et al. 2005).

The observed evolution of the line profiles of the Balmer lines of
NGC 4151 was studied by Sergeev et al.(2001) in 1988--1998 and was
well modeled  within the framework  of the two-component model,
where two variable components with fixed line profiles
(double-peaked and single-peaked) were used.

Although the AGN of NGC 4151 has been much observed and discussed,
there are still several questions concerning the BLR kinematics
(disk, jets or more complex BLR) and dimensions of the innermost
region. On the other hand, as was mentioned above, multi-wavelength
observations suggest both the presence of an accretion disk (with a
high inclination) and an outflow emission (absorption).
Consequently, further investigations of the NGC 4151 nucleus are
needed in order to constraint the kinematics, dimensions and
geometry of its BLR.

This work is subsequent to Paper I with the aim of studying the
variations of both the integrated profiles of the broad emission
lines and segments along the line profiles, during the (11-year)
period of monitoring of NGC 4151.

The paper is organized as follows: in \S2 observations and data
reduction are presented. In \S3 we study the averaged spectral line
profiles (over years, months and Periods I--III, see Paper I) of
H$\alpha$ and H$\beta$, the line asymmetries and their FWHM
variations, light curves of different line segments and the
line--segment to line--segment flux and continuum--line--segment
relations. In \S4 we analyze the Balmer decrements. The
results are discussed in \S6 and in \S7 we outline our conclusions.

\section{Observations and data reduction}

Optical spectra of NGC 4151 were taken with the 6-m and 1-m
telescopes of SAO, Russia (1996--2006), with the 2.1-m
telescope of the Guillermo Haro Astrophysical Observatory (GHAO) at
Cananea, Sonora, M\'exico (1998--2006), and with the 2.1-m telescope
of the Observatorio Astron\'omico Nacinal at San Pedro Martir
(OAN-SMP), Baja California, M\'exico (2005--2006). They were
obtained with a long--slit spectrograph equipped with CCDs. The
typical wavelength range was 4000 -- 7500 \AA , the spectral
resolution was R=5--15 \AA , and the S/N ratio was $>$ 50 in the
continuum near H$\alpha$ and H$\beta$. In total 180 blue and 137
red spectra were taken during 220 nights.

Spectrophotometric standard stars were observed every night. The
spectrophotometric data reduction was carried out either with the
software developed at the SAO RAS by Vlasyuk (1993), or with IRAF
for the spectra observed in M\'exico. The image reduction process
included bias subtraction, flat-field corrections, cosmic ray
removal, 2D wavelength linearization, sky spectrum subtraction,
addition of the spectra for every night, and  relative flux
calibration based on standard star observations. Spectra were scaled
to the constant flux F$([\ion{O}{iii}]\lambda\,5007)$. More details
about observations and data reduction are given in Paper I and will
not be repeated.

The observed fluxes of the emission lines were corrected for the
position angle (PA), seeing and aperture effects (see Paper I). The
mean error (uncertainty) in our { integral} flux determinations
for H$\alpha$ and H$\beta$ and for the continuum is $<$3\%. In order
to study the broad components of emission lines showing the main BLR
characteristics, we removed from the spectra the narrow components
of these lines and the forbidden lines. To this purpose, we
construct spectral templates using the blue and red spectra in the
minimum activity state (May 12, 2005). Both the broad and narrow
components of  H$\beta$ and H$\alpha$, were fitted with Gaussians
{ (see Fig. \ref{fig1}, available electronically only)}.
 The template spectrum contains the following lines:  for H$\beta$ the
narrow component of  H$\beta$ and [\ion{O}{iii}]\,$\lambda\lambda$
4959, 5007; for H$\alpha$ the narrow component of H$\alpha$,
[\ion{N}{ii}]\,$\lambda\lambda$\,6548, 6584,
[\ion{O}{i}]\,$\lambda\lambda$\,6300, 6364,
[\ion{S}{ii}]\,$\lambda\lambda$\,6717, 6731. Then, we scaled the
blue and red spectra according to our scaling scheme (see Appendix
in Shapovalova et al. 2004), using the template spectrum as a
reference. The template spectrum and any observed spectrum are thus
matched in wavelength, reduced to the same resolution, and then the
template spectrum is subtracted from the observed one. More details
can be found in Paper I and Shapovalova et al. (2004).

\onlfig{1}{\begin{figure*}
\centering
\includegraphics[width=7cm,angle=-90]{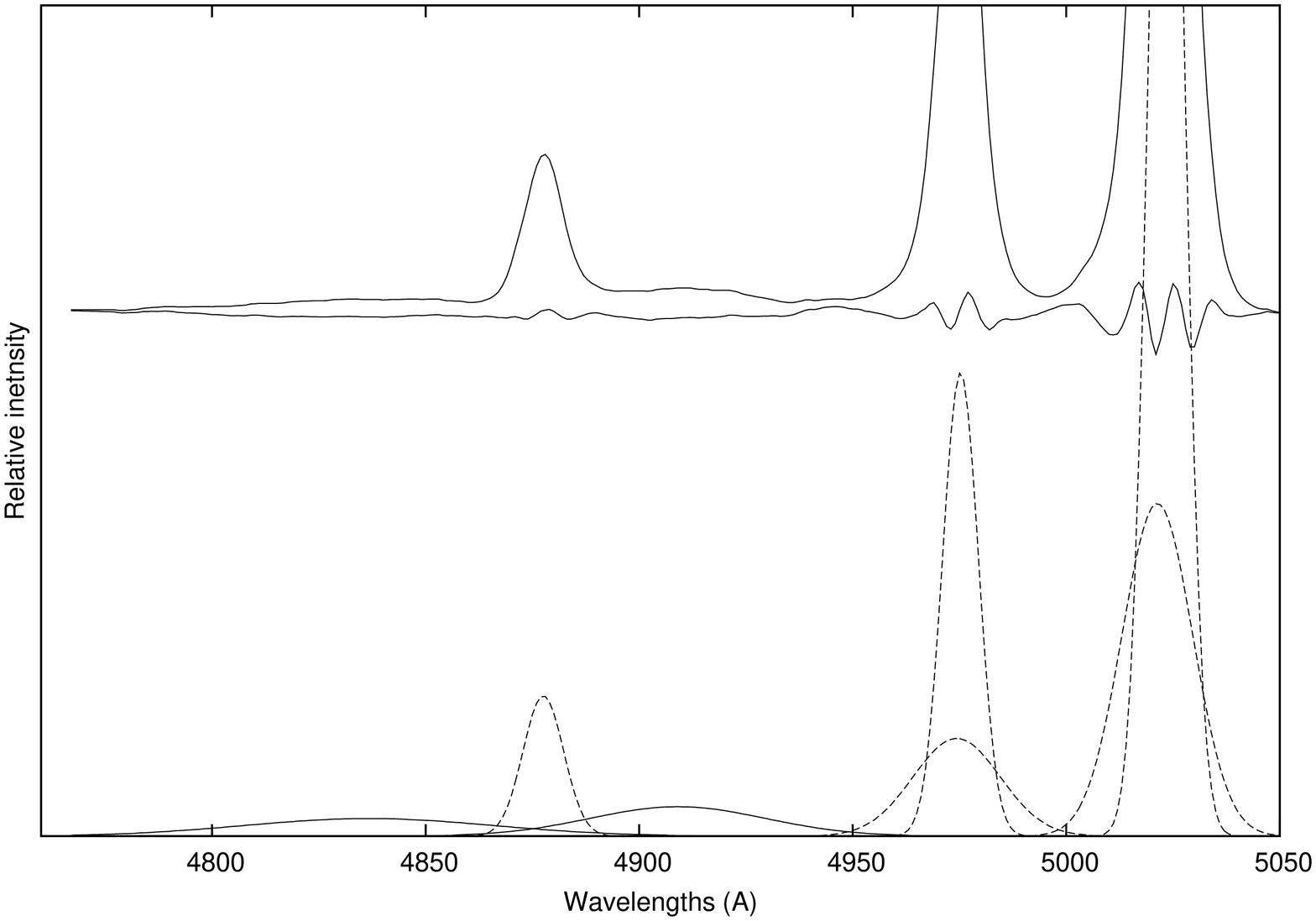}
\includegraphics[width=7cm,angle=-90]{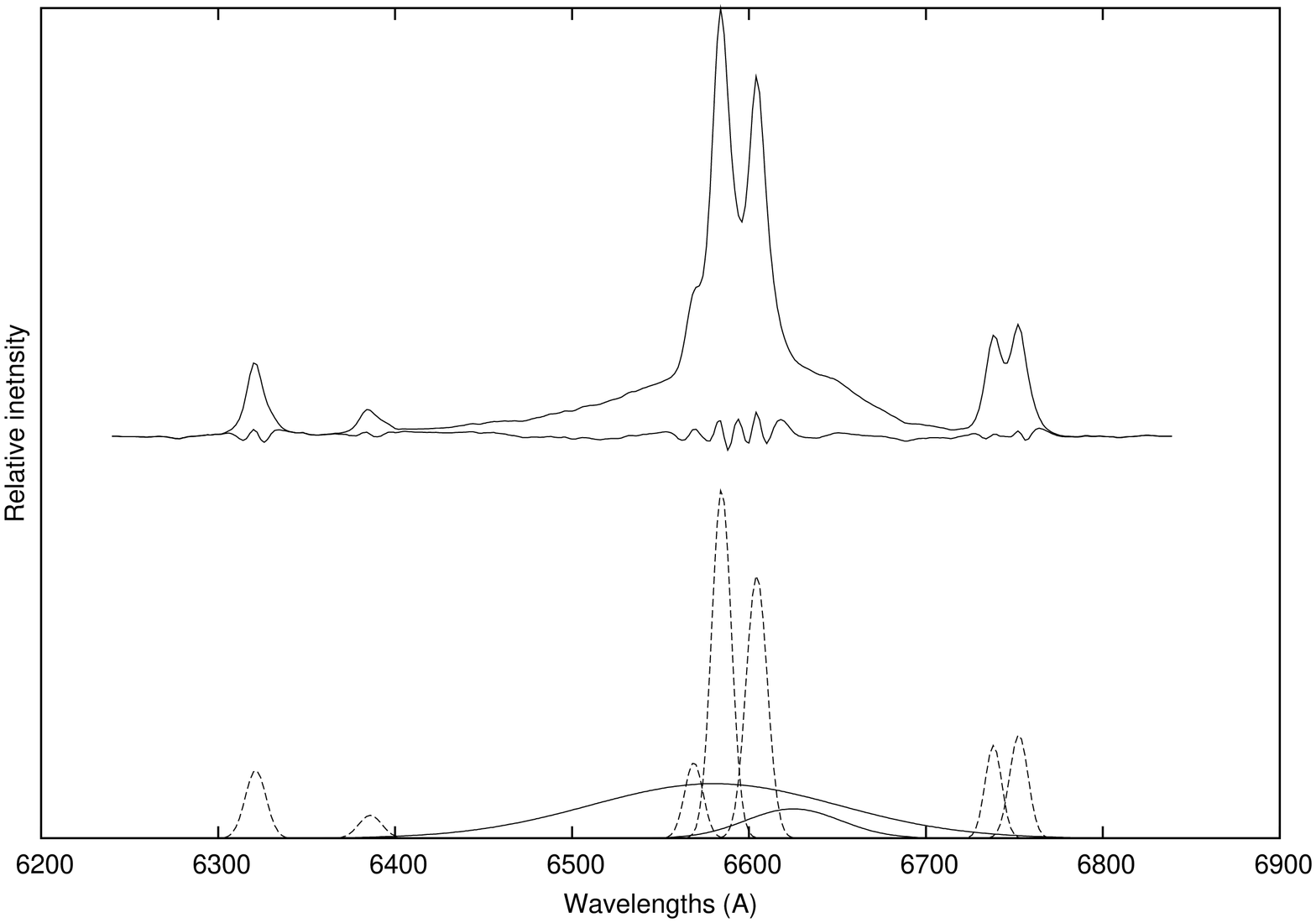}
\caption{The decomposition of H$\beta$ (upper panel) and H$\alpha$
(bottom panel) with Gaussians (below) in order to construct narrow
line templates. The dashed Gaussians correspond to the narrow
components taken for the narrow line templates, while the solid ones
correspond to the broad components (down). The observed spectra and
rms (up) are denoted with solid line. } \label{fig1}
\end{figure*}}

\onlfig{2}{\begin{figure*}
\centering
\includegraphics[width=8cm]{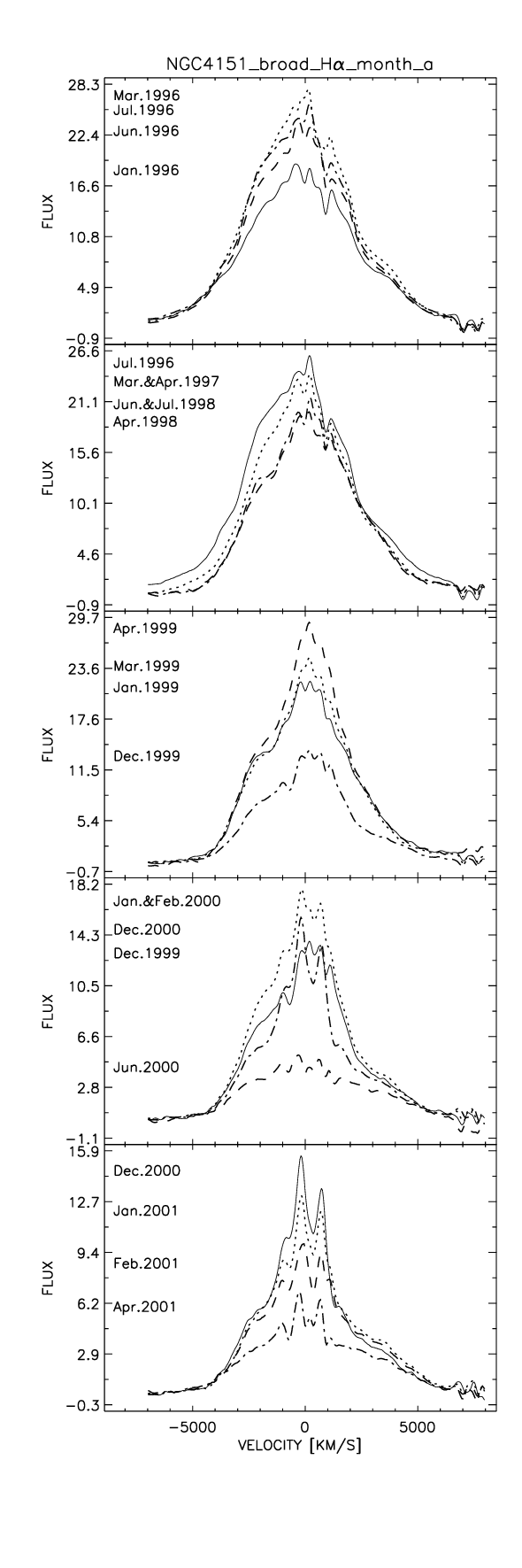}
\includegraphics[width=8cm]{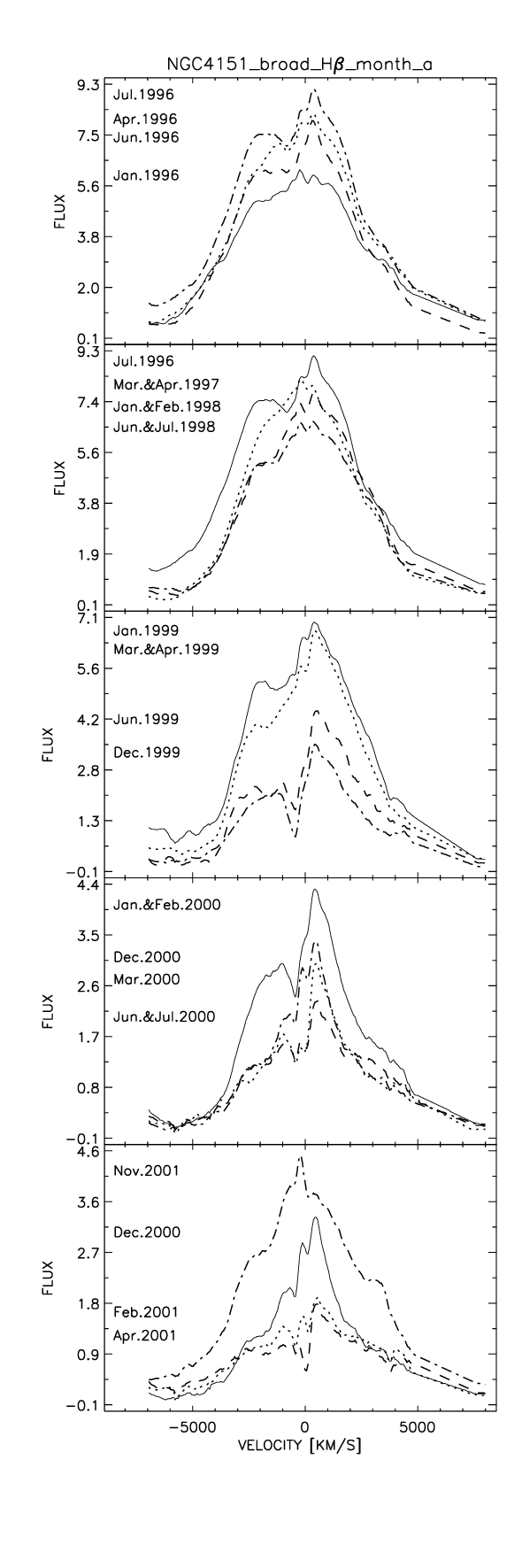}
\caption{The month-averaged profiles of the H$\alpha$ and H$\beta$
broad emission lines in the period 1996--2006. The abscissae  shows
the radial velocities relative to the narrow component of H$\alpha$
or H$\beta$. The ordinate shows the flux in units of
$10^{-14}$\,erg\,cm$^{-2}$\,s$^{-1}$\,\AA$^{-1}$.} \label{fig2}
\end{figure*}}

\addtocounter{figure}{-1}

\onlfig{2}{\begin{figure*}
\centering
\includegraphics[width=8cm]{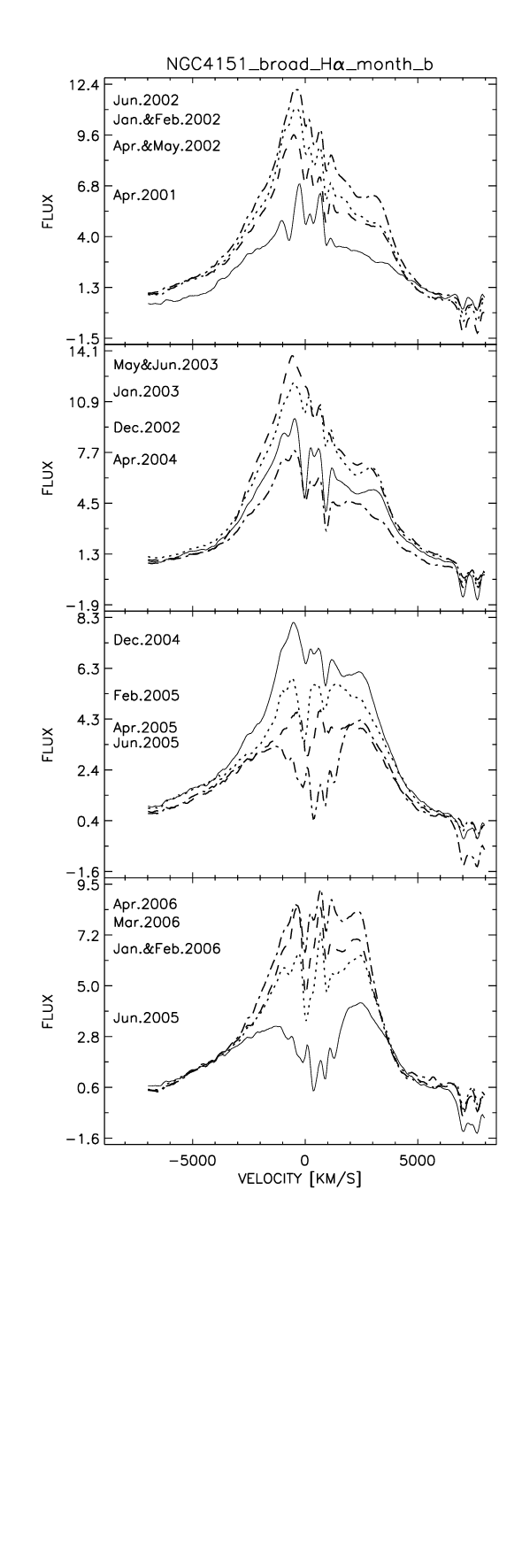}
\includegraphics[width=8cm]{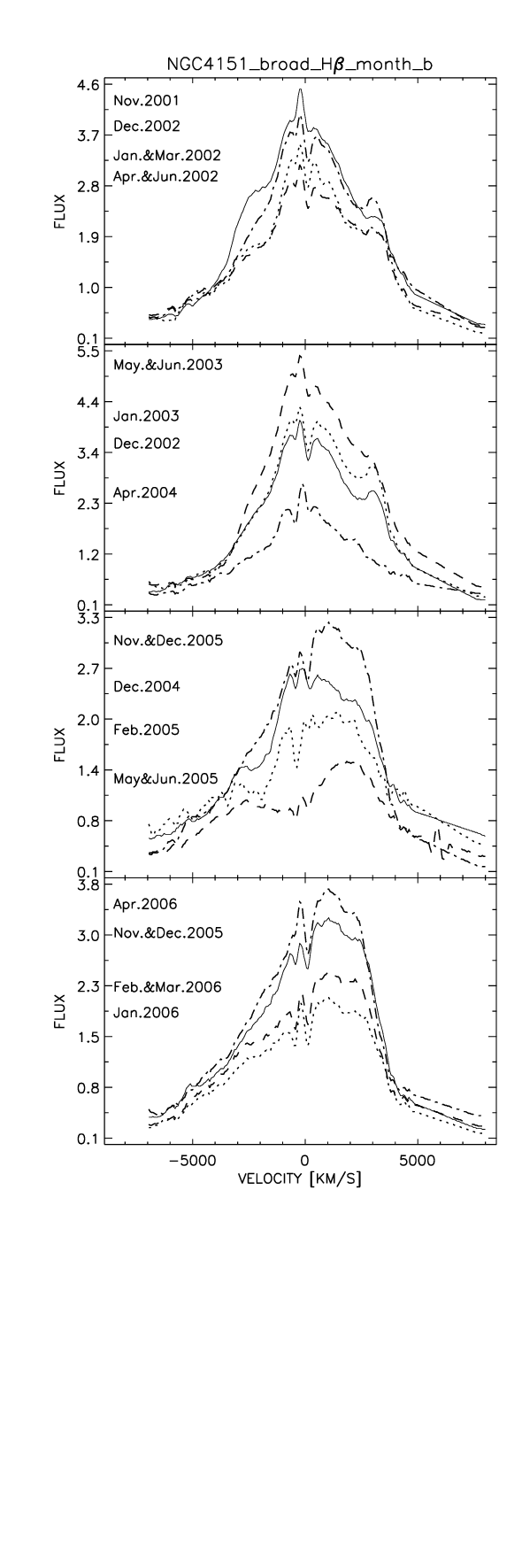}
\caption{Continued.} \label{fig2}
\end{figure*}}

\begin{figure*}
\centering
\includegraphics[width=8cm]{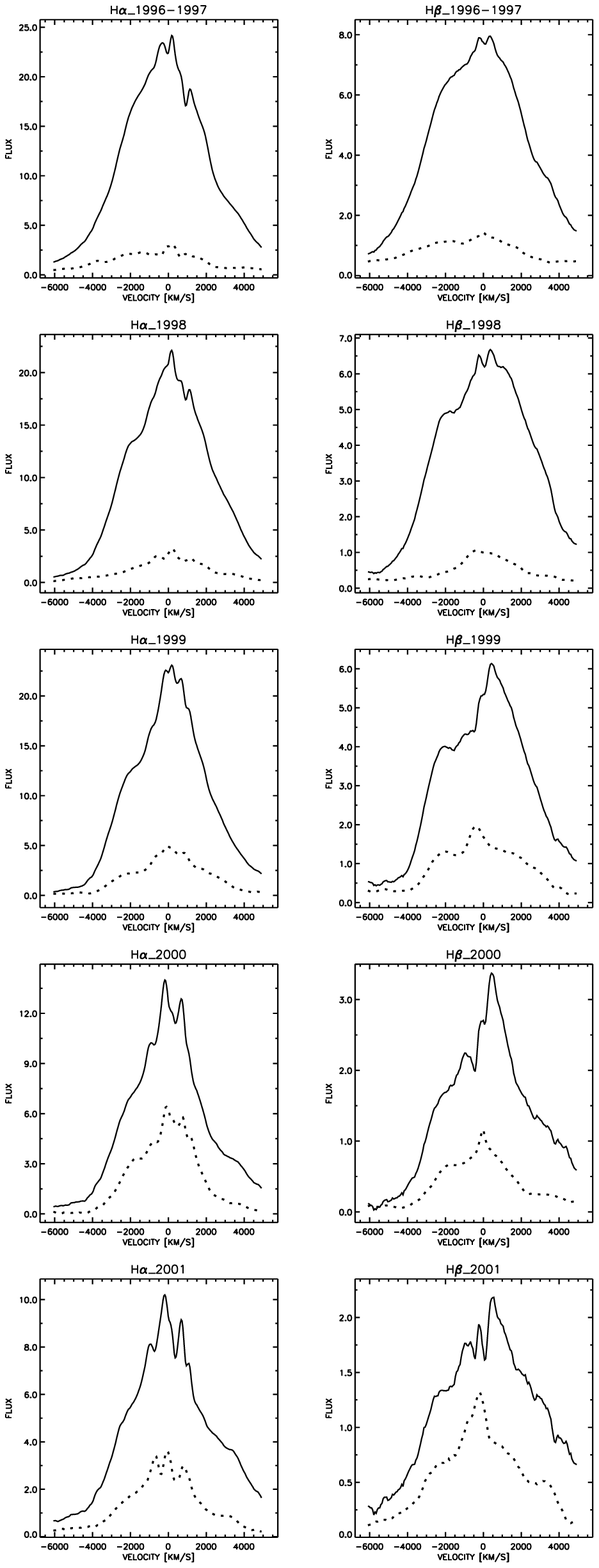}
\includegraphics[width=8cm]{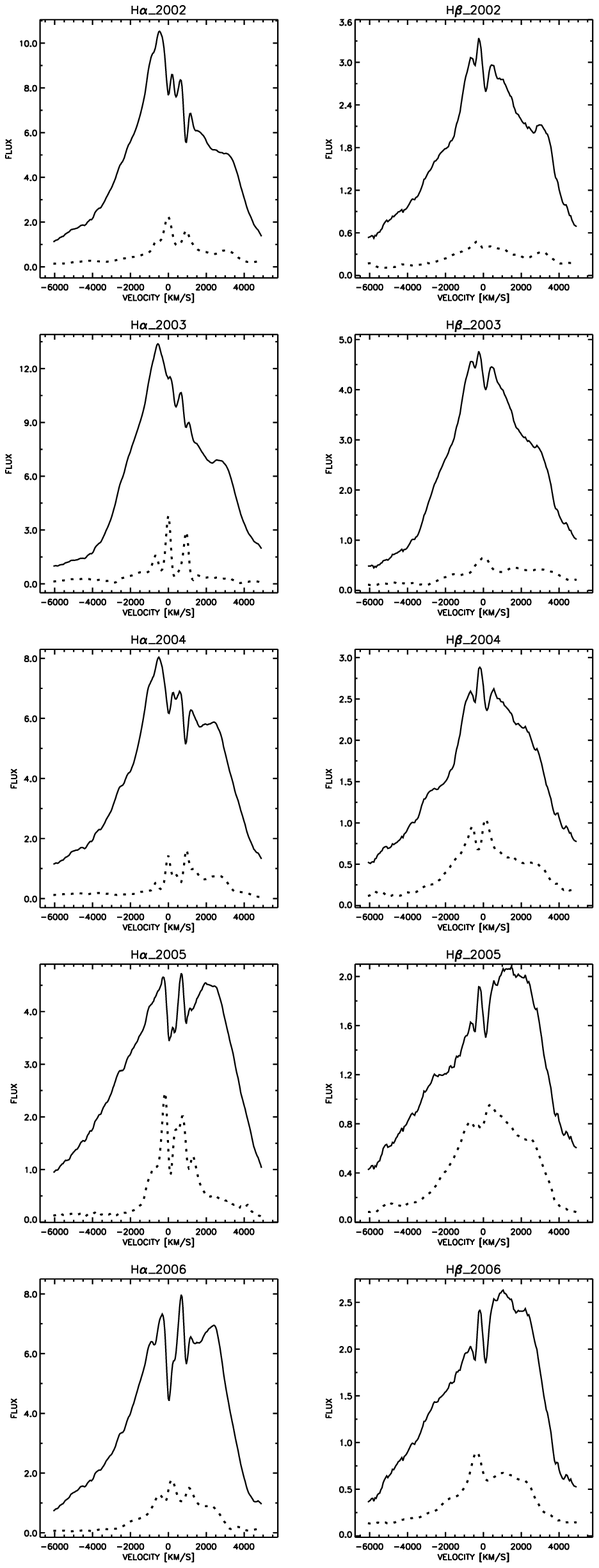}
\caption{The year-averaged  profiles (solid line) and their rms
(dashed line) of the H$\alpha$ and H$\beta$ broad emission lines in
1996-2006. The abscissae (OX) shows the radial velocities relative
to the narrow component of the H$\alpha$ or H$\beta$ line. The
ordinate (OY) shows the flux in units of
$10^{-14}$\,erg\,cm$^{-2}$\,s$^{-1}$\,\AA$^{-1}$.}\label{fig3}
\end{figure*}


\section{Line profile variations}

To investigate the broad line profile variations, we use the most
intense broad lines in the observed spectral range, i.e. H$\alpha$
and H$\beta$, only from spectra with the spectral resolution of
$\sim8\,\AA$. In Paper I we defined 3 characteristic time periods
(I: 1996--1999, II: 2000--2001, III: 2002--2006) during which the
line profiles of these lines were similar. Average values and rms
profiles of both lines were obtained for these periods.

Here we recall some of the most important results of Paper I. In the
first period (I, 1996--1999, JD=2450094.5--2451515.6), when the
lines were the most intense, a highly variable blue component was
observed, which showed two peaks or shoulders at $\sim-4000 \ {\rm
km \ s^{-1}}$ and $\sim-2000 \ {\rm km \ s^{-1}}$ in the rms
H$\alpha$ profiles and, to a less degree, in H$\beta$\footnote{Here
and after in the text the radial velocities are given with respect
to the corresponding narrow components of H$\alpha$ or H$\beta$,
i.e. it is accepted that $V_{\rm r}=0$ for the narrow components of
H$\alpha$ and H$\beta$}. In the second period (II, 2000--2001,
JD=2451552.6--2452238.0) the broad lines were much fainter; the
feature at $\sim-4000 \ {\rm km \ s^{-1}}$ disappeared from the blue
part of the rms profiles of both lines;  only the shoulder at
$\sim-2000 \ {\rm km \ s^{-1}}$ was present. A faint shoulder at
$\sim3500 \ {\rm km \ s^{-1}}$ was present in the red part of rms
line profiles (see Fig. 6 in Paper I). In the third period (III,
2002--2006, JD=2452299.4--2453846.4) a red feature (bump, shoulder)
at $\sim2500 \ \ {\rm km \ s^{-1}}$ was clearly seen in the red part
of both the mean and the rms line profiles (see Fig. 7 in Paper I).
In this paper we study the variations of the broad line profiles in
more details.

\subsection{Month- and year-averaged profiles of the broad H$\alpha$ and H$\beta$
lines}

A rapid inspection of spectra shows that the broad line profiles
vary negligibly within a one-month interval. On the other hand, in
this time-interval, a slight variation in the broad line flux is
noticed (usually around $\sim5-10$\%, except in  some cases up to
$30$\%). Therefore we constructed  the month-averaged line profiles
(see Fig. \ref{fig2}, available electronically only) of H$\alpha$
and H$\beta$.

Moreover, the broad H$\alpha$ and H$\beta$ line profiles were not
changing within a one-year period (or even during several years),
while the broad line fluxes sometimes varied by factors $\sim 2-2.5$
even during one year. The smallest flux variations (factors
$\sim1.1-1.3$) were observed in 1996--1998 (during the line flux
maximum). The largest line flux variations were observed in
2000--2001 and 2005 (factors $\sim 1.7-2.5$), during the minimum of
activity.

As it was mentioned in Paper I, specific line profiles are observed
during the three periods, but a more detailed inspection of the line
profiles shows that slight changes can also be seen between the
year-averaged profiles (see Fig.~\ref{fig3}). Note here that in the
central part of the H$\alpha$ profiles one can often see
considerable residuals (e.g., in 2001 in the form of peaks) due to a
bad subtraction of the bright narrow components of H$\alpha$ and
[\ion{N}{ii}]\,$\lambda\lambda$\,6548, 6584, (at $V_{\rm r} \sim-680
\ {\rm km \ s^{-1}} $ and $\sim960\ {\rm km \ s^{-1}}$). Therefore,
we cannot conclude on the presence of some absorption in the central
part of H$\alpha$. But, in the H$\beta$ profiles (in the central
part, at $V_{\rm r} \sim-430\ {\rm km \ s^{-1}} $; $\sim-370\ {\rm
km \ s^{-1}}$) an absorption, especially strong from June 1999 to
the end of 2000 (see Fig. \ref{fig2}), was detected. Note here that
Hutchings et al. (2002) also found absorbtion at $V_{\rm r} (-2000 -
0) \ {\rm km \ s^{-1}} $ in the H$\beta$ line.

Now let us point some noticeable features in month-averaged line
profiles:


\begin{figure}
\centering
\includegraphics[width=8cm]{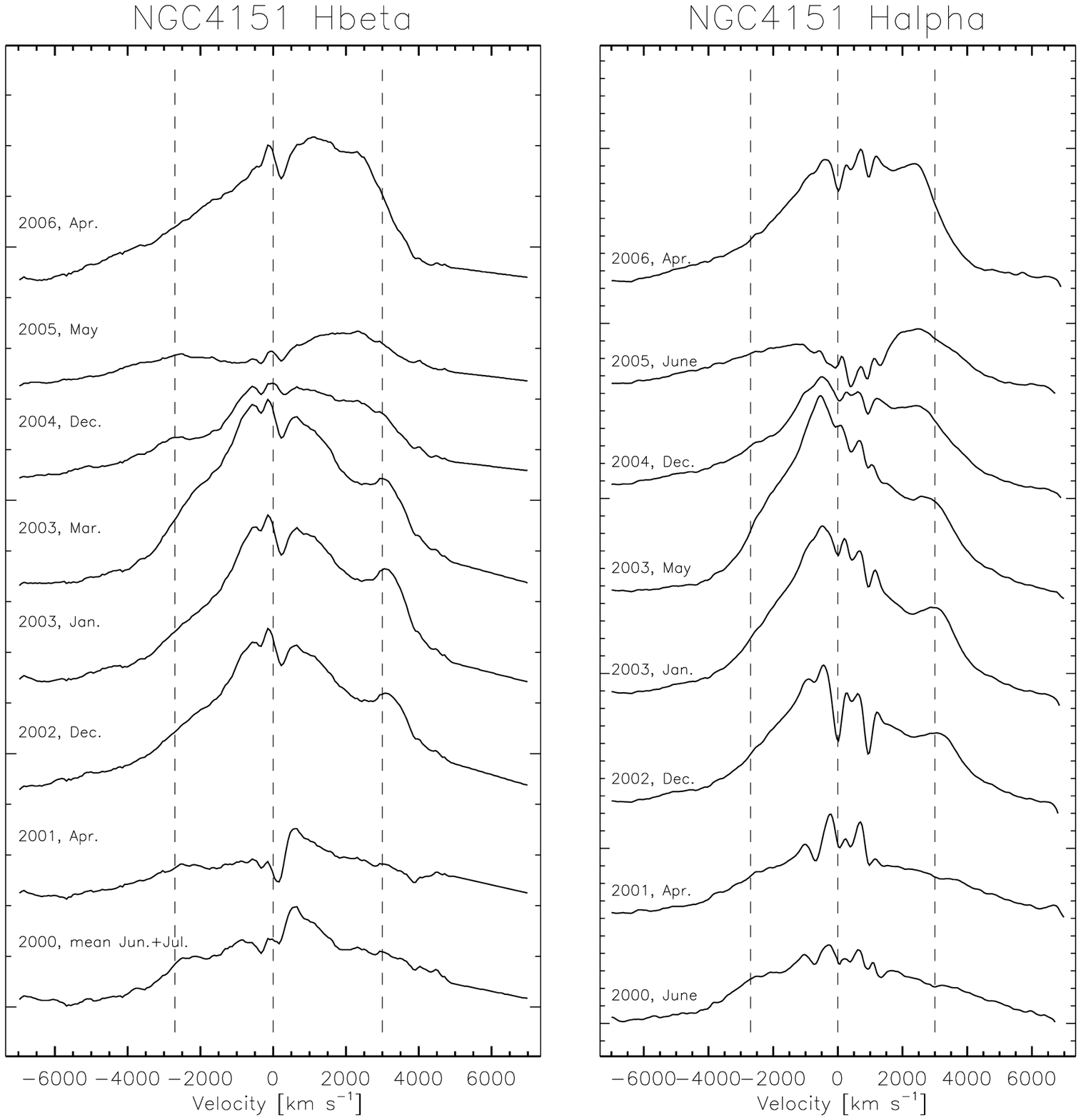}
\caption{Some examples of month-averaged  profiles of the H$\alpha$
and H$\beta$ broad emission lines from 2000 to 2006.
         The abscissae shows the radial velocities relative to the narrow
         component of H$\alpha$ or H$\beta$.
         The vertical  dashed lines correspond to radial velocities:
         -2600$\ {\rm km \ s^{-1}}$, 0$\ {\rm km \ s^{-1}}$  and 3000 $\ {\rm km \ s^{-1}}$.
         The profiles are shifted vertically by
         a constant value.}\label{fig4}
\end{figure}

\onlfig{5}{
\begin{figure*}
\centering
\includegraphics[width=8.5cm]{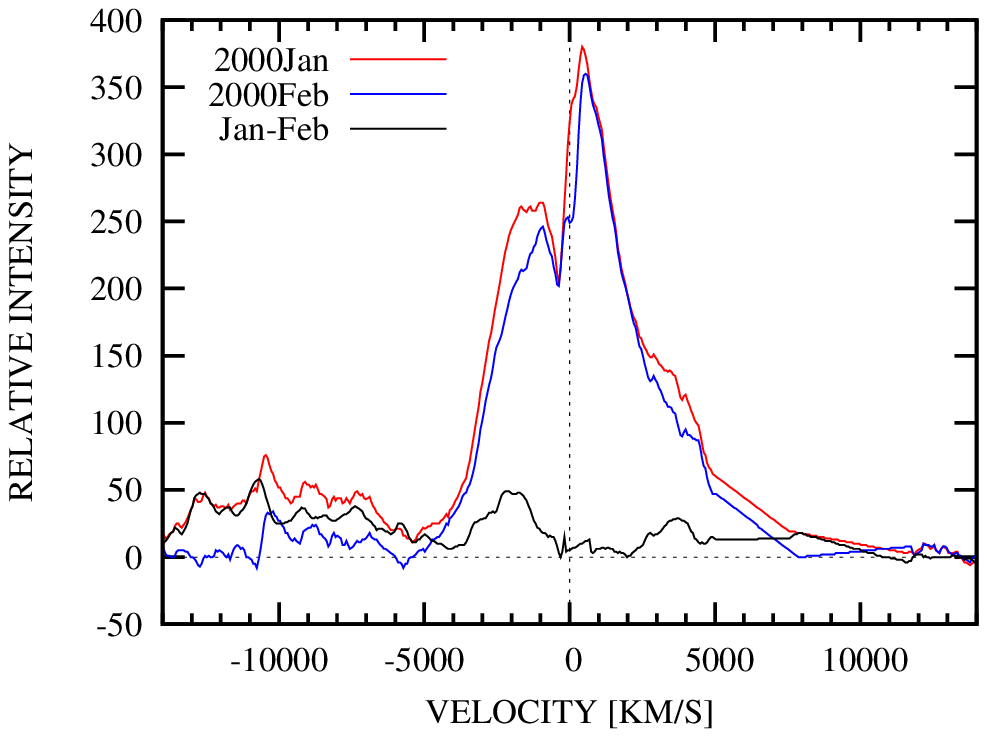}
\includegraphics[width=8.5cm]{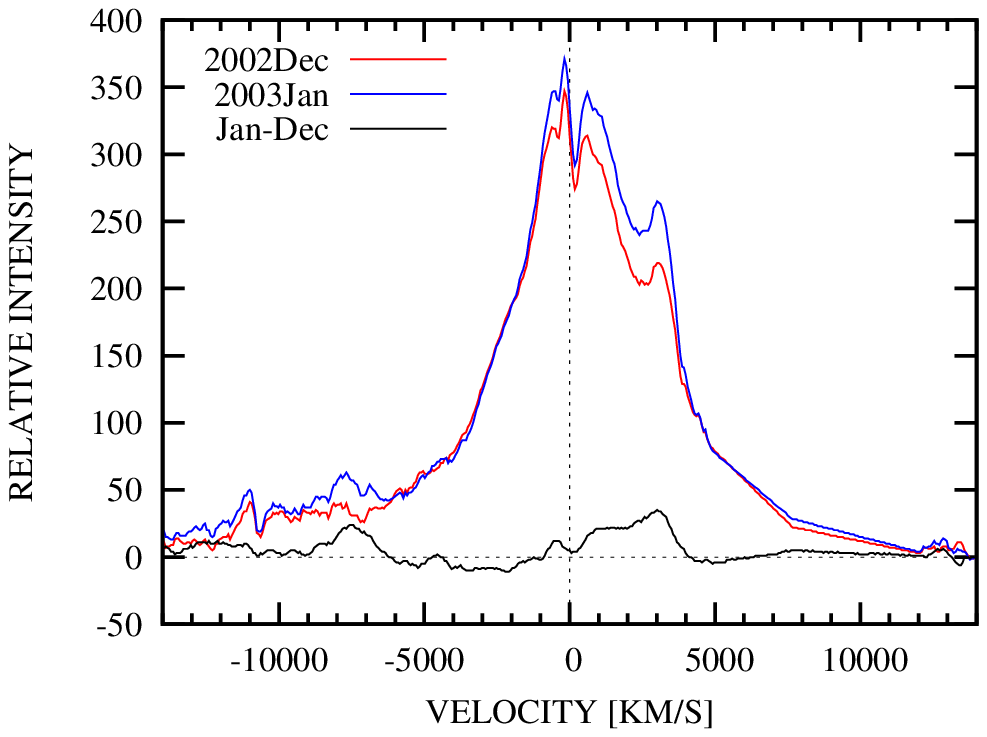}
\caption{The H$\beta$ broad emission lines of NGC 4151 in
particular two sequent months where bumps were strong (red and blue
line). Their residual is also given (black line below).} \label{fig5}
\end{figure*} }

1) In 1996--2001 a blue peak (bump) in H$\beta$ and a  shoulder in
H$\alpha$  were clearly seen at $\sim-2000\ {\rm km \ s^{-1}}$ (Fig.
\ref{fig2}). However, in 2002--2004, the blue wing of both lines
became steeper than the red one and it did not contain any
noticeable features.

2) In 2005 (May--June), when the nucleus of NGC 4151 was in the
minimum activity, the line profiles had a double-peak structure with
two distinct peaks (bumps) at radial velocities of ($-2586; +2027)\
{\rm km \ s^{-1}}$ in H$\beta$ and $(-1306; +2339) \ {\rm km \
s^{-1}}$ in H$\alpha$ (see Fig. \ref{fig2}; 2005\_05 and 2005\_06).
In principle, the two-peak structure in the H$\beta$ profiles is
also seen in spectra of 1999--2001: at $V_{\rm r}\sim-1500 \ {\rm km
\ s^{-1}}$ the blue and $V_{\rm r}\sim500 \ {\rm km \ s^{-1}}$ the
red peak. But in this case the blue peak may be caused by a broad
absorption line at the radial velocity ($\sim-400 \ {\rm km \
s^{-1}}$).

3) In 2006 the line profiles changed dramatically - the blue wing
became flatter than in previous period, while the red wing was very
steep without any  feature at $V_{\rm r}>2300\ {\rm km \ s^{-1}}$.

4) In 2002 a distinct peak (bump) appeared in the red wing of the
H$\alpha$ and H$\beta$ lines at the radial velocity  $V_{\rm
r}\sim3000\ {\rm km \ s^{-1}} $. The radial velocity of the red peak
decreases: in  2002-2003 it corresponds to $\sim3100\ {\rm km \
s^{-1}}$ and in  2006 to $\sim2100\ {\rm km \ s^{-1}}$.  This effect
is well seen in Fig. \ref{fig4}, especially  in the H$\alpha$ line
profile. Table \ref{tab1} gives the obtained radial velocities of
red peak measured in the H$\alpha$ and H$\beta$ profiles for which
the peak was clearly seen. Radial velocities of the H$\alpha$ and
H$\beta$ red peak, measured in the same periods, are similar (the
differences are within the error-bars of measurements). The mean
radial velocity of the red peak decreased by $\sim1000\ {\rm km \
s^{-1}}$ from 2002 to 2006 (Table \ref{tab1}). It is not clear if the
red peak is shifting along the line profile or it disappears and
again appears as a new red peak at another velocity.

\begin{table*}[t]
\begin{center}
\caption[]{The peak shifts in the red wing of the H$\alpha$ and
H$\beta$ lines. Columns: 1 - year and month for H$\alpha$ (e.g.,
2002\_1\_3 = 2002, January and March); 2 - $V_{\rm r}$(H$\alpha$),
H$\alpha$ peak velocity in the red wing ($ {\rm km \ s^{-1}}$); 3 -
year and month for H$\beta$; 4 - $V_{\rm r}$(H$\beta$), H$\beta$
peak velocity in the red wing ($ {\rm km \ s^{-1}}$). The line after
each year (or interval of years) gives the mean $V_{\rm
r}$(H$\alpha$) and $V_{\rm r}$(H$\beta$) and their standard
deviations. The last line gives the maximal shift of the red peak
($\Delta V_{\rm r}$) from 2002 to 2006.}\label{tab1}
\begin{tabular}{lccc}
\hline
\hline
H$\alpha$  & $V_{\rm r}$(H$\alpha$) & H$\beta$  &  $V_{\rm r}$(H$\beta$)\\
year\_month &  $ {\rm km \ s^{-1}}$      &year\_month &  $ {\rm km \ s^{-1}}$     \\
\hline
             &                  &           2001b\_11   &  3257              \\
2002\_1\_3   &     3068         &           2002\_1\_3   &  3257              \\
2002\_4\_5   &     3114         &           2002\_4\_6   &  3196              \\
2002r\_06   &     3068         &                      &                    \\
2002r\_12   &     2977         &           2002b\_12   &  3073              \\
\hline
mean 2002  &     3057$\pm$57     &           mean 2001\_11 + 2002  &  3197$\pm$94          \\
\hline
2003r\_01   &     2886         &           2003b\_01   &  3073              \\
2003r\_05   &     2613         &           2003b\_03   &  3012              \\
2003r\_06   &     2613         &           2003\_5\_6   &  2889              \\
\hline
mean 2003  &     2704$\pm$158    &           mean 2003  &  2991$\pm$94          \\
\hline
2004r\_12   &     2339         &           2004b\_12   &  2273              \\
2005r\_02   &     2294         &           2005\_1\_2   &  2027              \\
2005r\_04   &     2294         &           2005\_5\_6   &  2027              \\
2005r\_06   &     2339         &           2005\_11\_12 &  2338              \\
2006\_1\_2   &     2430         &           2006b\_01   &  2335              \\
           &                  &           2006b\_02   &  2335              \\
2006r\_03   &     2202         &           2006b\_03   &  2150              \\
2006r\_04   &     2294         &           2006b\_04   &  2027              \\
\hline
mean 2004-2006  & 2313$\pm$69     &     mean 2004-2006  & 2189$\pm$147    \\
\hline
\multicolumn{4}{l}{maximal shift $\Delta V_{\rm r}$(H$\alpha$+H$\beta$)=(1072$\pm$226)$ \ {\rm km \ s^{-1}}$} \\
\hline

\end{tabular}
\end{center}
\end{table*}

\begin{figure}
\centering
\includegraphics[width=9cm]{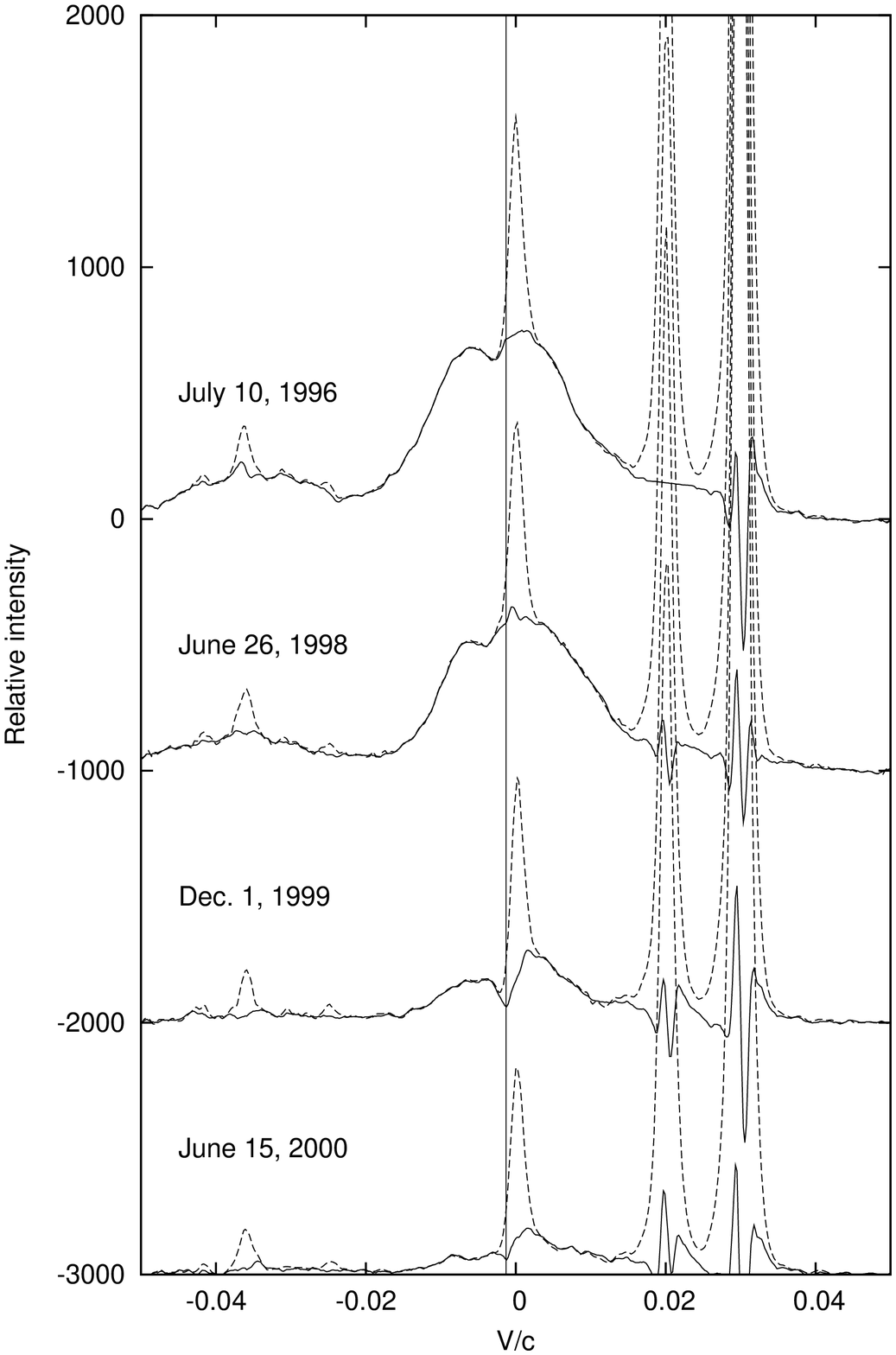}
\caption{ The absorption seen in years 1996, 1997, 1998 and 2000
(from top to bottom). The H$\beta$ line after subtraction of the
narrow lines is shown with solid lines. The vertical line marks the
radial velocity of -400 km/s. }\label{fig6}
\end{figure}


\subsection{The absorption and emission features in H$\alpha$ and H$\beta$ line profiles}

As mentioned above (\S 3.1), the  H$\alpha$ and H$\beta$ line
profiles have the absorption and emission features. Distinguishing
between these features is very difficult. The question is are the
"bumps" an intrinsic property of the broad H$\alpha$ and H$\beta$
lines or is it just the strengthening of the absorption features?
Furthermore, an open question is the origin of these features, i.e.
is there an intrinsic mechanism in the BLR which creates these
"bumps"?  We should mentioned here that in the central part of
both lines, the residuals of the narrow components can affect these relatively
weak absorption/emission features.

To study it, the residuals between some H$\beta$ broad lines with
prominent (noticeable) bumps from particular two successive months
were obtained. An example of these residuals are presented in
Fig.~\ref{fig5} (available electronically only). The noticeable
residual bumps (without an absorption-like feature) at $V_{\rm r}$
from (-2000) $ {\rm km \ s^{-1}}$ to (-1000) $ {\rm km \ s^{-1}}$ in
1999-2001, and at $V_{\rm r}$ from 3500 $ {\rm km \ s^{-1}}$ to 2500
$ {\rm km \ s^{-1}}$ in 2002-2006 are seen well. It seems  that
the absorption changes slowly and during several next months it
remains constant. Therefore this absorption disappears in profile
residuals (see Fig. \ref{fig5}). Consequently, it seems that
emission bumps observed in the H$\beta$ and H$\alpha$ line profiles
(seen in the residuals in Fig. \ref{fig5}) are mostly an intrinsic
property of the broad emission-line profile.

 The strong absorption features are also present in 1996-2001
H$\beta$ spectra. In this period we observed the dip and broad
absorbtion  at radial velocity that changes from $\sim$-1000 km/s in
(1996-1998) to $\sim$-400 km/s in 1999-2000 (see Fig. \ref{fig2},
available electronically only). This velocity corresponds to the
minimum of the absorption band, but its blue edge extends to higher
velocities ($\sim$-1800 km/s in 1998 and $\sim$-1170 in 1999-2000).
Fig. \ref{fig6}  shows  some observed individual spectra  and their
broad component where the blue absorbtion is well resolved. The blue
shifted absorption is probably coming from an outflowing material.

It is interesting to note that the higher radial velocity ($\sim$
-1000 km/s, observed in 1996-1998) is appearing for a higher continuum
flux level, while the smaller velocity ($\sim$ -400 km/s) has been
detected when the continuum flux decreases 3 - 6 times, i.e. we
confirm the results reported by Hutchings et al. (2002), who found
the same trend that the outflow velocity increases with the
continuum flux.

\onlfig{7}{
\begin{figure*}
\centering
\includegraphics[width=15.5cm]{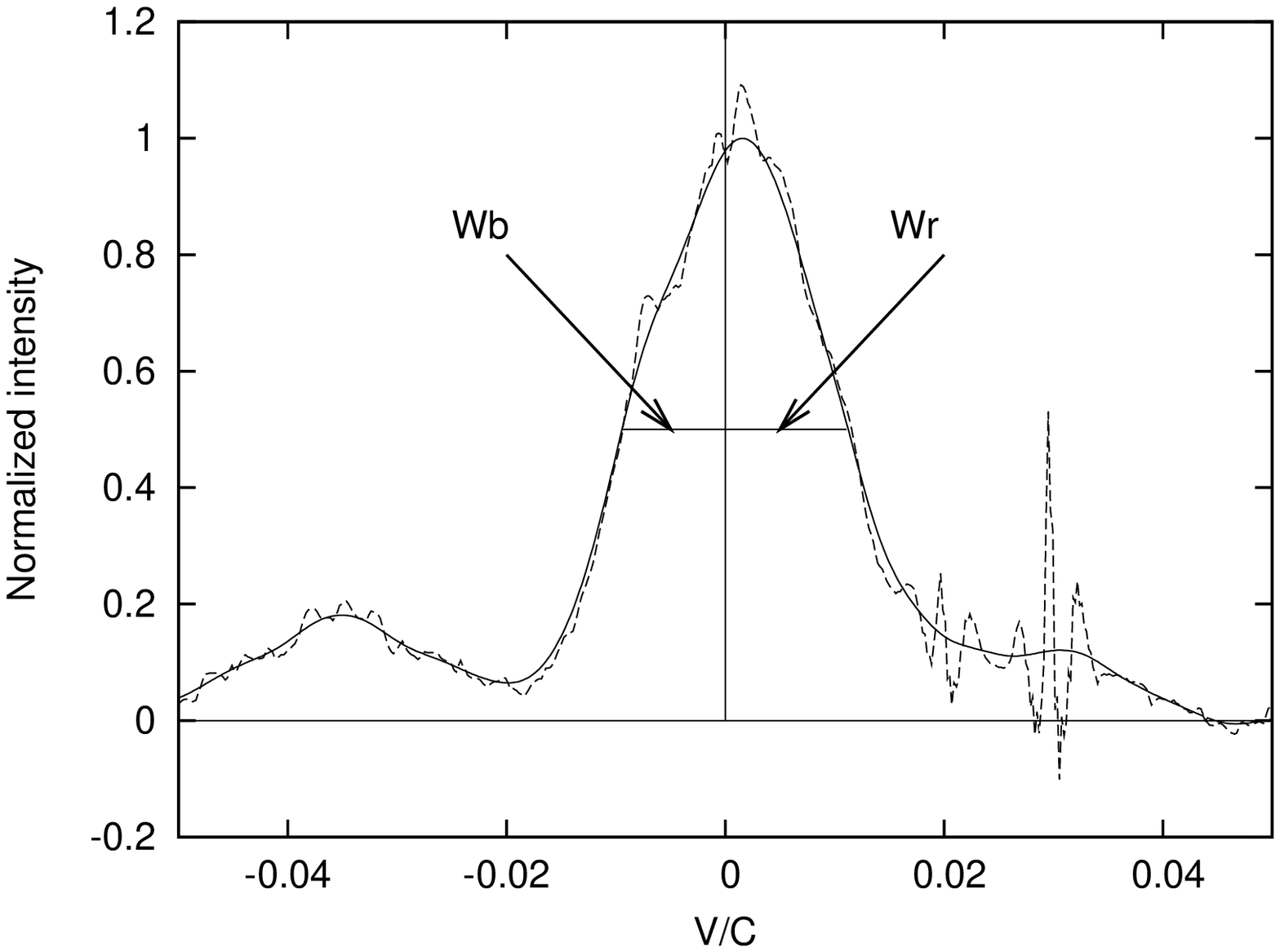}
\caption{An example of the FWHM  and asymmetry measurements. The
observed spectra is denoted with the dashed line and the smoothed
spectra with solid line. }\label{fig7}
\end{figure*}
}

\begin{figure}
\centering
\includegraphics[width=10cm]{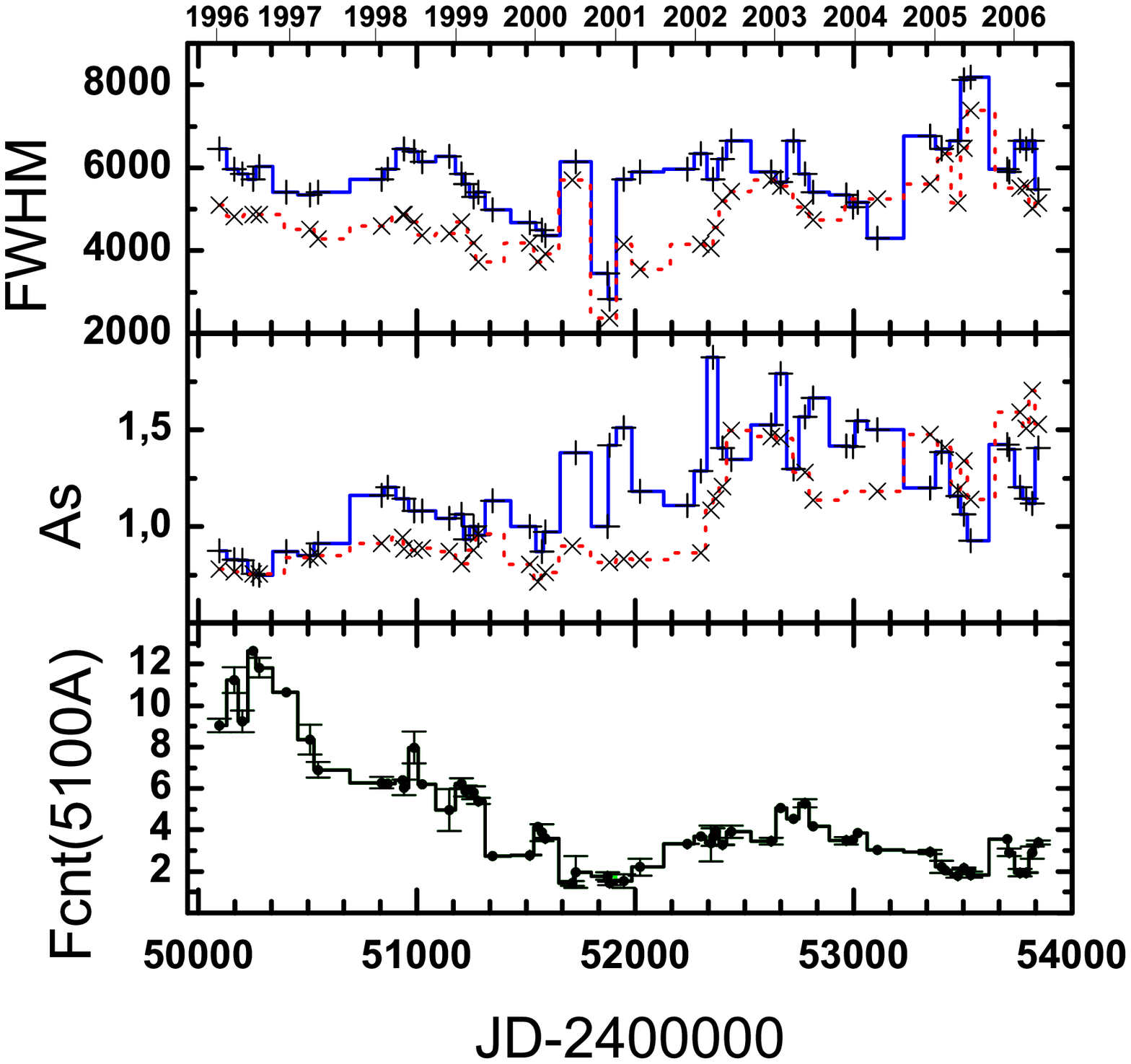}
\caption{Variations of the FWHM (upper panel), asymmetry (middle
panel) in H$\alpha$ (denoted with crosses) and H$\beta$ (denoted
with plus) broad lines, and of the continuum flux at $\lambda 5100\,
\AA$ (bottom panel) in 1996--2006. The abscissae shows the Julian
date (bottom) and the corresponding year (up). The continuum flux is
given in $10^{-14}\rm erg \ cm^{-2} \ s^{-1} \AA^{-1}$.
}\label{fig8}
\end{figure}

\subsection{Asymmetry of the broad H$\alpha$ and H$\beta$ line profiles}

We measured the full width at half maximum (FWHM) of the broad lines
from their month-averaged profiles and determined the asymmetry (A)
as a ratio of the red and blue parts of FWHM, i.e. A=W$_{\rm
red}$/W$_{\rm blue}$, { where W$_{\rm red}$ and W$_{\rm blue}$
are the red and blue half-widths at maximal intensity (see Fig.
\ref{fig7}, available electronically  only) with respect to the
position of the narrow component of H$\alpha$ and H$\beta$. As we
mentioned above, there are residuals in the center of the H$\alpha$
and H$\beta$ lines due to residuals from the subtraction of narrow
components, that can affect the FWHM and A measurements. Therefore
we first smoothed the line profiles, in order to avoid artificial
peaks from the residuals (see Fig. \ref{fig7}), before measuring the
FWHM and asymmetry. The two independent measurements of FWHM and A
were performed.} We determined the averaged continuum and its
dispersion for each month and the averaged Julian date from the
spectra which were used to construct the month-averaged profiles.
The measurements of the FWHM, asymmetry and continuum during the
whole period of monitoring (1996--2006) are presented in Table
\ref{tab2} (available electronically only) and in Fig. \ref{fig8}.
{ In Table \ref{tab2} we give averaged values of the FWHM and A
from two independent measurements and their dispersions.}

\onllongtab{2}{
\begin{longtable}{cccccccc}
\caption[]{\label{tab2} The FWHM and asymmetry of the H$\alpha$ and
H$\beta$ broad emission lines in the period of 1996--2006. Columns:
1 - year and month (e.g., 1996\_01 = 1996\_January); 2 - Julian date
(JD); 3 - the FWHM of H$\beta$; 4 - the asymmetry of H$\beta$; 5 -
the FWHM of H$\alpha$ ; 6 - the asymmetry of H$\alpha$; 7 - F(5100),
the continuum flux at $\lambda 5100\, \AA$ in units of
$10^{-14}$\,erg\,cm$^{-2}$\,s$^{-1}$\,\AA$^{-1}$ and $\sigma$(cnt),
the estimated continuum flux error in the same units.}\\
\hline \hline
year\_month &  JD      & FWHM H$\beta\pm\sigma$& A H$\beta\pm\sigma$& FWHM H$\alpha\pm\sigma$ &A H$\alpha\pm\sigma$&    \multicolumn{2}{c}{F(5100)$\pm \sigma$}\\
           & 2400000+ & $ {\rm km \ s^{-1}}$ &         &  ${\rm km \ s^{-1}}$ &    & \multicolumn{2}{c}{$10^{-14} \ \rm erg \ cm^{-2} \ s^{-1} \AA^{-1}$}  \\
\hline1&2&3&4&5&6&7\\
\hline
\endfirsthead
\caption{Continued.}\\
\hline
year\_month &  JD      & FWHM H$\beta\pm\sigma$& A H$\beta\pm\sigma$& FWHM H$\alpha\pm\sigma$ &A H$\alpha\pm\sigma$&    \multicolumn{2}{c}{F(5100)$\pm \sigma$}\\
           & 2400000+ & $ {\rm km \ s^{-1}}$ &         &  ${\rm km \ s^{-1}}$ &    & \multicolumn{2}{c}{$10^{-14} \ \rm erg \ cm^{-2} \ s^{-1} \AA^{-1}$}  \\
\hline
1&2&3&4&5&6&7\\
\hline
\endhead
\hline
\endfoot
\hline
\endlastfoot
1996b\_01&  50096.0  & 6860 $\pm$  566  & 0.904 $\pm$ 0.043 &  5332 $\pm$  324 & 0.799 $\pm$ 0.030  &  9.031 $\pm$  0.322\\
1996b\_03 & 50163.4 &  6182  $\pm$ 303&   0.877 $\pm$ 0.069&   5036 $\pm$  291 & 0.782 $\pm$ 0.021 &  11.219 $\pm$ 0.617\\
1996b\_04 & 50200.8  & 6183 $\pm$  479 &  0.877 $\pm$ 0.072 &  --     --   &--     --    &   9.226 $\pm$ 0.527\\
1996b\_06 & 50249.3 &  6028 $\pm$  434 &  0.797 $\pm$ 0.060 &  --     --  & --     --   &   12.641  --\\
1996b\_07 & 50277.8  & 6367 $\pm$  479  & 0.783 $\pm$ 0.048 &  5058 $\pm$  259 & 0.762 $\pm$ 0.011 &  11.822 $\pm$ 0.473\\
1996b\_11 & 50402.6  & 5659 $\pm$  348  & 0.896 $\pm$ 0.035 &  --     -- &  --     --    &  10.649  --\\
1997b\_03 & 50510.9  & 5568 $\pm$  305  & 0.884 $\pm$ 0.048  & 4808 $\pm$  420 & 0.853 $\pm$ 0.019   & 8.350 $\pm$ 0.716\\
1997b\_04 & 50547.9  & 5567 $\pm$  218 &  0.946 $\pm$ 0.046 &  4512 $\pm$  323 & 0.854 $\pm$ 0.005  &  6.897 $\pm$ 0.375\\
\hline
1996-1997&& 6051 $\pm$ 449&0.871 $\pm$ 0.054 & 4959  $\pm$ 307&0.810 $\pm$ 0.042& 9.979$\pm$1.950\\
\hline
1998b\_01 & 50838.0 &  5844 $\pm$  173   &1.159 $\pm$ 0.005  & 4831 $\pm$  323 & 0.947 $\pm$ 0.050   & 6.280 $\pm$ 0.277\\
1998b\_02 & 50867.4  & 6059 $\pm$  130  & 1.189 $\pm$ 0.021  & --     --  & --     --   &    6.219  --\\
1998r\_04&  50934.5 &  --     --  &  --     --   &   5058  $\pm$ 259 & 0.981 $\pm$ 0.051  &  6.387  --\\
1998b\_05&  50940.4  & 6489 $\pm$   42  & 1.131 $\pm$ 0.016  & 5150 $\pm$  388 & 0.933 $\pm$ 0.071  &  6.049 $\pm$ 0.385\\
1998b\_06&  50988.3  & 6458 $\pm$   86 &  1.059 $\pm$ 0.029  & 4922 $\pm$  323 & 0.931 $\pm$ 0.073  &  7.974 $\pm$ 0.776\\
1998b\_07 & 51025.3 &  6214 $\pm$   88 &  1.082 $\pm$ 0.000  & 4762 $\pm$  549 & 0.935 $\pm$ 0.067  &  6.205  --\\
1998\_11-12 & 51148.6  & 6275 $\pm$    1  & 1.062 $\pm$ 0.028  & 4694 $\pm$  387 & 0.909 $\pm$ 0.053  &  4.975 $\pm$ 1.011 \\
\hline
1998&& 6223 $\pm$ 245&1.114 $\pm$ 0.054 & 4903  $\pm$ 176&0.939 $\pm$ 0.024&6.298$\pm$0.880\\
\hline
1999b\_01&  51202.6 &  5967 $\pm$  174  & 1.063 $\pm$ 0.001 &  4831 $\pm$  194  &0.863 $\pm$ 0.079  &  6.245 $\pm$ 0.243 \\
1999b\_02 & 51223.1 &  5721 $\pm$  174  & 0.957 $\pm$ 0.031 &  --     --  & --     --  &     5.873 $\pm$ 0.258  \\
1999b\_03 & 51260.5 &  5475 $\pm$  261  & 1.045 $\pm$ 0.064  & 4398 $\pm$  291  &0.910 $\pm$ 0.045  &  5.793 $\pm$ 0.296\\
1999b\_04 & 51281.5 &  5567 $\pm$  218  & 1.011 $\pm$  0.078 &  4010  $\pm$ 388 & 0.959 $\pm$ 0.001  &  5.397 $\pm$ 0.146 \\
1999b\_06 & 51346.4 &  5352 $\pm$  522 &  1.173 $\pm$ 0.058  & --     -- &  --     --  &     2.753  --\\
1999b\_12 & 51515.6  & 5106  $\pm$ 609  & 1.023 $\pm$ 0.032 &  4352  $\pm$ 226 & 0.836 $\pm$ 0.045  &  2.782 $\pm$ 0.016  \\
\hline
1999&& 5531 $\pm$ 298&1.045 $\pm$ 0.072 & 4903  $\pm$ 337&0.892 $\pm$ 0.054&4.807$\pm$1.603\\
\hline
2000b\_01 & 51553.6 &  --     --  &  --     --  &    4010 $\pm$  388  &0.744 $\pm$ 0.041 &   4.155  --\\
2000\_1\_2 & 51571.6 &  4890 $\pm$  566  & 0.960 $\pm$ 0.124 &  --     --  & --     --      & 3.868 $\pm$ 0.406  \\
2000b\_02 & 51589.5  & 4767 $\pm$  566  & 1.036 $\pm$ 0.090  & 4056 $\pm$  194 & 0.801 $\pm$ 0.055   & 3.581 $\pm$  0.029  \\
2000b\_06 & 51711.3 &  --     --  &  --     --    &  6244 $\pm$  775 & 1.014$\pm$  0.162  &  1.434 $\pm$ 0.115 \\
2000\_6\_7 & 51728.6 &  6336 $\pm$  262  & 1.423 $\pm$ 0.060 &  --     --  & --     --    &   1.974 $\pm$ 0.764 \\
2000b\_11 & 51874.1 &  3968 $\pm$  739  & 1.014 $\pm$ 0.020  & --     --  & --     --     &  1.778 $\pm$ 0.177 \\
2000b\_12 & 51883.6  & 3383 $\pm$  783   &1.353 $\pm$ 0.096 &  2985  $\pm$ 871 & 0.805 $\pm$ 0.013  &  1.427 $\pm$  0.09\\
2001b\_02 & 51947.9  & 6367 $\pm$  914  & 1.523 $\pm$ 0.015  & 4603 $\pm$  646 & 0.907 $\pm$ 0.106  &  1.538 $\pm$ 0.320 \\
2001b\_11 & 52238.0  & 6152 $\pm$  261  & 1.127 $\pm$ 0.027  & --     --  & --     --  &     3.331 $\pm$ 0.007  \\
\hline
2000-2001&& 5123 $\pm$ 1199&1.205 $\pm$ 0.224 & 4380  $\pm$ 1195&0.854 $\pm$ 0.107&2.565$\pm$1.193\\
\hline
2002b\_01 & 52299.4 &  6183 $\pm$  217  & 1.284 $\pm$ 0.006  & 4785 $\pm$  903 & 0.957 $\pm$ 0.134   & 3.680  --\\
2002r\_03 & 52345.8  & --     --   & --     --     & 4603  $\pm$ 775 & 1.173 $\pm$ 0.124   & 3.348 $\pm$ 0.862 \\
2002\_3\_4 & 52357.3  & 6060 $\pm$  479  & 1.706 $\pm$ 0.239 &  --     --  & --     -- &      3.645 $\pm$ 0.420 \\
2002r\_04 & 52368.7 &  --     --   & --     --     & 5150 $\pm$  838 & 1.196 $\pm$ 0.076 &   3.942  --\\
2002b\_05 & 52398.7  & 6336 $\pm$  173  & 1.369 $\pm$ 0.050  & 5605  $\pm$ 581 & 1.203 $\pm$ 0.005 &   3.283 $\pm$ 0.021 \\
2002b\_06 & 52439.0  & 7012 $\pm$  521  & 1.285 $\pm$ 0.088  & 5582 $\pm$  226  &1.414 $\pm$ 0.115  &  3.903 $\pm$ 0.292  \\
2002b\_12 & 52621.0  & 6121 $\pm$  305 &  1.489 $\pm$ 0.051  & 5970 $\pm$  387 & 1.374 $\pm$  0.129  &  3.467 $\pm$ 0.157 \\
\hline
2002&& 6490 $\pm$ 465&1.427 $\pm$ 0.177 & 5283  $\pm$ 528&1.220 $\pm$ 0.163&3.629$\pm$0.243\\
\hline
2003b\_01 & 52665.9  & 5905 $\pm$  348  & 1.710 $\pm$ 0.112  & 5651  $\pm$ 129 & 1.373 $\pm$ 0.117  &  5.072  --\\
2003b\_03 & 52723.8  & 6336 $\pm$  434  & 1.344 $\pm$ 0.066  & --     --  & --     --      & 4.541  --\\
2003b\_05 & 52777.1  & 5906 $\pm$   87  & 1.559 $\pm$ 0.009  & 5287 $\pm$  324  &1.260 $\pm$ 0.027   & 5.283 $\pm$ 0.205 \\
2003b\_06 & 52813.7  & 5660$ \pm$  349  & 1.630 $\pm$ 0.050  & 5013 $\pm$  388  &1.163 $\pm$ 0.038 &   4.184  --\\
2003b\_11 & 52966.8  & 5505 $\pm$  217  & 1.358 $\pm$ 0.082  & --     -- &  --     --   &    3.478 $\pm$ 0.193  \\
2003b\_12 & 52995.6  & 5414 $\pm$  523  & 1.352 $\pm$ 0.083  & --     --  & --     --     &  3.515  --\\
\hline
2003&& 5788 $\pm$ 336&1.492 $\pm$ 0.162 & 5317  $\pm$ 320&1.265 $\pm$ 0.105&4.346$\pm$0.763\\
\hline
2004b\_01 & 53019.5  & 5351 $\pm$  261  & 1.456 $\pm$ 0.124  & --     --  & --     --     &  3.851  --\\
2004b\_12 & 53349.0  & 6889 $\pm$  173 &  1.240 $\pm$ 0.057  & 5879 $\pm$  389  &1.422 $\pm$ 0.076  &  2.932 $\pm$ 0.105 \\
\hline
2004&& 6120 $\pm$ 1088&1.348 $\pm$ 0.153 & 5879  $\pm$ 389&1.422 $\pm$ 0.076&3.392$\pm$0.650\\
\hline
2005\_1\_2 & 53402.8 &  7013 $\pm$  783  & 1.332 $\pm$ 0.076  & --     --  & --     --      & 2.218 $\pm$ 0.308 \\
2005b\_02 & 53417.5  & --     --   & --     --     & 6586 $\pm$  356 & 1.359 $\pm$ 0.072   & 2.040  --\\
2005b\_05 & 53505.3  & 8305 $\pm$  262  & 1.093 $\pm$ 0.043  & 6882 $\pm$  582 & 1.265 $\pm$ 0.105   & 2.140 $\pm$ 0.037\\
2005b\_06 & 53535.0  & 8427 $\pm$  348  & 0.957 $\pm$ 0.041   &7635 $\pm$  357 & 1.100 $\pm$ 0.058  &  1.841 $\pm$ 0.142 \\
2005b\_11 & 53704.0 &  6059  $\pm$ 130  & 1.402 $\pm$ 0.031  & --     --  & --     --      & 3.567  --\\
2005b\_12 & 53711.5  & 5967 $\pm$   88  & 1.395 $\pm$ 0.007  & --     --  & --     --      & 2.911 $\pm$ 0.177\\
\hline
2005&& 7154 $\pm$ 1180&1.236 $\pm$ 0.200 & 7034  $\pm$ 540&1.241 $\pm$ 0.131&2.953$\pm$0.656\\
\hline
2006b\_01 & 53762.4 &  6797 $\pm$  218  & 1.188 $\pm$ 0.021   &6015 $\pm$  710 & 1.449 $\pm$  0.202  &  1.939 $\pm$ 0.179 \\
2006b\_02 & 53788.0  & 6582 $\pm$  174  & 1.161 $\pm$ 0.027  & 5925  $\pm$ 518 & 1.403 $\pm$ 0.145   & 1.926 $\pm$ 0.016\\
2006b\_03 & 53816.9 &  6704 $\pm$   86  & 1.117 $\pm$ 0.002  & 5446 $\pm$  613 & 1.546 $\pm$ 0.222  &  2.894 $\pm$ 0.293 \\
2006b\_04 & 53845.1  & 5691 $\pm$  305  & 1.373 $\pm$ 0.044   &5401 $\pm$  356 & 1.480 $\pm$ 0.069  &  3.390 $\pm$ 0.103 \\
\hline
2006&& 6444 $\pm$ 509&1.210 $\pm$ 0.113 & 5697  $\pm$ 318&1.470 $\pm$ 0.060&2.537$\pm$0.727\\
\hline
\hline
\hline
mean 96-06&& 6021 $\pm$ 851&1.187 $\pm$ 0.237 & 5164  $\pm$ 873&1.072 $\pm$ 0.245&\\
\hline
\hline
Paper I&& 6110 $\pm$ 440&1.056 $\pm$ 0.018 & 4650  $\pm$ 420&1.00 $\pm$ 0.023&\\
\hline
\end{longtable}
}

\onlfig{9}{
\centering
\begin{figure*}[]
\includegraphics[width=14cm]{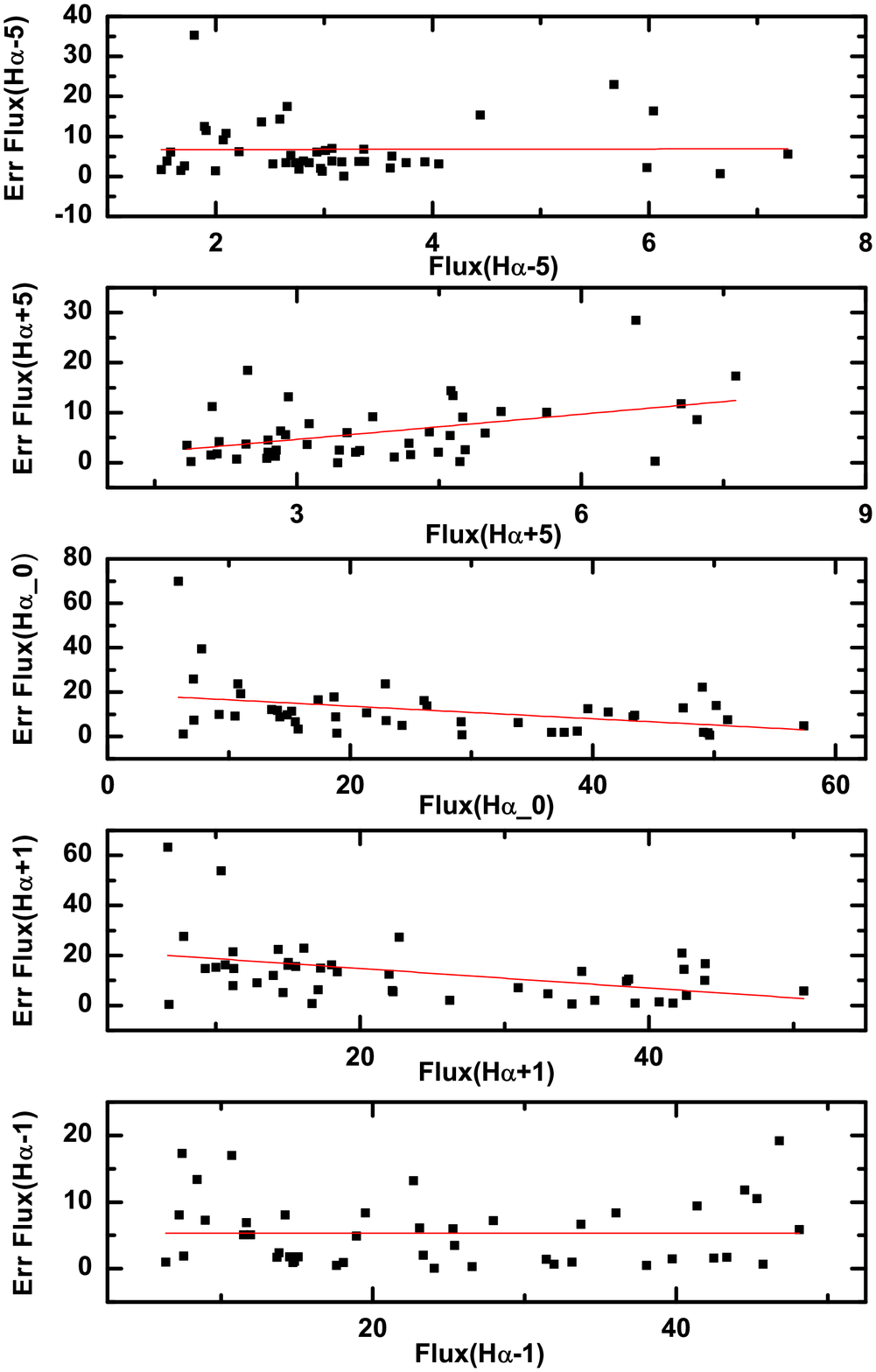}
\caption{The flux error measurements (in percent of part line flux) against flux for -5, -1, 0, 1, 5 segments of the
H$\alpha$ line. The flux is in units of 10$^{-13}\rm erg cm^{-2}sec^{-1}$.} \label{fig9}
 \end{figure*}
}

\begin{figure*}
\centering
\includegraphics[width=9.5cm]{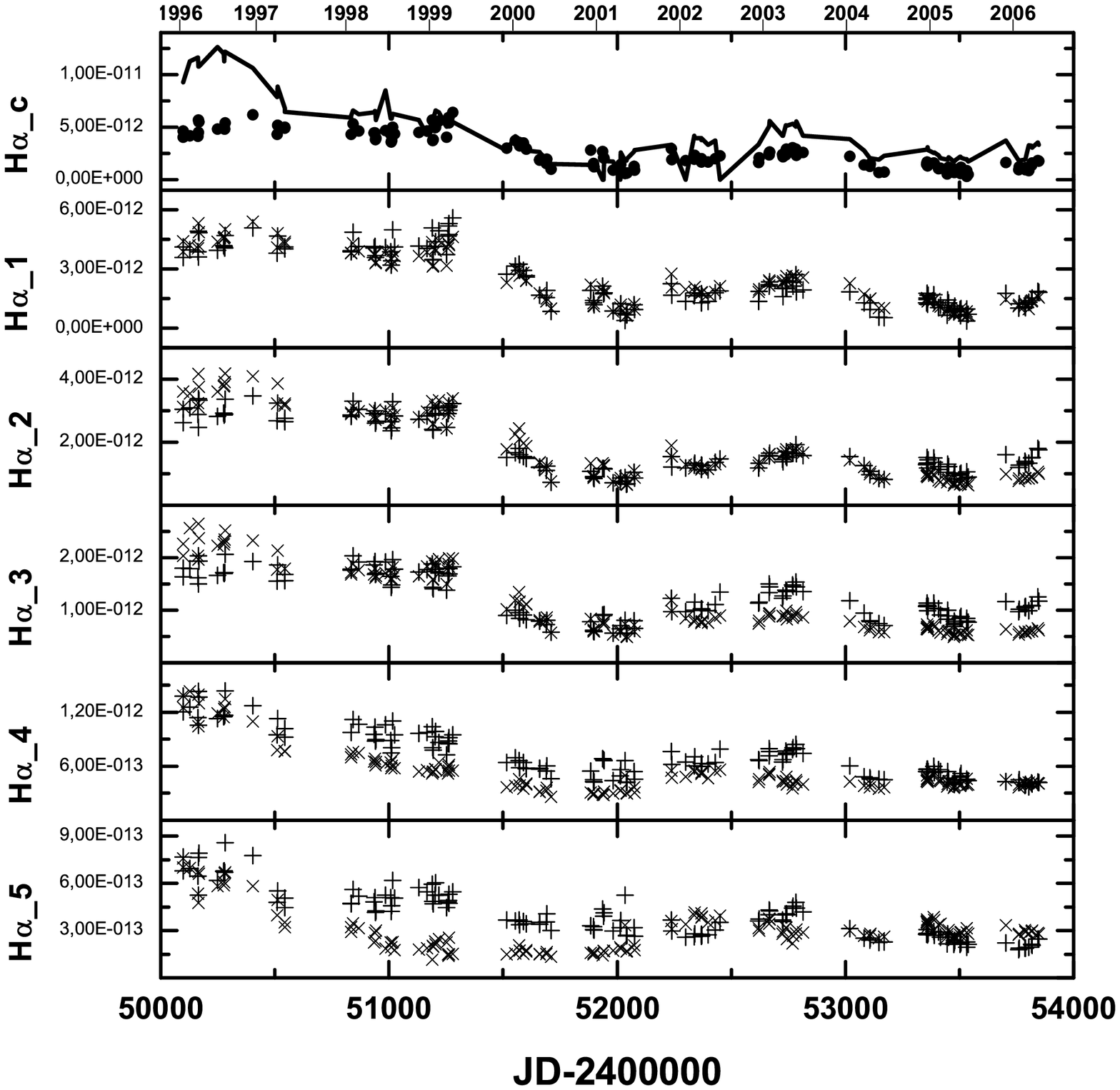}
\includegraphics[width=9.5cm]{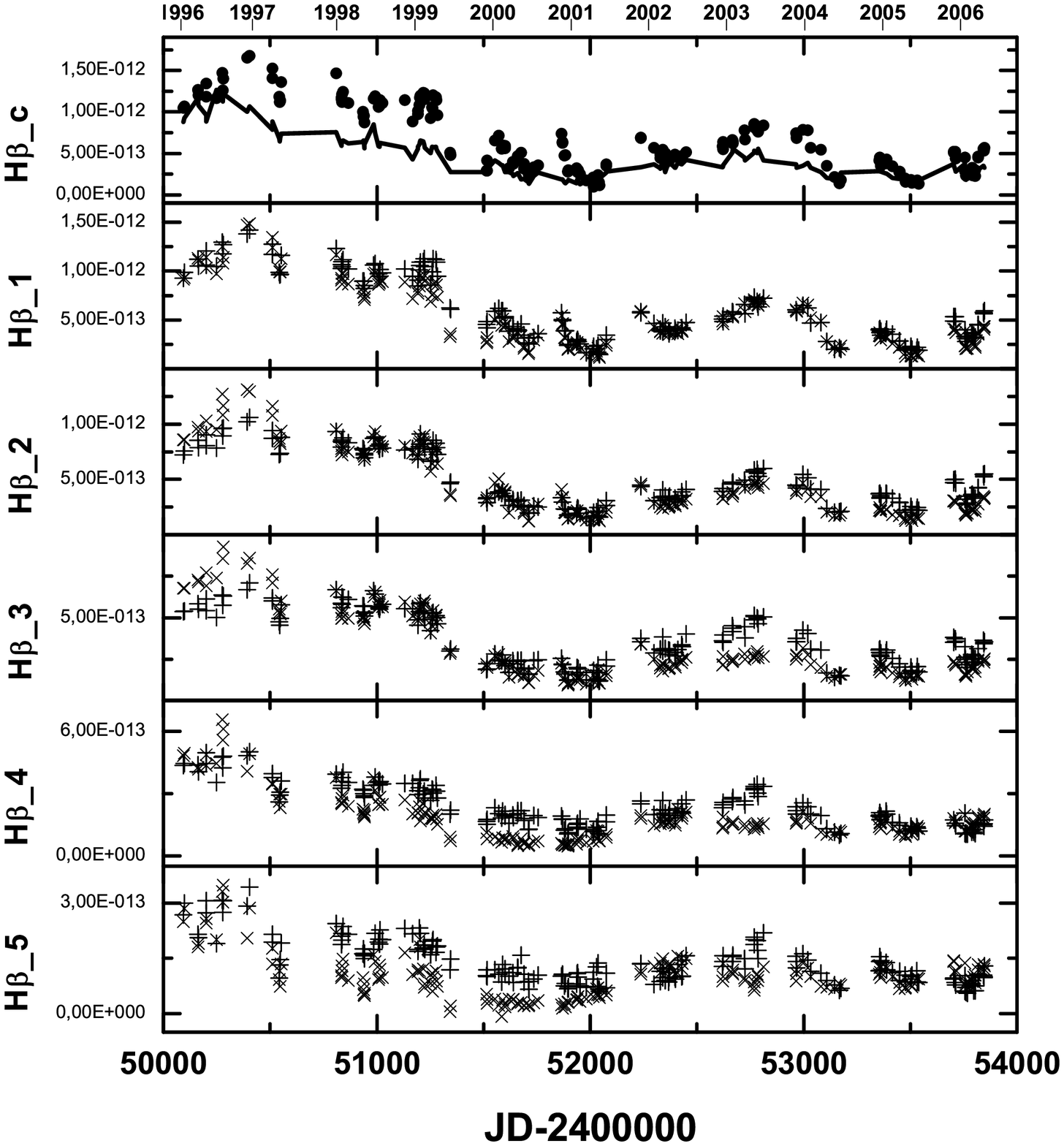}
\caption{Light curves of the  different H$\alpha$ (upper panel) and
H$\beta$ (bottom panel) line segments.
          The segments in the blue wing (numbers from -5 to -1 from Table \ref{tab3}) are
          marked with crosses ($\times$) and in the red wing (numbers from +5 to +1
          from Table \ref{tab3}) with plus ($+$). The abscissae gives the Julian date
          (bottom) and the corresponding year (top). The ordinate gives the flux in units
           $\rm erg\,cm^{-2}$\,s$^{-1}$.
           The variation of the continuum flux, that is scaled by different factors
           (100 and 10 times for H$\alpha$ and H$\beta$, respectively) in order to be
comparable with the variation of the central part of the line, is
presented with the solid line. }\label{fig10}
\end{figure*}

\begin{figure*}[]
\includegraphics[width=6cm]{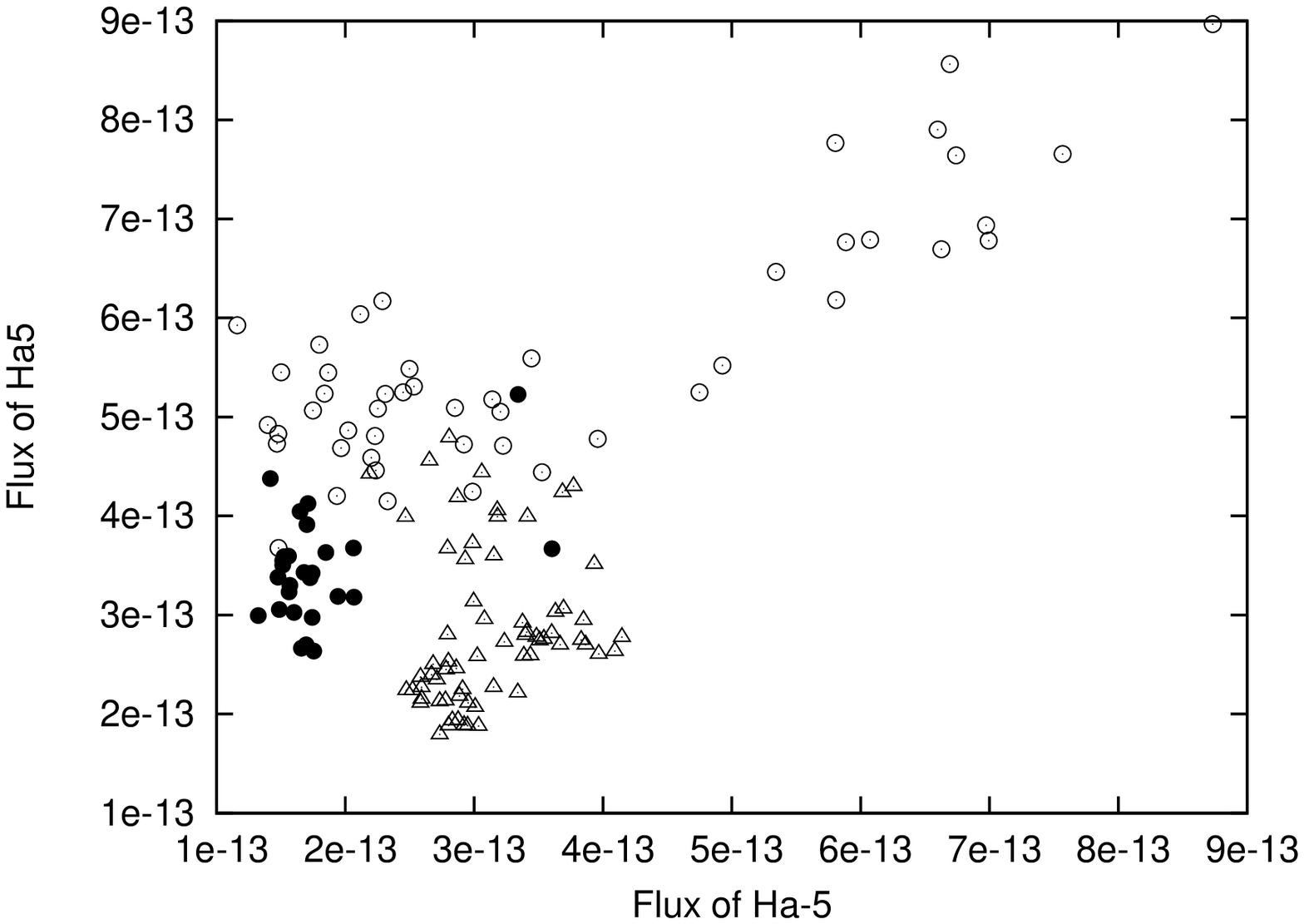}
\includegraphics[width=6cm]{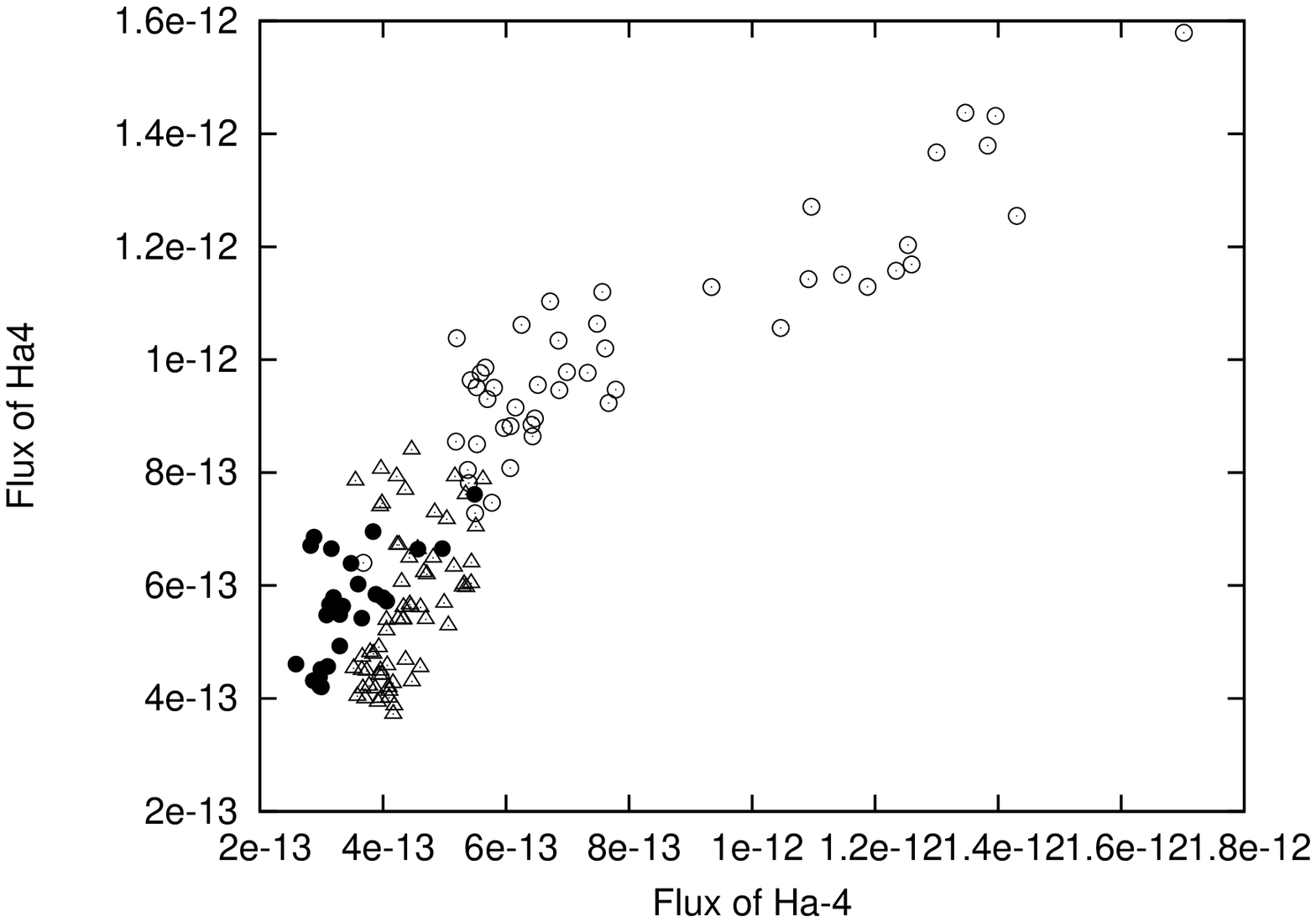}
\includegraphics[width=6cm]{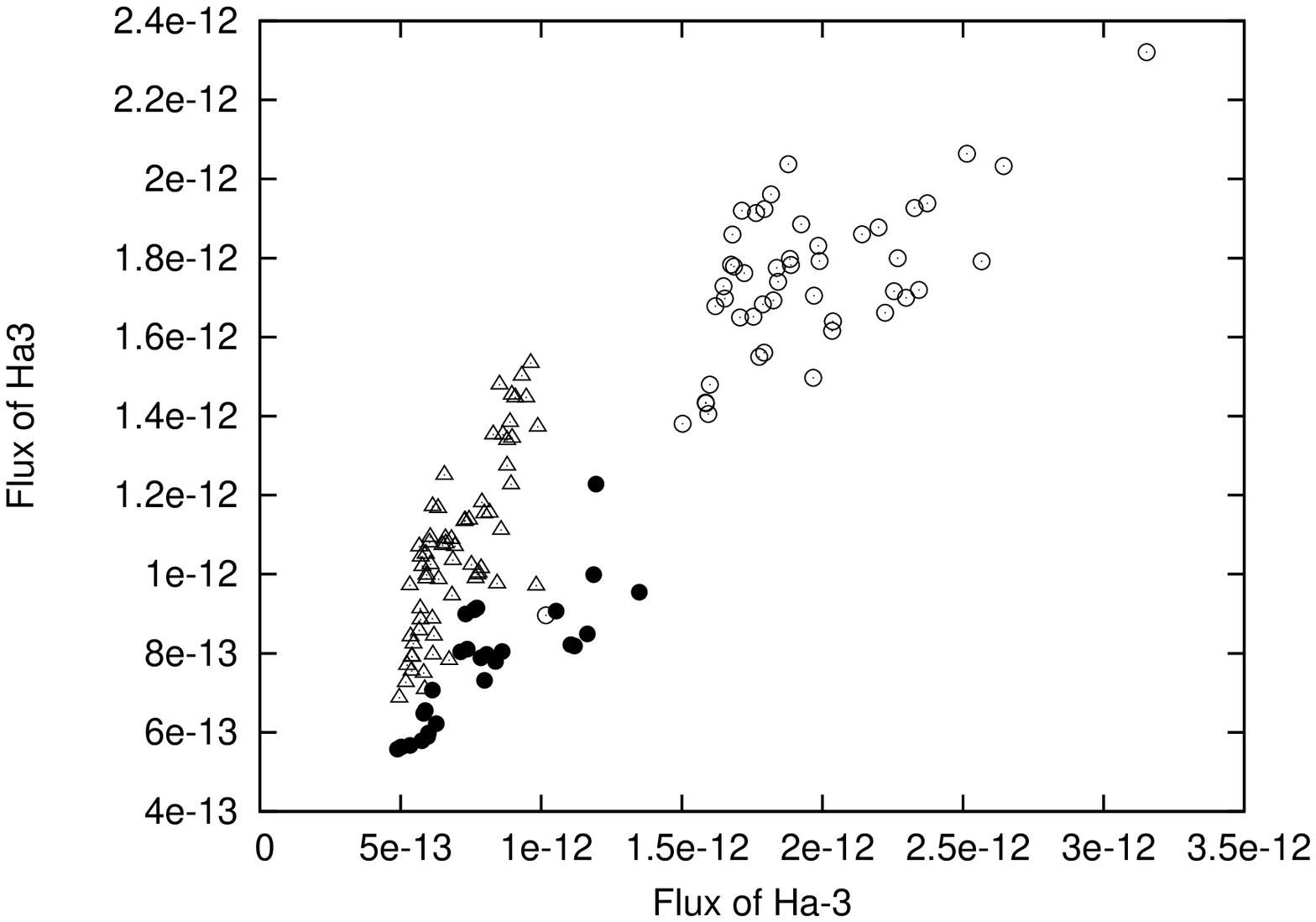}
\includegraphics[width=8cm]{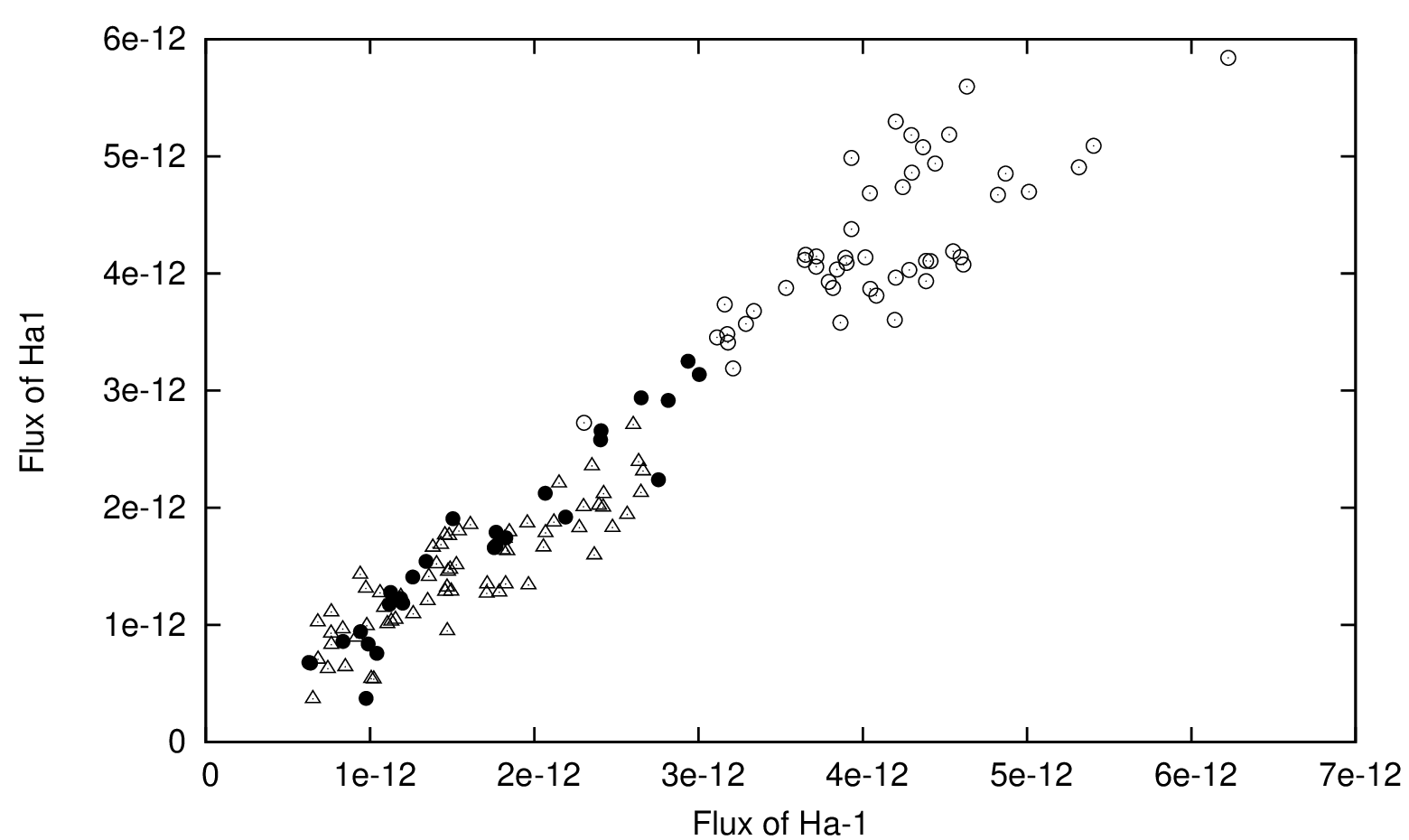}
\includegraphics[width=8cm]{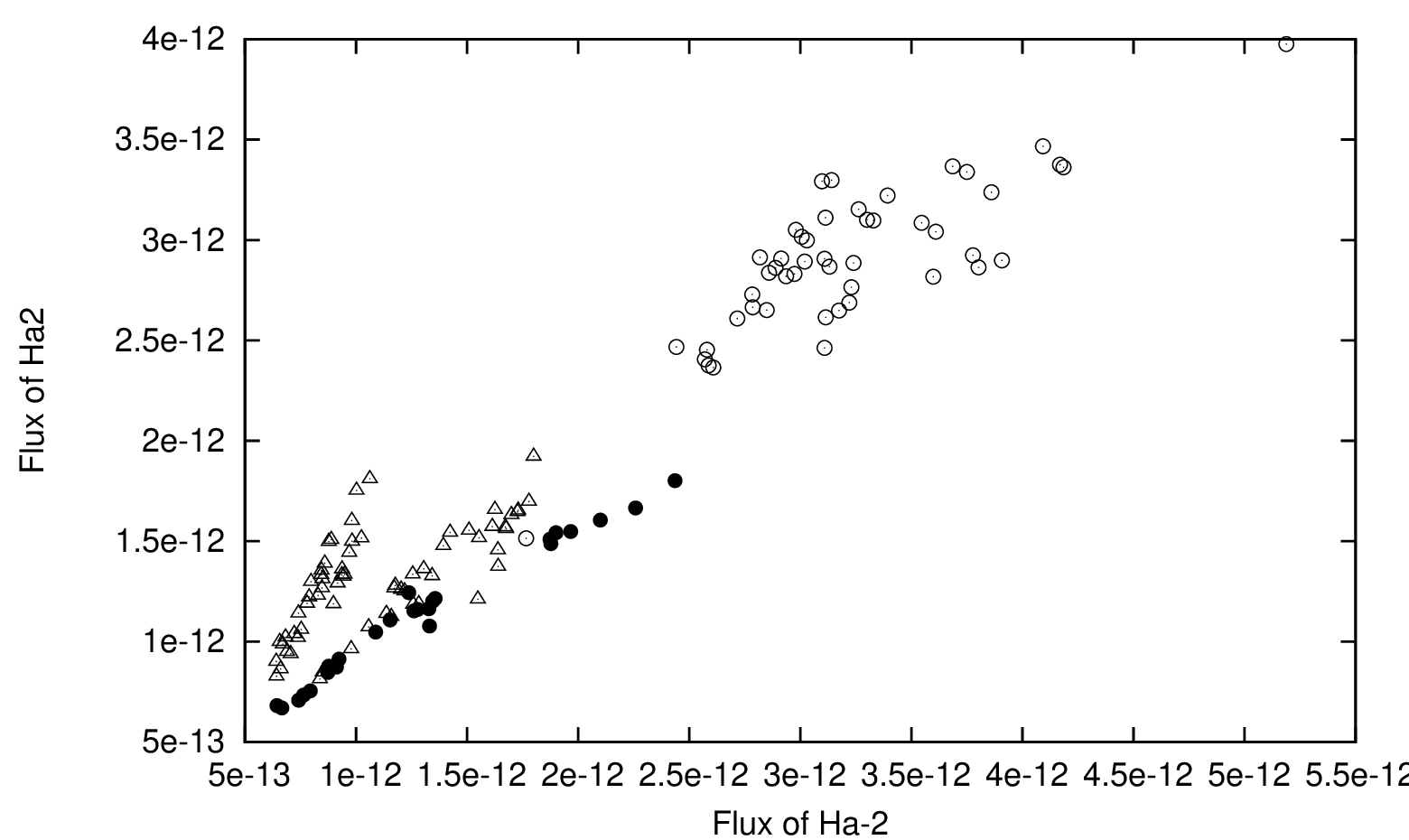}
\caption{The segment to segment response, where the first period
(Period I, 1996--1999) is denoted with open circles, the second
period (Period II, 2000--2001) with full ones, and the third period
(Period III, 2002--2006) with open triangles (Paper I). The flux is
given in $\rm erg \ cm^{-2} \ s^{-1}$.} \label{fig11}
 \end{figure*}


\begin{table*}
\begin{center}
\caption[]{The beginning and ending radial velocities, $V_{\rm beg}$
and $V_{\rm end}$, in km/s for different segments in the line
profiles.} \label{tab3}
\begin{tabular}{llllllllllll}
\hline
\hline
\backslashbox[0pt][l]{Vr}{segment} & -5    &   -4  &  -3   &  -2   &  -1  &  0(C) & +1  &   +2  &   +3  &  +4  &  +5   \\
\hline
$V_{\rm beg}$   &  -5500  & -4500 & -3500 & -2500 & -1500 & -500 &  500 & 1500 &  2500 &  3500 & 4500   \\
$V_{\rm end}$   &  -4500  & -3500 & -2500 & -1500 &  -500 &  500 & 1500 & 2500 &  3500 &  4500 & 5500    \\
\hline

\end{tabular}
\end{center}
\end{table*}

The FWHM of H$\beta$ was almost always larger than the H$\alpha$ one
(see upper panel of Fig. \ref{fig8}). The asymmetry of both lines
(middle panel of Fig. \ref{fig8}) has been gradually increasing from
1996 to 2006 and it slightly anticorrelates with the variations of
the continuum (bottom panel in Fig. \ref{fig8}). The largest values
of the FWHM and an outstanding red asymmetry of both lines (A$>$1.2)
was observed in 2002--2006. We calculated average values of the FWHM
and A for each year, as well as for the whole monitoring period
(they are given in Table \ref{tab2}). FWHMs and asymmetries obtained
in this work from measurements of the month-average profiles are
similar to the results given in  Paper I and difference between them
is  within the error bars. As one can see from Table \ref{tab2} the
FWHM of both lines was varying rather considerably from year to year
($\Delta$FWHM$\sim 500-1500 \ {\rm km \ s^{-1}}$ ). These lines were
the narrowest in 2000-2001 (FWHM $\sim4000-5000\ {\rm km \ s^{-1}}$)
and the broadest in 2005 ($\sim7000\ {\rm km \ s^{-1}}$) (see Table
\ref{tab2}). At the same time the H$\beta$ FWHM was always broader
than  H$\alpha$ by  $\sim1000 \ {\rm km \ s^{-1}}$ on average. The
asymmetry was varying in different ways: in 1996--1997 the blue
asymmetry was observed in H$\beta$ (A$\sim$0.85) when the continuum
flux was maximum; in 1998--2000 H$\beta$ was almost symmetric, and
from 2001 to 2006 the red asymmetry appeared (A$>$1.2). In
1996--2001, the blue symmetry (A$\sim$0.8) was observed in
H$\alpha$, and in 2002--2006 it was a red asymmetry (A$>$1.2).

{  Also, we tried to find correlations between the FWHM and A with
the continuum flux and found that the FWHM practically does not
correlate with the continuum (r$\sim0.0;-0.1$). In the case of the
asymmetry A, there is an indication of anticorrelation, but it
should be also taken with caution since there is a large scatter of
points on A vs. continuum flux plane, especially in the case of low
continuum fluxes $ F_{\rm c}\ < \ 5.3 \times 10^{-14} \rm erg \
cm^{-2}s^{-1}\AA^{-1}$ when the measured asymmetry reaches the
highest values.} Note here that photoinoization model predicts that
the Balmer lines should be broader in lower continuum states and
narrower in the higher continuum states (see Korista \& Goad 2004),
as due to larger response in the line cores one can expect that
Balmer lines became narrower in higher continuum states. As one can
see from Table \ref{tab2} there is no trend that FWHM is
significantly narrower in the high continuum state.

\subsection{Light curves of different line segments}

  In Paper I we obtained light curves for the integrated flux in lines
and continuum. To study the BLR in more details, we are starting in
this paper from the fact that a portion of the broad line profile
can respond to variations of the continuum in different ways.
Therefore, we divided the line profiles into 11 identical profile
segments, each of width $\sim$1000 $\ {\rm km \ s^{-1}}$ (see Table
\ref{tab3}).

The observational uncertainties  were determined for each segment of
the H$\beta$ and H$\alpha$ light curves. In the uncertainties we
include the uncertainties due to: the position angle correction,
seeing correction procedure and aperture correction. The methods to
evaluate these uncertainties (error-bars) are given in Paper I. The
effect of the subtraction of the template spectrum (or the narrow
components) has been studied by comparing the flux of pairs of
spectra obtained in the time interval from 0 to 2 days. In Table
\ref{tab4} (available electronically only) we presented the
year-averaged uncertainties (in percent) for each segment of
H$\alpha$ and H$\beta$ and mean values for all segments {  and
corresponding mean-year flux}. To determine the error-bars we used
44 pairs of H$\alpha$ and 68 pairs of H$\beta$. As one can see from
Table \ref{tab4}, for the far wings (segments $\pm$5) the error-bars
are greater ($\sim$10\%) in H$\beta$ than in H$\alpha$ ($\sim$6\%).
But when comparing the error-bars in the far red and blue wings, we
find that the error-bars are similar. Also, higher error-bars can be
seen in the central part of the H$\alpha$ due to the narrow line
subtraction. Fig. \ref{fig9} (available electronically only) shows
the distributions of the error-bars as a function of the line flux
for segments ha$\pm$5 and ha0 and ha$\pm$1. It can be seen that in
the case of ha0 and ha+1 there is a slight anticorrelation with flux
and two points, corresponding to the very small flux
(F$<4\cdot$10$^{-13}\rm erg/cm^{-2}sec^{-1}$ in 2005), have the
highest error-bar of (40-70)\%.

We constructed light curves for each segment of  the
H$\alpha$ and H$\beta$ lines. Fig. \ref{fig10} presents light curves
of profile segments in approximately identical velocity intervals in
the blue and red line wings (segments from 1 to 5, where larger
number corresponds to higher velocity, see Table \ref{tab3}) and for
the central part (0 in Table \ref{tab3} or H$\alpha$\_c, H$\beta$\_c
in Fig. \ref{fig10}, corresponding to the interval $\sim\pm 500\ {\rm
km \ s^{-1}}$).  To compare the segment variation with the continuum
we plot (as a solid line) the continuum flux
variation into central part (see Fig. \ref{fig10}).

\onltab{4}{
\begin{table*}
\caption[]{\label{tab4}The errors of measurements (e$\pm \sigma$)
for {  all line segments (see Table \ref{tab3}) of H$\alpha$ and
H$\beta$ given in percents. Also, for each segment the mean-year
flux is given in units $10^{-13} \ \rm erg \ cm^{-2} s^{-1} $.}}
\begin{tabular}{ccccccccccccc}
\hline\hline
 Year & \multicolumn{2}{c}{Ha(-5)}  &  \multicolumn{2}{c}{Ha(+5)}  & \multicolumn{2}{c}{Ha(-4)}   & \multicolumn{2}{c}{Ha(+4)} &
\multicolumn{2}{c}{Ha(-3)} & \multicolumn{2}{c}{Ha(+3)}\\
   & Flux & (e$\pm \sigma$) & Flux  &(e$\pm \sigma$)& Flux & (e$\pm \sigma$)& Flux & (e$\pm \sigma$)& Flux & (e$\pm \sigma$) & Flux & (e$\pm \sigma$)\\
 \hline
 1996  & 6.327  &  9.6$\pm$ 9.7  & 7.052  & 13.3$\pm$10.5  & 12.456  &  9.9$\pm$5.9  & 12.494   & 11.7$\pm$7.1  & 22.736 &  9.1$\pm$6.8  & 17.722  & 10.9$\pm$7.1\\
 1997  & 3.904  & 11.1$\pm$ 6.1  & 4.950  &  9.7$\pm$ 0.8  &  8.100  &  6.7$\pm$8.8  &  10.048  & 9.7$\pm$3.8   & 18.741 &  6.7$\pm$9.2  & 16.634  & 9.1$\pm$5.3\\
 1998  & 2.497  & 11.7$\pm$ 4.2  & 4.516  &  7.2$\pm$ 6.6  &  6.381  &  5.4$\pm$2.4  &  8.904   & 7.4$\pm$4.8   & 16.570 &  3.2$\pm$2.9  & 16.518  & 4.6$\pm$4.7\\
 1999  & 2.030  & 14.3$\pm$18.3  & 5.133  &  6.2$\pm$ 3.8  &  5.649  &  2.1$\pm$3.0  &  8.776   & 3.3$\pm$1.1   & 16.851 &  2.2$\pm$2.0  & 16.470  & 2.1$\pm$1.0\\
 2000  & 1.647  &  5.3$\pm$ 4.3  & 3.460  &  4.3$\pm$ 3.5  &  3.561  &  4.4$\pm$4.4  &  5.570   & 4.2$\pm$3.5   &  9.204 &  4.2$\pm$3.8  &  7.770  & 3.8$\pm$3.6\\
 2001  & 2.251  &  5.5$\pm$ 5.3  & 2.906  &  7.3$\pm$ 6.2  &  3.671  &  4.7$\pm$6.0  &  5.220   & 5.9$\pm$6.1   &  7.249 &  7.3$\pm$6.0  &  7.760  & 8.1$\pm$7.0\\
 2002  & 3.686  &  3.5$\pm$ 0.4  & 3.016  &  3.0$\pm$ 1.3  &  5.041  &  2.7$\pm$2.4  &  6.443   & 4.9$\pm$5.3   &  7.792 &  3.6$\pm$1.4  & 10.742  & 4.1$\pm$5.1\\
 2003  & 3.004  &  5.7$\pm$ 5.0  & 4.186  &  2.9$\pm$ 1.7  &  4.447  &  4.9$\pm$2.4  &  7.560   & 3.4$\pm$1.3   &  9.104 &  4.0$\pm$2.3  & 14.034  & 2.9$\pm$2.1\\
 2004  & 3.261  &  3.2$\pm$ 0.8  & 2.722  &  3.3$\pm$ 1.8  &  4.340  &  6.1$\pm$2.3  &  5.299   & 1.9$\pm$1.3   &  6.619 &  5.7$\pm$2.4  &  9.922  & 3.9$\pm$2.1\\
 2005  & 2.908  &  4.3$\pm$ 1.3  & 2.350  &  7.1$\pm$ 6.7  &  4.122  &  4.4$\pm$3.4  &  4.687   & 5.9$\pm$6.2   &  5.756 &  4.5$\pm$3.1  &  8.575  & 5.5$\pm$4.3\\
 2006  & 2.906  &  1.7$\pm$ 0.4  & 1.938  &  1.7$\pm$ 1.7  &  3.979  &  3.0$\pm$2.7  &  4.115   & 3.0$\pm$1.9   &  5.784 &  2.7$\pm$2.8  & 10.459  & 1.9$\pm$1.3\\
\hline
mean  && 6.9$\pm$4.1 & &  6.0$\pm$3.4 & & 4.9$\pm$2.2 & &5.6$\pm$3.0 & &4.8$\pm$2.1 & &5.2$\pm$3.0\\
\hline
&&&&&&\\
 \hline
Year & \multicolumn{2}{c}{Hb(-5)}  &  \multicolumn{2}{c}{Hb(+5)}  & \multicolumn{2}{c}{Hb(-4)}   & \multicolumn{2}{c}{Hb(+4)} &
\multicolumn{2}{c}{Hb(-3)} & \multicolumn{2}{c}{Hb(+3)}\\
& Flux & (e$\pm \sigma$) & Flux  &(e$\pm \sigma$)& Flux & (e$\pm \sigma$)& Flux & (e$\pm \sigma$)& Flux & (e$\pm \sigma$) & Flux & (e$\pm \sigma$)\\

\hline
 1996  & 2.511 &   3.2$\pm$0.17  & 2.535 &    6.5$\pm$ 5.3  & 4.965 &  7.0$\pm$ 5    & 4.470 &  7.1$\pm$1.5  & 8.003 &  7.6$\pm$6.1 & 5.838 &  6.6$\pm$2.5\\
 1997  & 1.303 &  17.8$\pm$ 2.1  & 1.507 &  11.6$\pm$ 2.5   & 3.134 &   4.8$\pm$ 5.9 & 3.813 &  4.7$\pm$2.0  & 6.391 &  3.5$\pm$2.0 & 5.381 &  2.4$\pm$0.6\\
 1998  & 0.923 &   13.0$\pm$ 9.7 & 1.735 &    5.4$\pm$ 2.4  & 2.393 &   4.7$\pm$ 3.9 & 3.318 &  4.1$\pm$2.5  & 5.268 &  3.0$\pm$1.6 & 5.648 &  2.1$\pm$1.9\\
 1999  & 1.215 &  19.3$\pm$14.0  & 1.594 &   12.1$\pm$10.2  & 1.674 &  11.1$\pm$10.4 & 2.786 &  6.1$\pm$3.2  & 4.403 &  3.1$\pm$1.9 & 4.322 &  5.5$\pm$3.8\\
 2000  & 0.255 &   6.5$\pm$ 7.1  & 0.822 &  14.9$\pm$ 5.7   & 0.626 &  7.5$\pm$ 3.2  & 1.593 &  3.0$\pm$1.3  & 1.654 &  3.2$\pm$1.9 & 1.977 &  2.2$\pm$1.6\\
 2001  & 0.642 &  10.5$\pm$ 8.9  & 1.068 &   15.2$\pm$11.7  & 1.064 &  12.8$\pm$11.4 & 1.620 &  9.2$\pm$9.6  & 1.923 &  3.2$\pm$4.2 & 2.210 &  7.8$\pm$8.1\\
 2002  & 1.253 &   10.3$\pm$ 6.0 & 1.011 &  19.2$\pm$10.5   & 1.684 &   8.3$\pm$ 6.1 & 2.217 &  7.4$\pm$4.8  & 2.165 &  5.7$\pm$5.6 & 3.235 &  5.7$\pm$5.3\\
 2003  & 1.000 &   9.7$\pm$ 6.1  & 1.568 &    7.8$\pm$ 2.6  & 1.515 &   4.6$\pm$ 2.1 & 2.715 &  5.2$\pm$3.4  & 2.627 &  4.1$\pm$2.5 & 4.347 &  4.5$\pm$3.3\\
 2004  & 1.161 &    9.5$\pm$ 6.5 & 1.427 &   12.4$\pm$ 7.0  & 1.568 &   8.1$\pm$ 6.2 & 1.861 &  5.1$\pm$3.7  & 2.062 &  6.7$\pm$5.7 & 2.841 &  3.3$\pm$1.9\\
 2005  & 1.062 &    3.8$\pm$ 3.8 & 0.969 &   10.5$\pm$13.0  & 1.458 &   3.5$\pm$ 2.3 & 1.535 &  3.4$\pm$2.3  & 1.818 &  3.5$\pm$3.2 & 2.609 &  8.2$\pm$2.8\\
 2006  & 1.012 &   6.6$\pm$ 5.5  & 0.886 &    9.2$\pm$11.5  & 1.536 &   4.2$\pm$ 3.2 & 1.323 &  5.2$\pm$4.1  & 2.024 &  3.1$\pm$3.1 & 2.841 &  2.6$\pm$2.1\\
\hline
mean & &10.0$\pm$ 5.2& & 11.3$\pm$ 4.1& & 7.0$\pm$3.0& &5.5$\pm$1.8& & 5.2$\pm$3.0& & 4.6$\pm$2.3\\
\hline
&&&&&&\\
\hline
Year  &   \multicolumn{2}{c}{Ha(-2)}  &  \multicolumn{2}{c}{Ha(+2)}  &  \multicolumn{2}{c}{Ha(-1)}  &  \multicolumn{2}{c}{Ha(+1)}   &
 \multicolumn{2}{c}{Ha(0)}\\
 & Flux & (e$\pm \sigma$) & Flux  &(e$\pm \sigma$)& Flux & (e$\pm \sigma$)& Flux & (e$\pm \sigma$)& Flux & (e$\pm \sigma$) \\
\hline
 1996 & 36.664  & 9.4$\pm$6.8   & 29.766   &   11.0$\pm$7.0   & 45.469 &  9.1$\pm$6.8  & 42.018 &  11.6$\pm$7.6  & 48.583 &  11.1$\pm$7.6  &   & \\
 1997 & 33.714  & 7.0$\pm$8.2   & 28.349   &    8.1$\pm$7.1   & 43.930 &  6.8$\pm$7.1  & 41.554 &  7.9$\pm$9.2   & 48.475 &   7.4$\pm$8.0  &   & \\
 1998 & 27.717  & 3.5$\pm$3.2   & 26.397   &   4.6$\pm$4.0    & 34.577 &  3.5$\pm$3.8  & 36.438 &  6.4$\pm$5.5   & 39.903 &  7.3$\pm$4.9   &   & \\
 1999 & 29.420  & 1.7$\pm$2.1   & 27.907   &    1.4$\pm$0.8   & 37.880 &  1.5$\pm$0.1  & 42.658 &  3.5$\pm$2.6   & 48.223 &   2.5$\pm$2.2  &   & \\
 2000 & 15.153  & 3.9$\pm$4.2   & 12.928   &   4.4$\pm$3.7    & 19.099 &  4.6$\pm$3.9  & 20.783 &   6.6$\pm$5.7  & 22.577 &   7.5$\pm$6.8  &   & \\
 2001 & 11.216  & 8.1$\pm$6.3   &  9.774   &    8.0$\pm$6.2   & 13.240 &  10.4$\pm$8.4 & 11.934 &  10.1$\pm$8.4  & 13.292 &   16.2$\pm$13.0&   & \\
 2002 & 12.285  & 4.1$\pm$1.1   &  12.612  &     6.0$\pm$2.1  & 18.827 &  4.7$\pm$3.8  & 16.231 &  14.9$\pm$8.3  & 19.611 &  12.4$\pm$4.7  &   & \\
 2003 & 16.800  & 2.5$\pm$2.2   &  16.071  &     4.6$\pm$4.5  & 24.724 &  3.6$\pm$2.5  & 21.440 &  13.5$\pm$9.0  & 25.761 &    9.8$\pm$4.9 &   & \\
 2004 &  9.552  & 4.8$\pm$1.5   &  12.490  &    9.1$\pm$3.2   & 14.567 &  1.4$\pm$0.4  & 13.621 &  18.0$\pm$4.9  & 14.724 &   9.7$\pm$2.5  &   & \\
 2005 &  7.416  & 4.3$\pm$3.2   &  10.705  &    9.8$\pm$5.3   & 9.008  &  8.3$\pm$6.2  & 9.852  &  32.0$\pm$21.3 & 8.791  &   26.0$\pm$25.4&   & \\
 2006 &  8.489  & 2.6$\pm$2.9   &  13.613  &    2.6$\pm$2.0   & 12.402 &  4.2$\pm$1.6  & 12.362 &  9.3$\pm$4.9   & 12.101 &  12.8$\pm$5.6  &   & \\
\hline
mean& &4.7$\pm$2.4 & &  6.3$\pm$3.1 & & 5.3$\pm$3.0& & 12.2$\pm$7.8 & & 11.2$\pm$6.1 &&\\
\hline
&&&&&\\
\hline
Year  &   \multicolumn{2}{c}{Hb(-2)}  &  \multicolumn{2}{c}{Hb(+2)}  &  \multicolumn{2}{c}{Hb(-1)}  &  \multicolumn{2}{c}{Hb(+1)}   &
 \multicolumn{2}{c}{Ha(0)}\\
 & Flux & (e$\pm \sigma$) & Flux  &(e$\pm \sigma$)& Flux & (e$\pm \sigma$)& Flux & (e$\pm \sigma$)& Flux & (e$\pm \sigma$) \\
\hline
 1996  & 10.383 &   7.0$\pm$4.5    & 8.673   &  6.2$\pm$1.5   & 11.245  & 6.3$\pm$4.5  & 11.531 & 6.7$\pm$2.2  & 12.884  &    7.8$\pm$3.9  &   &    \\
 1997  &  9.883&  3.7$\pm$2.3      & 8.188   &  3.3$\pm$3.1   & 11.509  & 4.1$\pm$2.5  & 10.986 & 4.0$\pm$3.1  & 13.094  &   4.9$\pm$1.1   &   &    \\
 1998  &  7.688&   2.7$\pm$2.5     & 7.905   &    2.1$\pm$1.4 &  8.515  & 2.7$\pm$2.9  &  9.689 & 1.7$\pm$1.4  & 10.530  &    1.8$\pm$2.1  &   &    \\
 1999  &  6.392&    4.1$\pm$1.9    & 6.609   &   5.1$\pm$2.9  &  6.829  & 6.3$\pm$3.9  &  8.664 & 4.8$\pm$3.2  &  9.542  &   6.6$\pm$6.8   &   &   \\
 2000  &  2.499&    2.6$\pm$2.0    & 2.698   &   1.7$\pm$1.0  &  3.105  & 3.9$\pm$2.5  &  3.844 & 3.0$\pm$1.7  &  4.193  &    5.5$\pm$4.6  &   &    \\
 2001  &  2.613&     4.4$\pm$2.5   & 2.716   &  5.6$\pm$5.6   &  3.220  & 4.5$\pm$4.0  &  3.367 & 2.5$\pm$1.3  &  3.772  &    8.0$\pm$8.5  &   &    \\
 2002  &  2.848&    5.0$\pm$4.2    & 3.575   &   4.9$\pm$3.8  &  4.058  & 5.3$\pm$4.2  &  4.222 & 5.3$\pm$4.2  &  4.873  &   6.8$\pm$4.6   &   &   \\
 2003  &  4.171&    4.2$\pm$2.1    & 4.950   &   3.0$\pm$1.5  &  6.619  & 2.9$\pm$2.9  &  6.203 & 3.4$\pm$3.0  &  7.277  &   4.2$\pm$3.2   &   &   \\
 2004  &  2.280&    5.4$\pm$2.5    & 3.476   &  3.3$\pm$2.7   &  3.402  & 4.6$\pm$4.6  &  3.801 & 3.5$\pm$3.1  &  4.153  &  3.9$\pm$5.3    &   &  \\
 2005  &  2.111&    4.0$\pm$3.4    & 3.379   &   5.0$\pm$2.4  &  2.790  &  4.4$\pm$2.5 &  3.395 & 7.7$\pm$2.7  &  3.193  &   9.1$\pm$4.1   &   &   \\
 2006  &  2.604&   2.5$\pm$2.5     & 3.930   &  1.2$\pm$0.9   &  3.007  & 2.0$\pm$1.7  &  4.105 & 2.3$\pm$1.8  &  3.686  &   3.5$\pm$4.0   &   &   \\
\hline
mean &  &  4.1$\pm$1.3 & & 3.8$\pm$1.7 & &   4.3$\pm$1.4 & & 4.1$\pm$1.9 & & 5.6$\pm$2.2 &&\\
\hline
\end{tabular}
\end{table*}
}

In 1996--1997 the blue segments 2 and 3 were slightly brighter than
the red ones, while the segments 1, 4 and 5 were similar to the red
ones. In 1998--2001 the blue segments 4 and 5 (3500--5500 $\ {\rm km
\ s^{-1}}$, regions the closest to the BH) and the segment 3 (in
2002--2006) of both lines were essentially fainter than the red
ones. In 1998--2004 the blue segments 1 and 2 (500 km/s -- 3500
km/s) were close to the corresponding red ones or slightly fainter
(see Fig. \ref{fig10}).

\subsection{The line-segment to line-segment flux
and continuum-line-segment relations}

The line-segment to line-segment flux and continuum-line-segment
relations for the H$\alpha$  and H$\beta$ lines are practically the
same, therefore here we present only results for the H$\alpha$ line.
First, we are looking for relations between the H$\alpha$ segments
which are symmetric with respect to the center (i.e. segments -1,
and 1; -2 and 2, ... -5 and 5). In Fig. \ref{fig11} we present the
response of symmetrical segments of H$\alpha$ to each other for the
three periods given in Paper I. {  As can be seen from Fig.
\ref{fig11} the symmetric segments are pretty well correlated, with
some notable exceptions: a) weaker response of the red to the blue
wing in the II and III period and b) the apparent bifurcation in the
Ha2/Ha-2 and Ha3/Ha-3 plots. This appears to be associated with
Period III,  related to the appearance of the ~3000 km/s "red
bump"}. This also supports that lines are probably formed in a
multi-component BLR and that the geometry of the BLR is changing
during the monitoring period.

\begin{figure*}[]
\includegraphics[width=6cm]{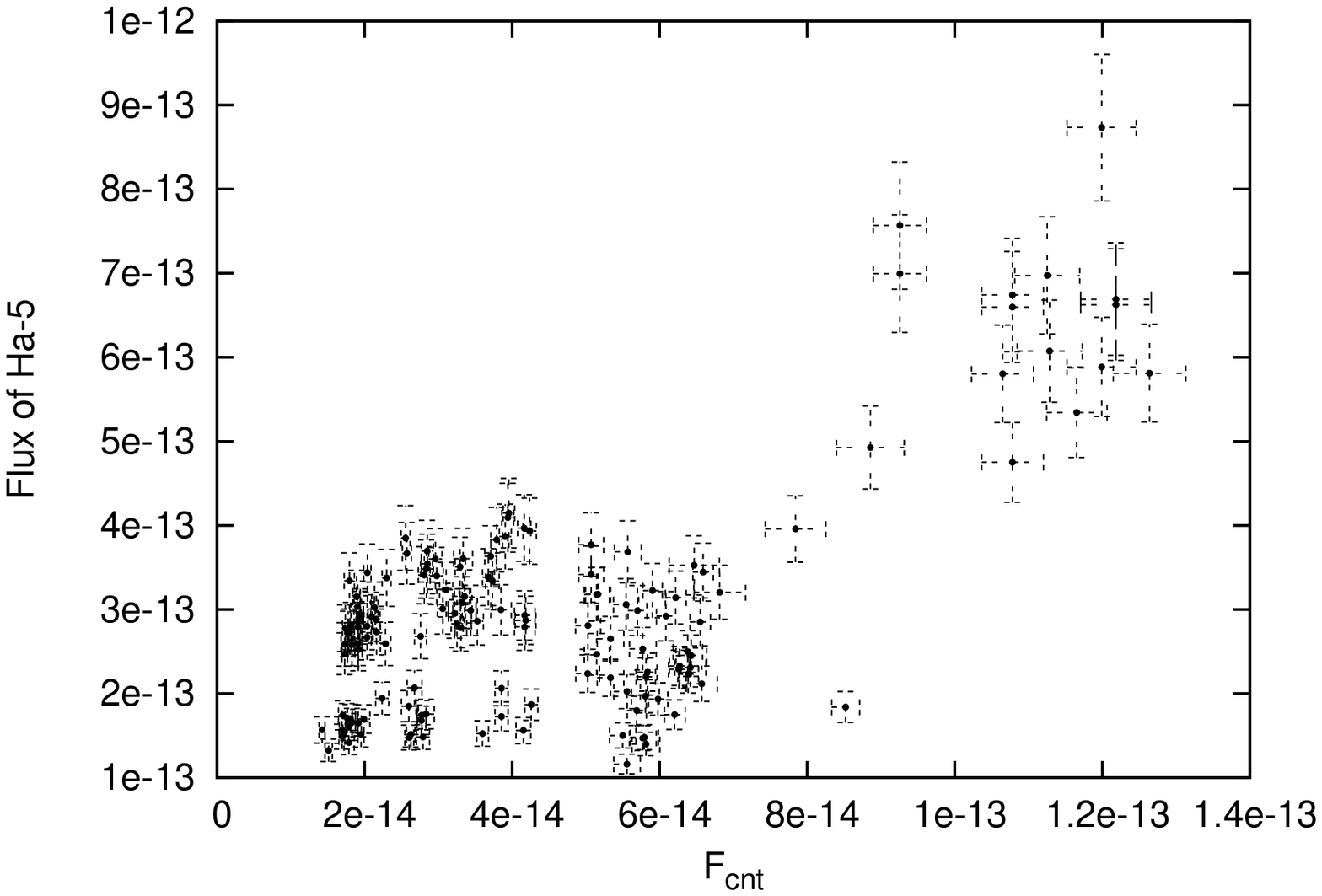}
\includegraphics[width=6cm]{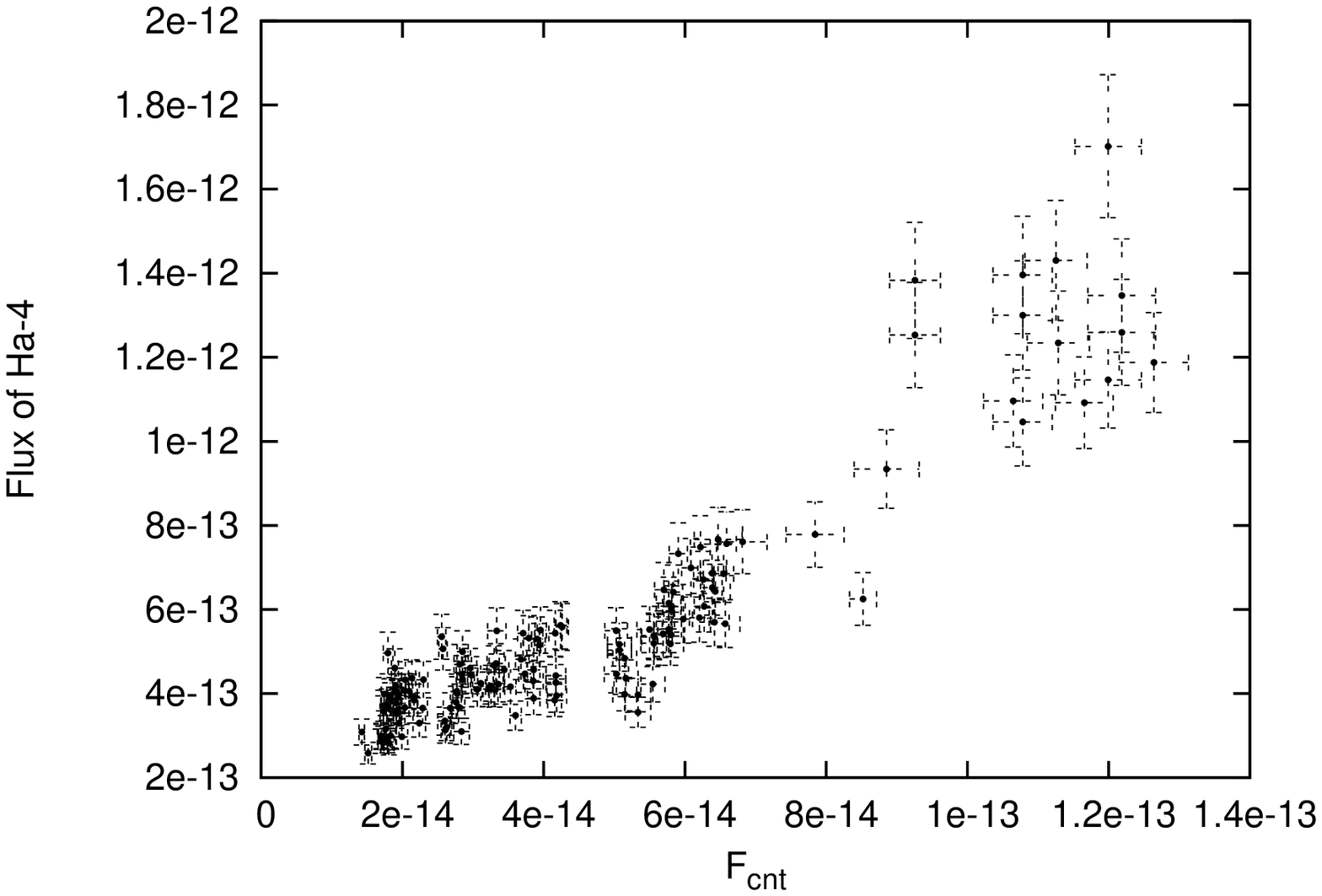}
\includegraphics[width=6cm]{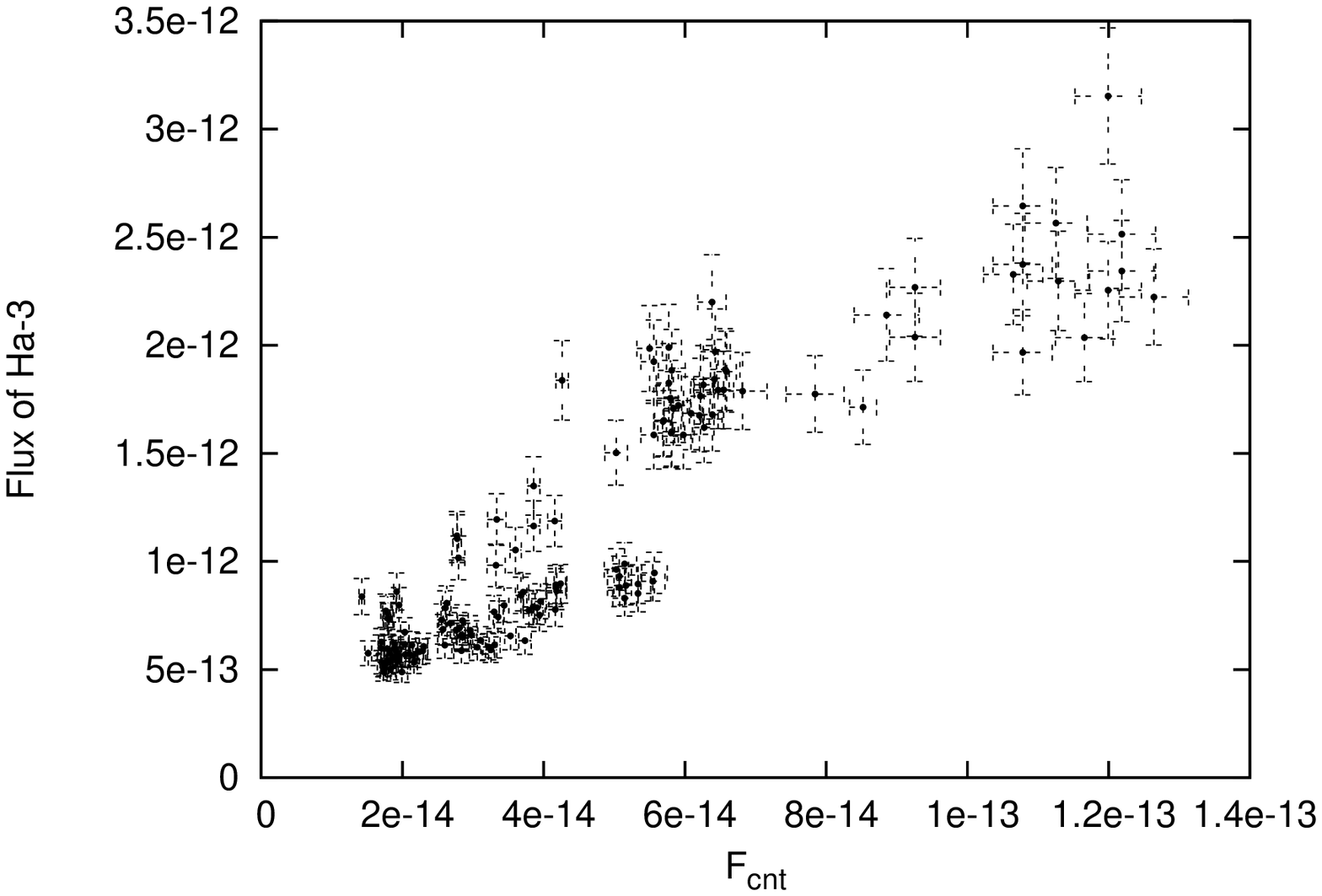}
\includegraphics[width=6cm]{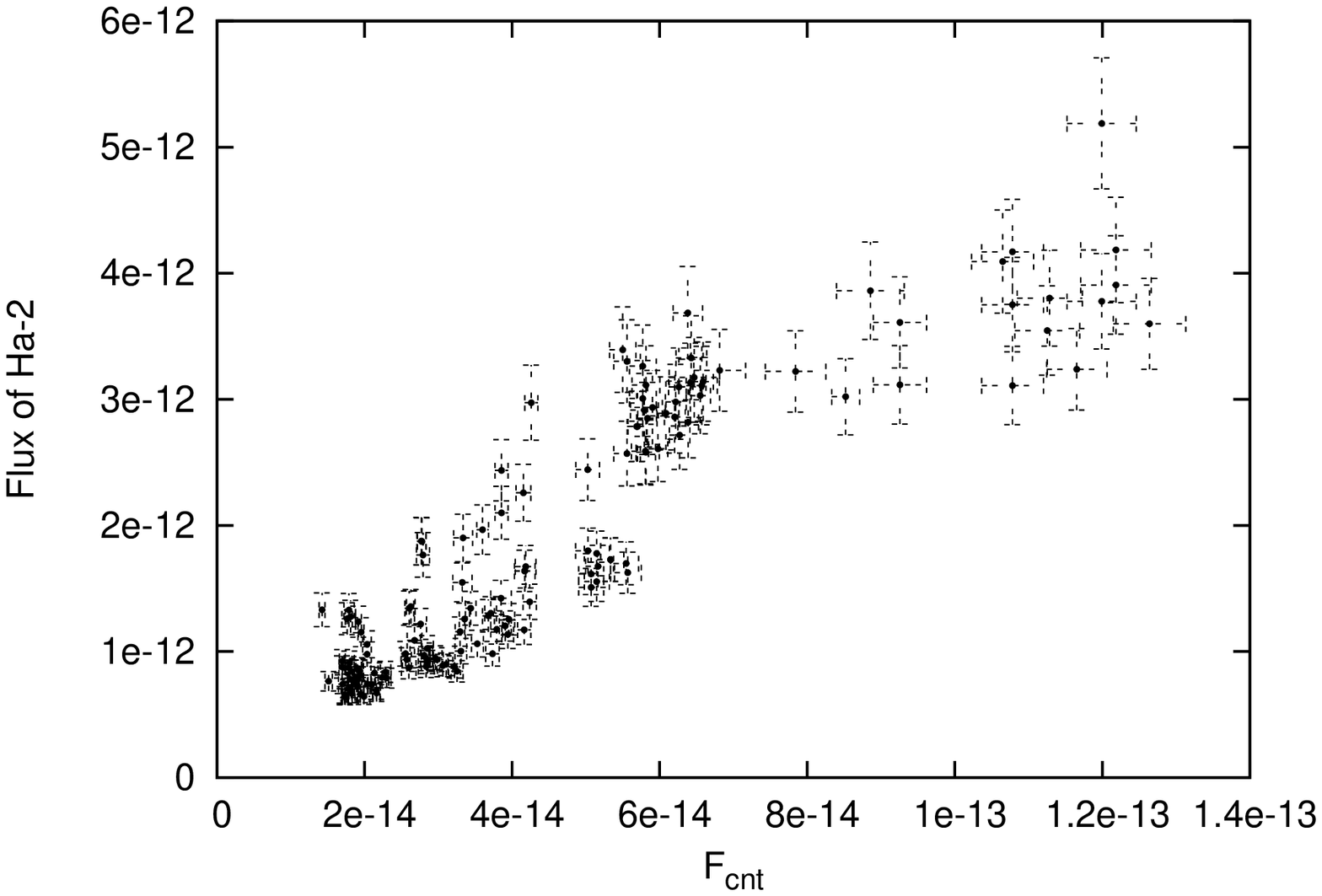}
\includegraphics[width=6cm]{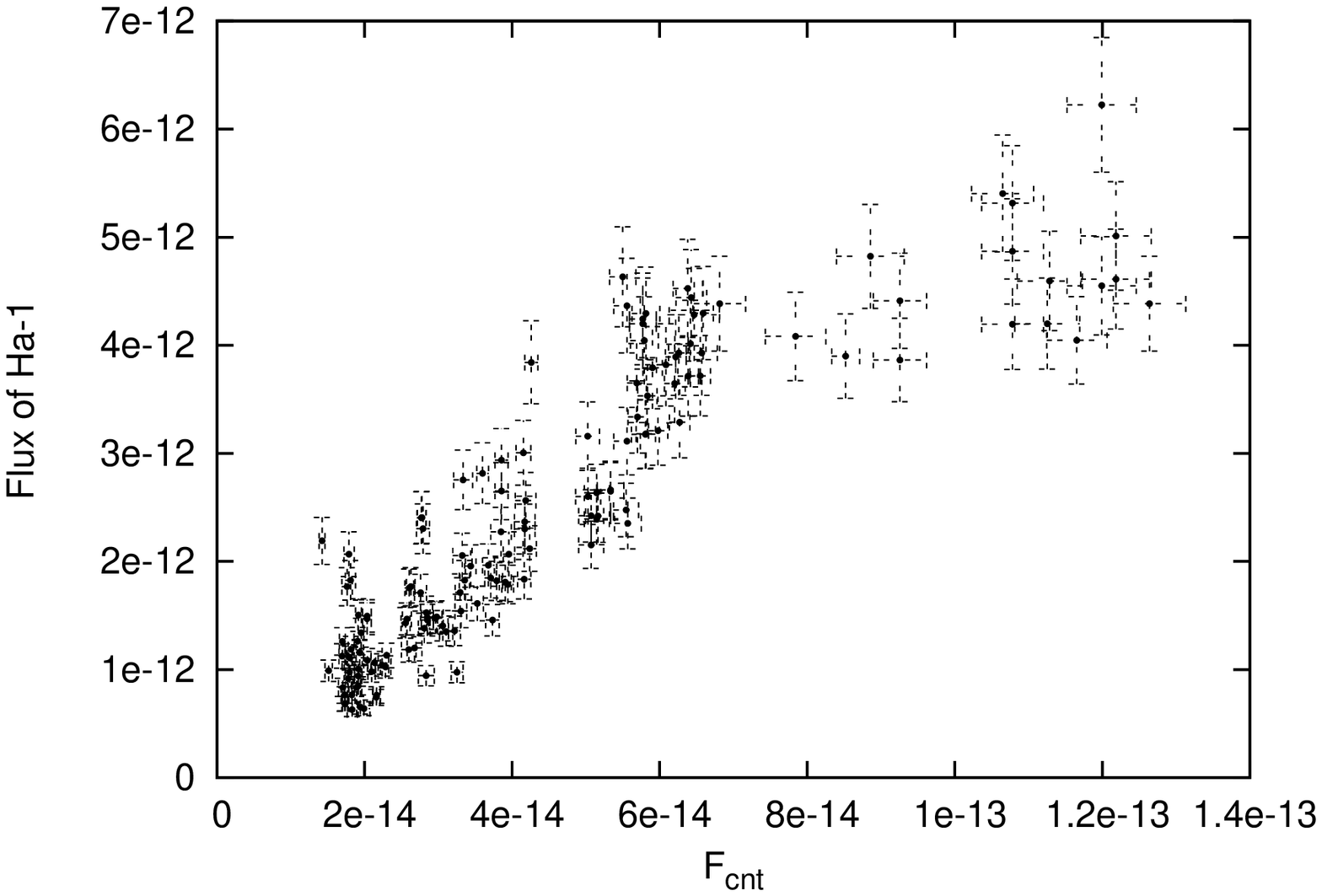}
\includegraphics[width=6cm]{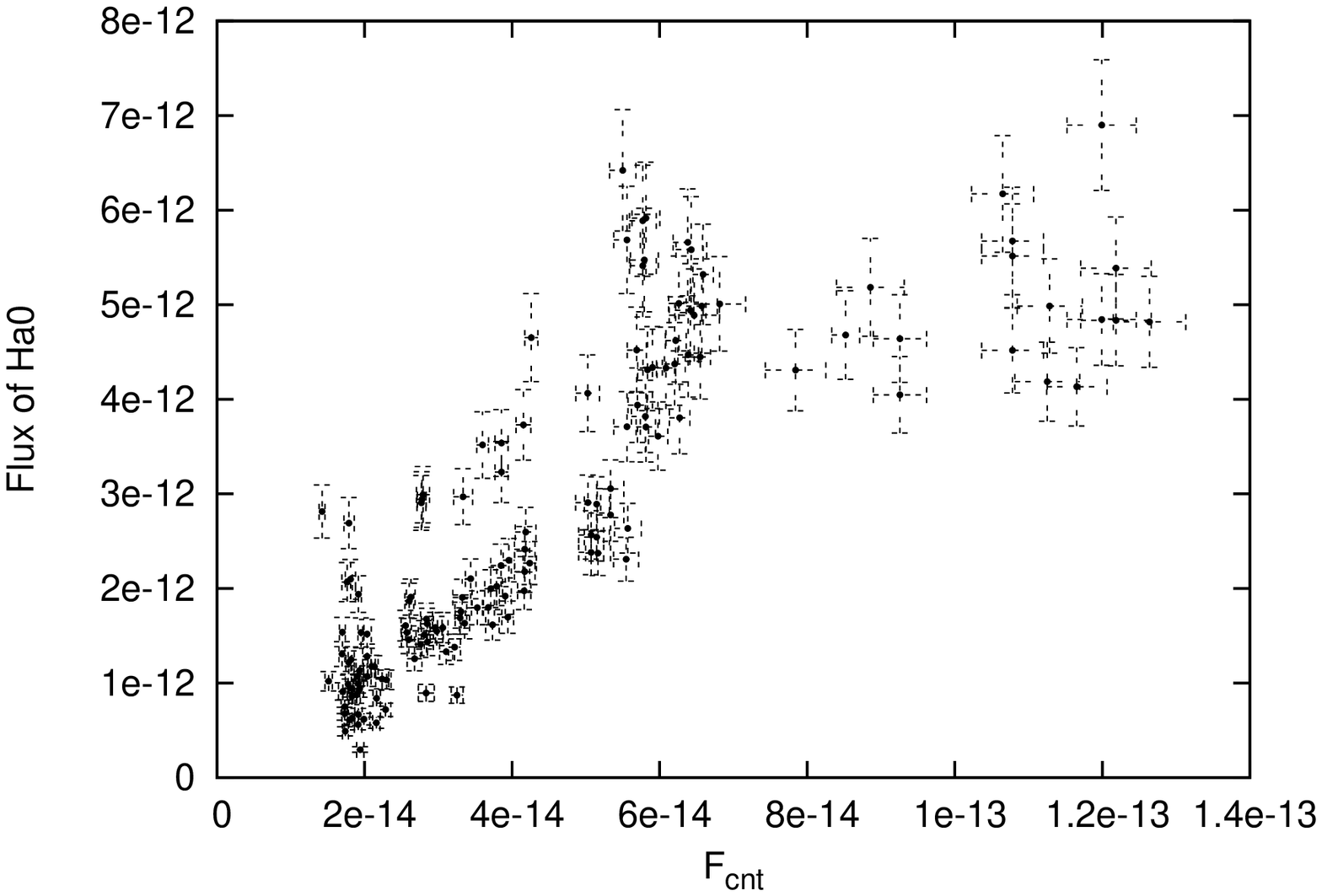}
\includegraphics[width=6cm]{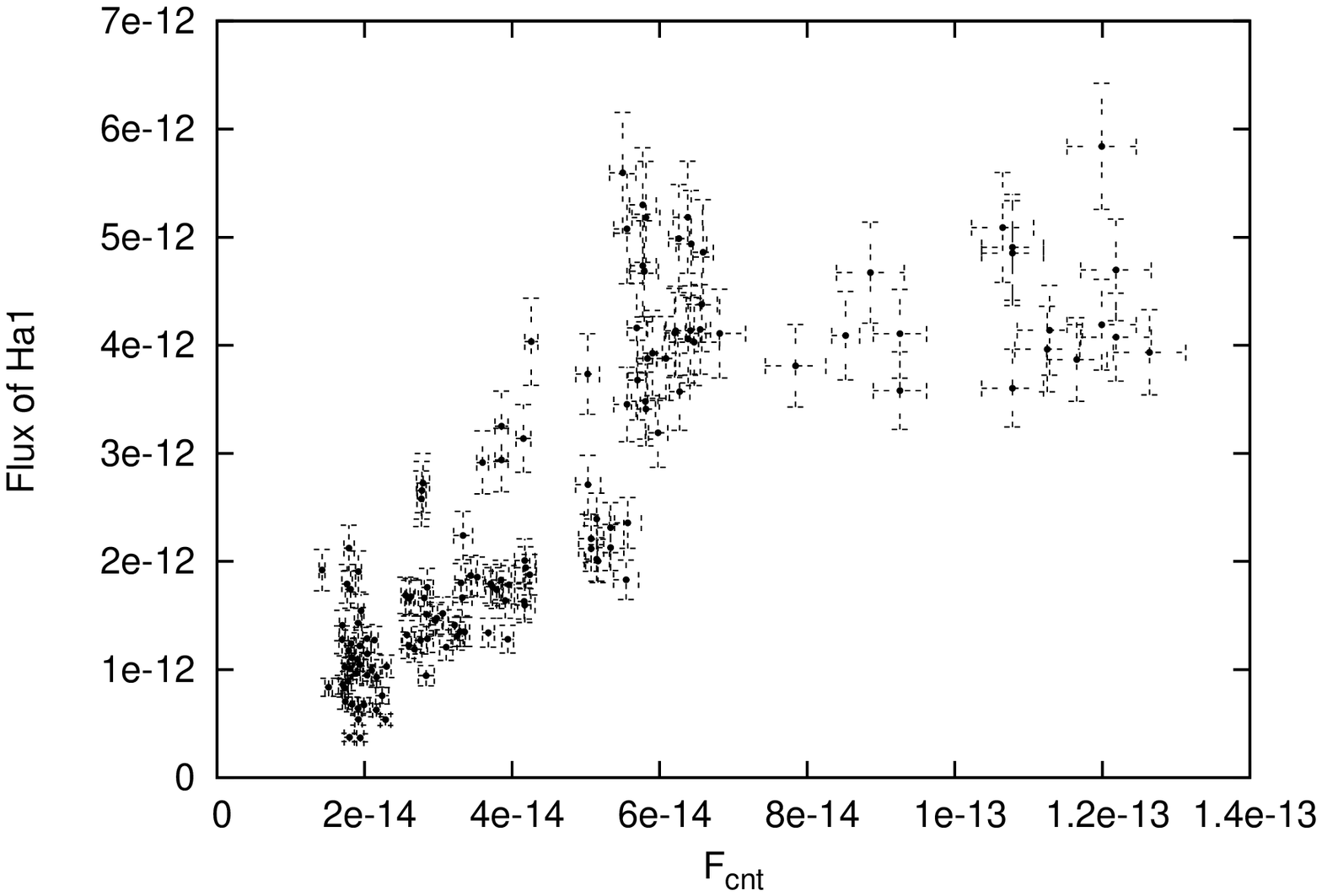}
\includegraphics[width=6cm]{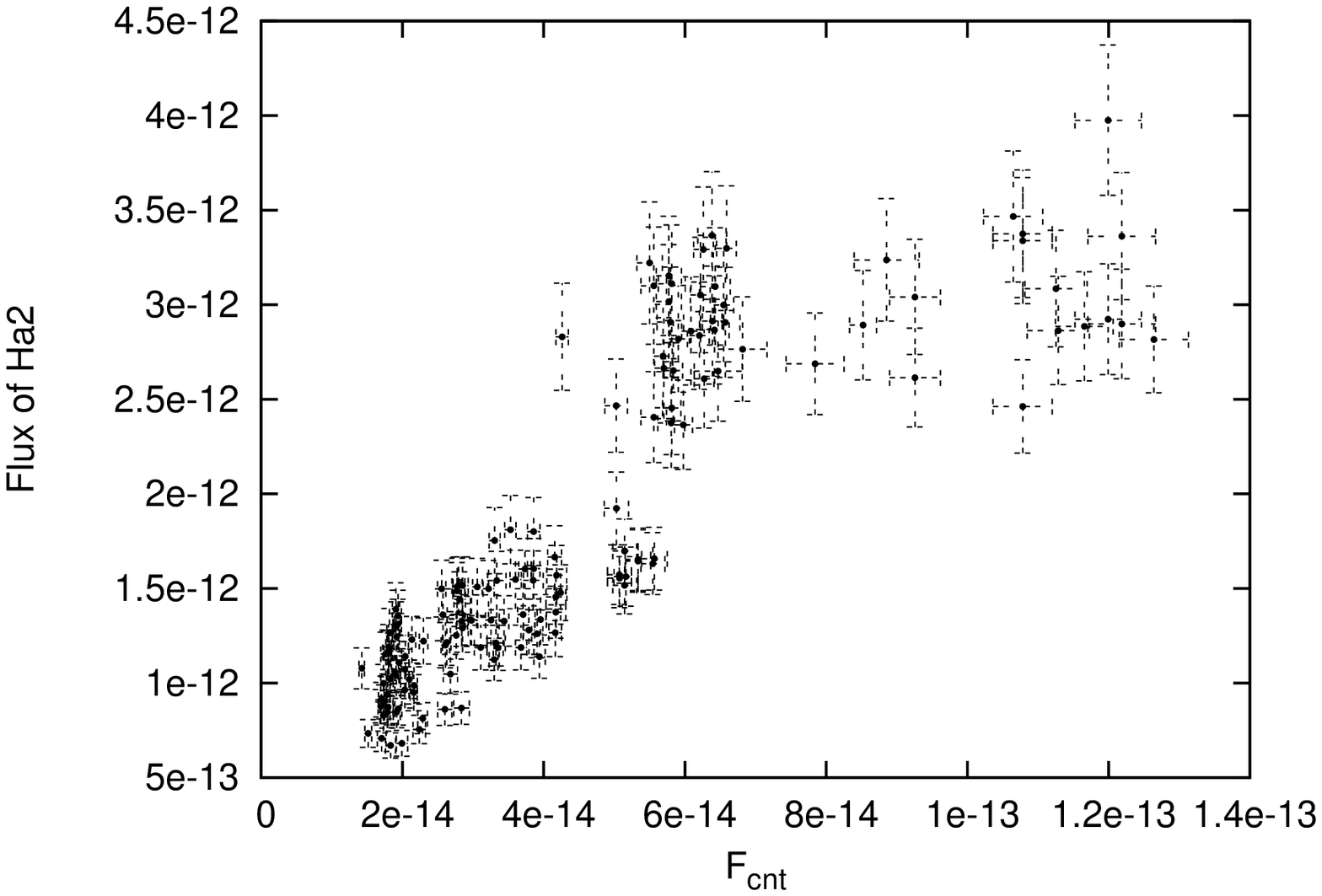}
\includegraphics[width=6cm]{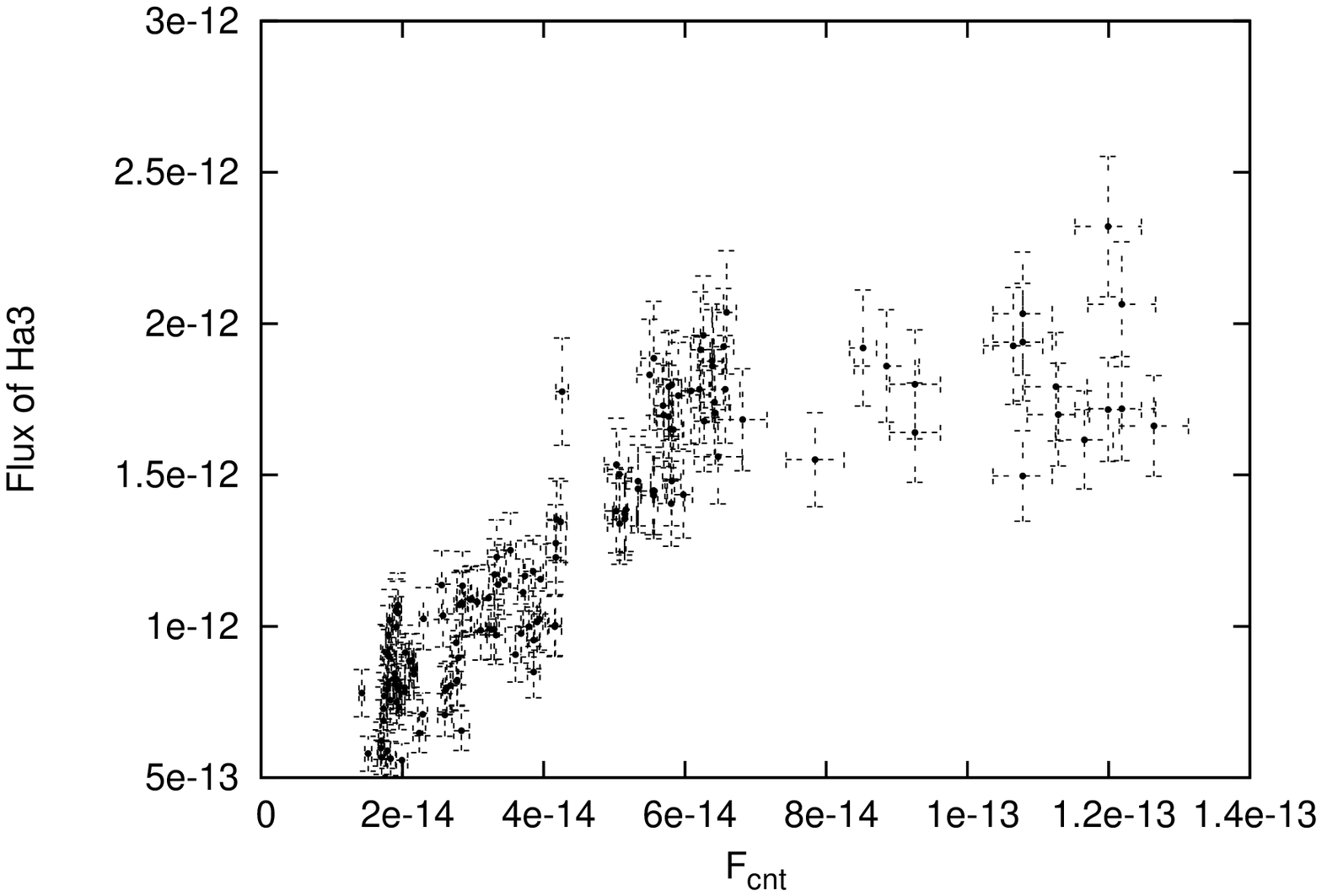}
\includegraphics[width=6cm]{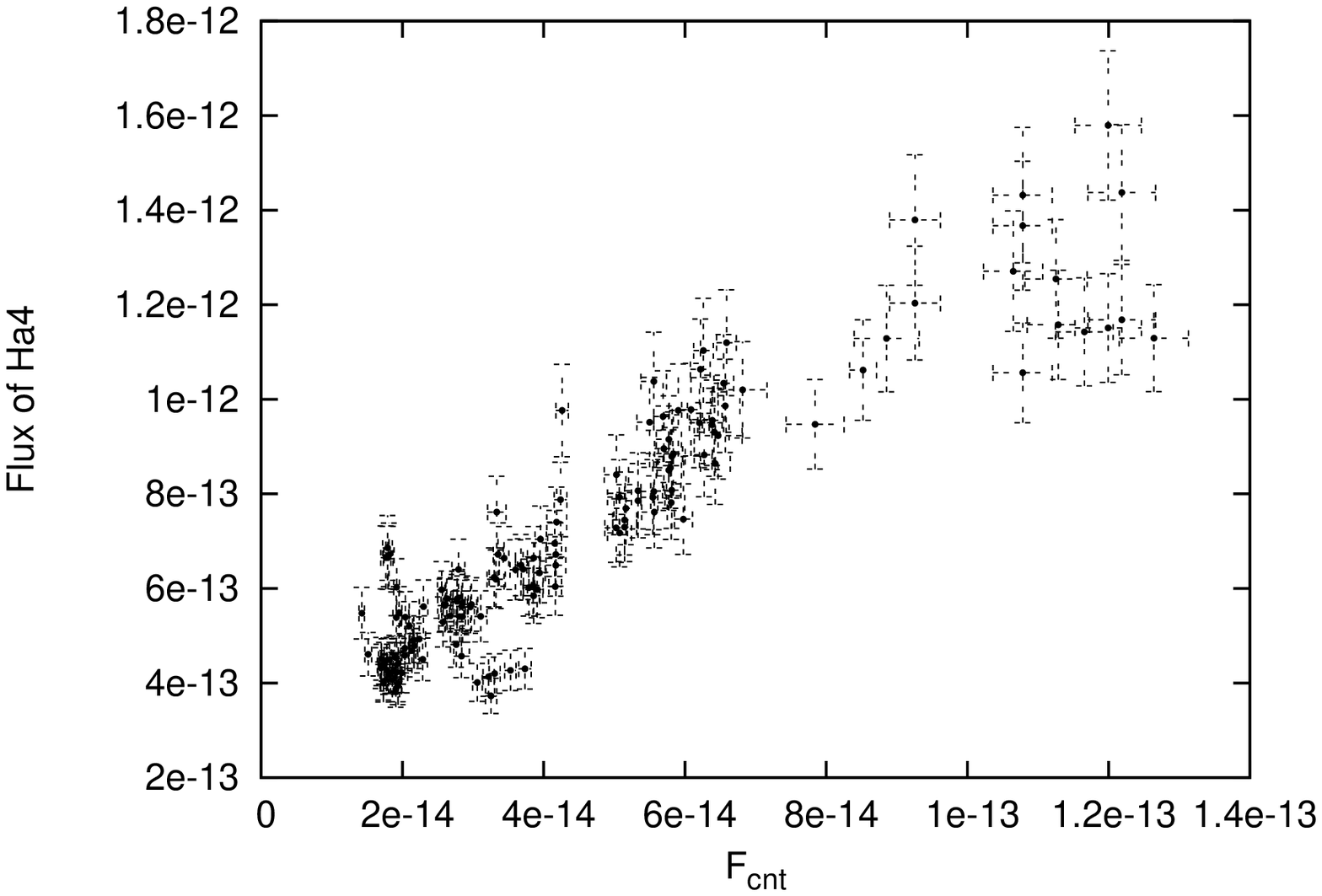}
\includegraphics[width=6cm]{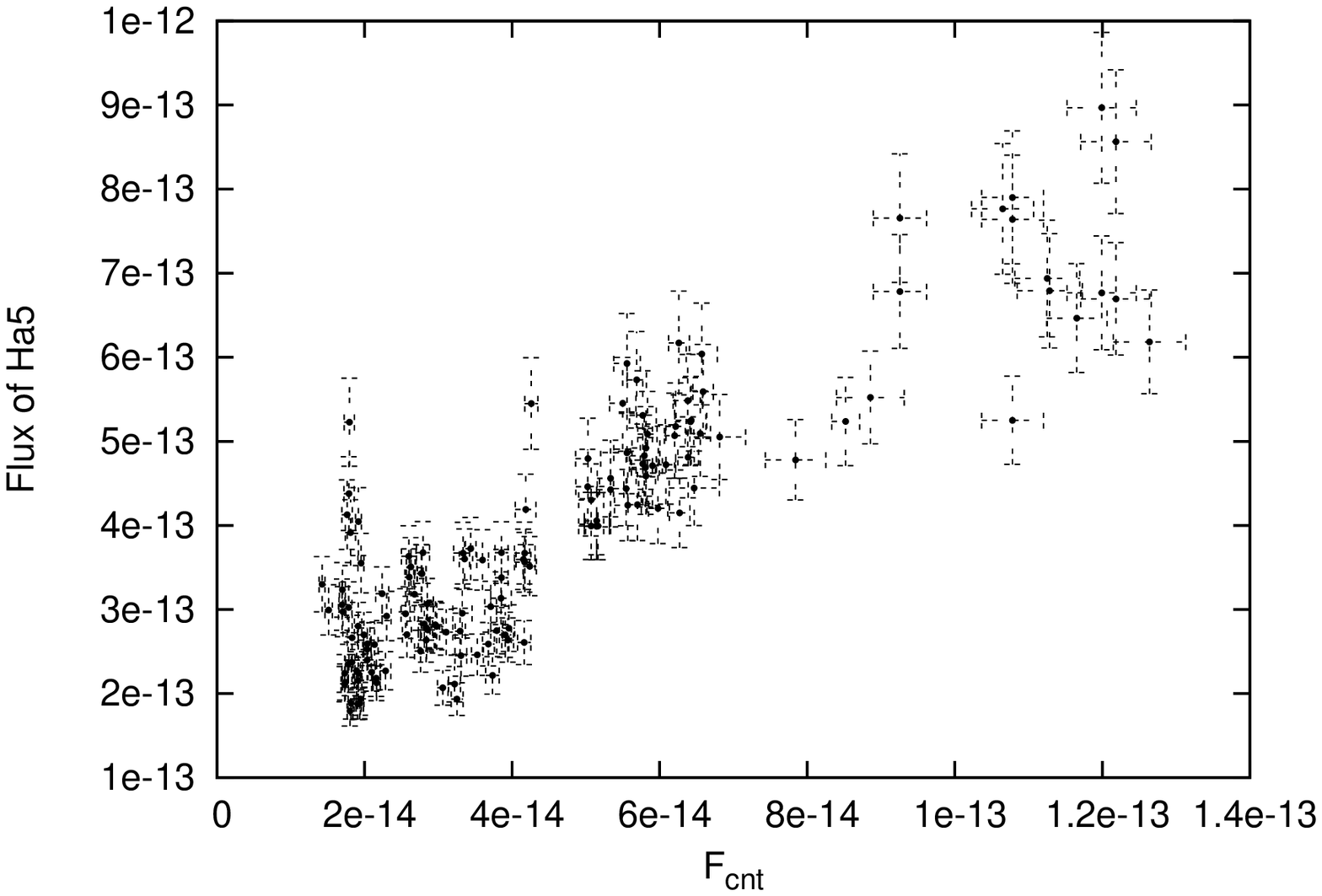}
\caption{The flux of different segments of the line as a function of
the continuum flux. The line flux is given in $ \rm erg \
cm^{-2}s^{-1}$ and the continuum flux in $ \rm erg \
cm^{-2}s^{-1}\AA^{-1}$.} \label{fig12}
 \end{figure*}

Additionally, in Fig. \ref{fig12} we present the response of
different H$\alpha$ segments to the continuum flux. As it can be
seen the responses are different: in the far wings, the response to
the continuum is almost linear for the red wing (segments 4 and 5)
and for a fraction of the blue wing (segment -4), but for the far
blue wing (-5500 to -4500 $\ {\rm km \ s^{-1}}$) there is
practically no response for $F_{\rm c}<7 \times 10^{-14} \rm erg \
cm^{-2}s^{-1}\AA^{-1}$. For higher continuum fluxes, the far blue
wing has a higher flux, but it seems also that there is no linear
relation between the line wing and the continuum flux. On the other
hand, the  central segments (from - 3500 to 3500 $\ {\rm km \
s^{-1}}$) have a similar response to the continuum like the H$\beta$
and H$\alpha$ total line fluxes (see Paper I) - a linear response
for the low continuum flux $F_{\rm c}<7 \times 10^{-14} \rm erg \
cm^{-2}s^{-1}\AA^{-1}$) and no linear response for the high
continuum flux $F_{\rm c}>7 \times 10^{-14} \rm erg \
cm^{-2}s^{-1}\AA^{-1}$) The linear response indicates that this part
of line (red wings and central segments at the low continuum flux)
is coming from the part of the BLR ionized by the AGN source, while
the blue and partly central part of the line could partly
originating from a substructure out of this BLR (and probably not
photoionized).  Note here that the photoionization in the case of a
mix of a thin (fully ionized) and thick (ionization-bounded) clouds
can explain the observed non-linear response of emission lines to
the continuum in variable as seen in the central parts of broad
lines (see Shields et al. 1995), i.e. in the case of the optically
thick BLR the detailed photoionization models show that the response
of the Balmer lines declines as the continuum flux increases (see
Goad at al. 2004, Korista \& Goad 2004). {  In our case we found
that the flux in wings (except the far blue wing) have almost linear
response to the continuum flux, while the central parts show
non-linear response to the higher continuum flux. Also the response
to the continuum flux of the far blue wing (-5) and far red wing
(+5) is very different. It may  indicate different physical
conditions in sub-regions or across the BLR.}

\onlfig{13}{
\centering
\begin{figure*}[]
\includegraphics[width=6cm]{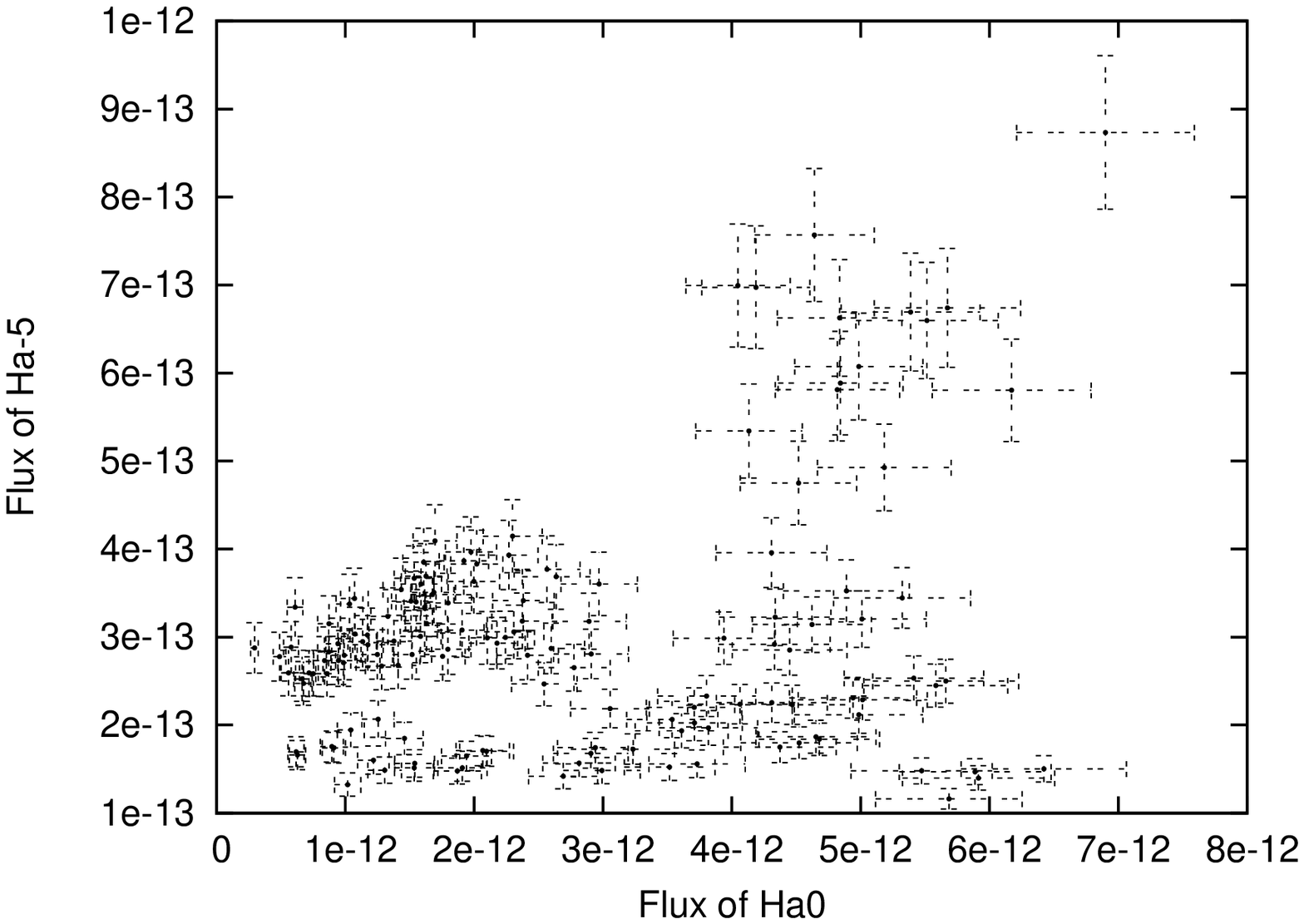}
\includegraphics[width=6cm]{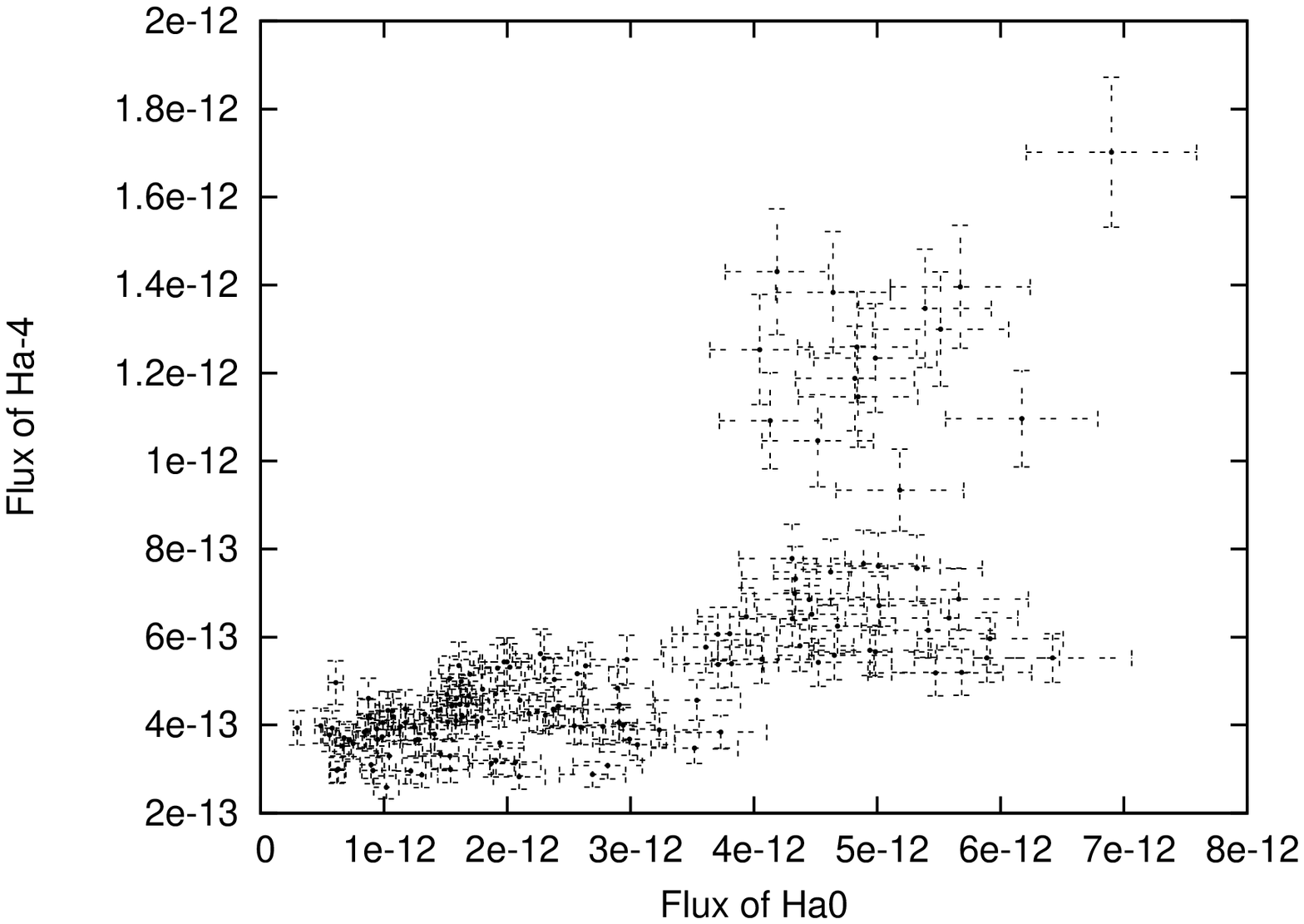}
\includegraphics[width=6cm]{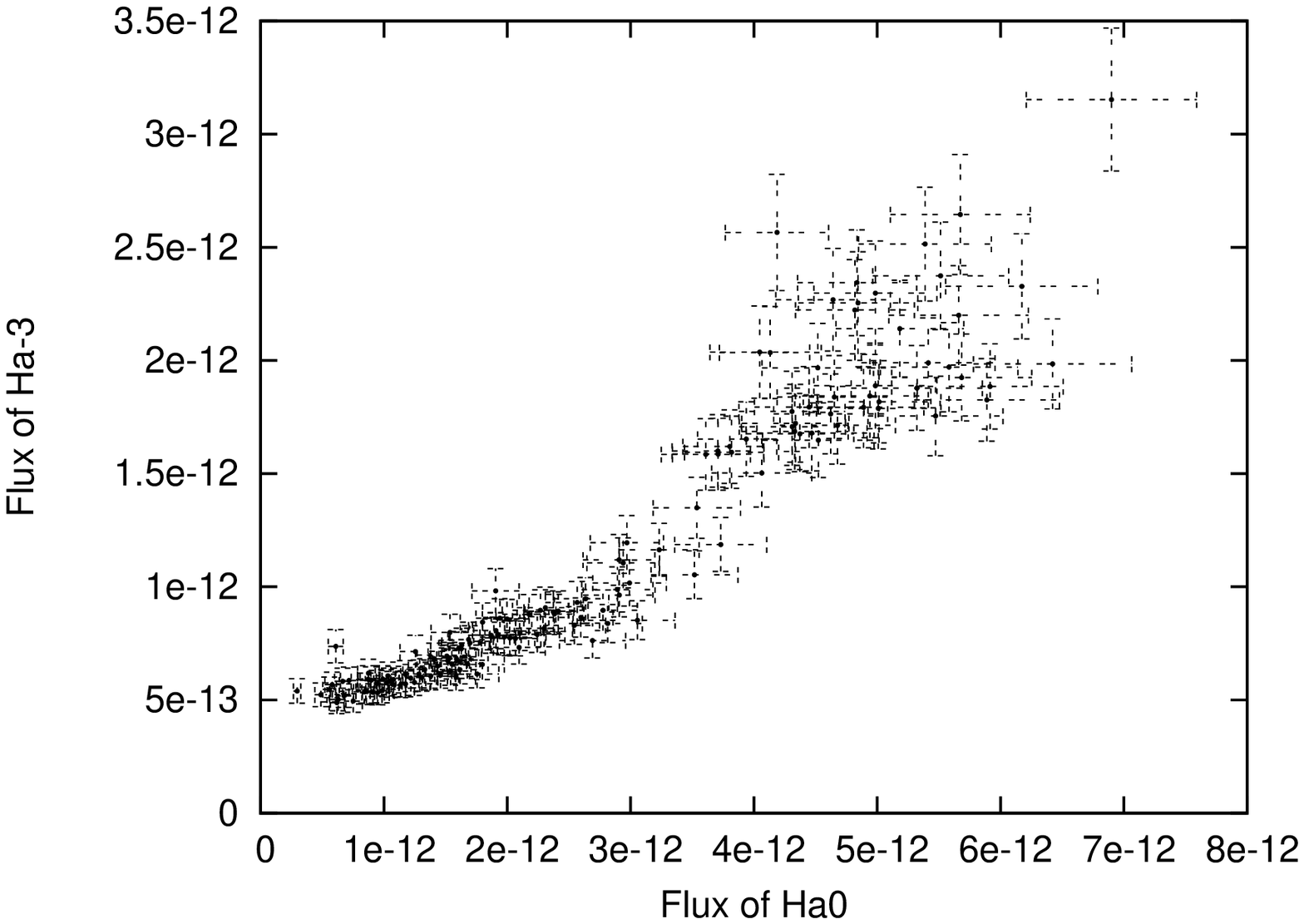}
\includegraphics[width=6cm]{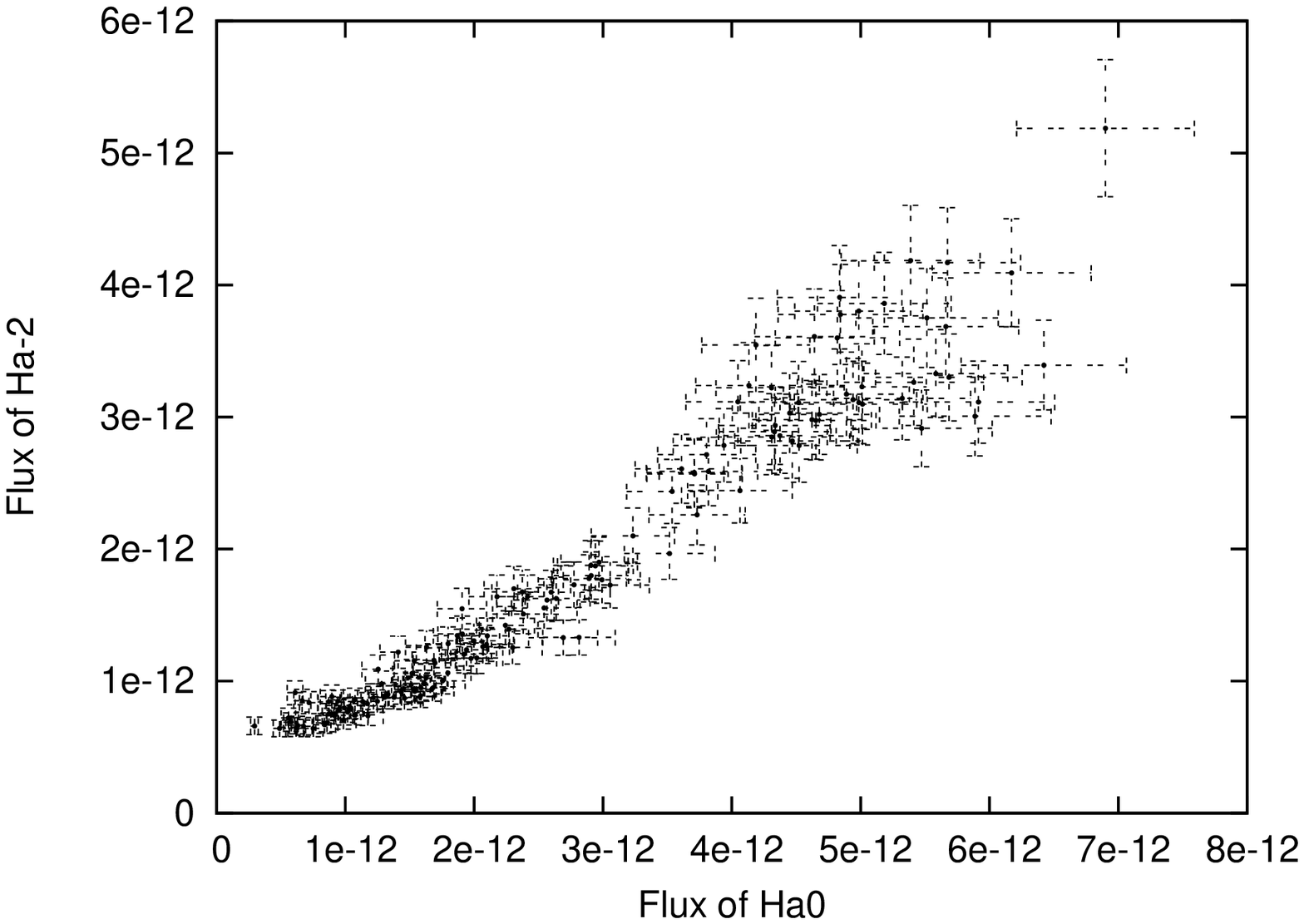}
\includegraphics[width=6cm]{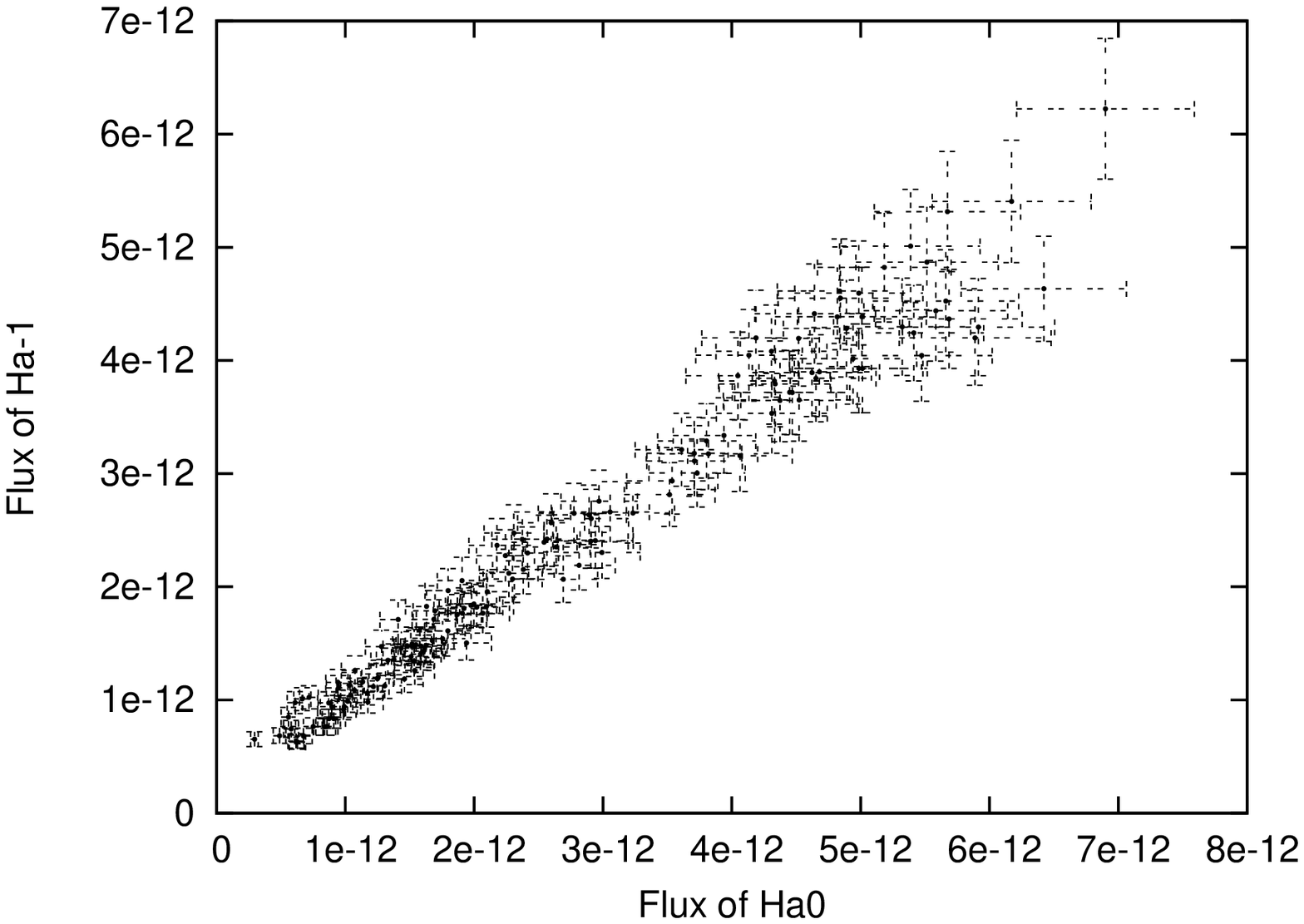}
\includegraphics[width=6cm]{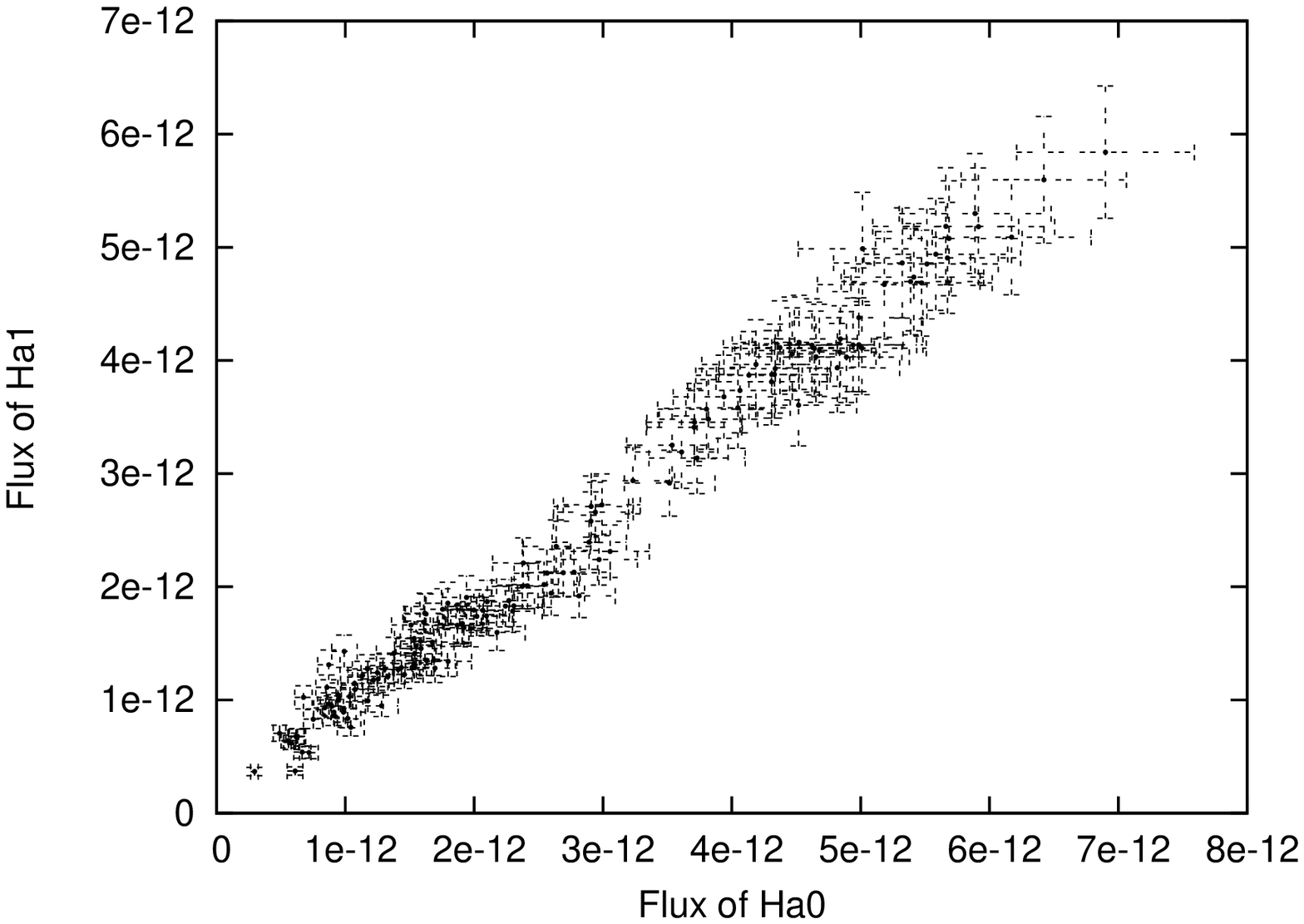}
\includegraphics[width=6cm]{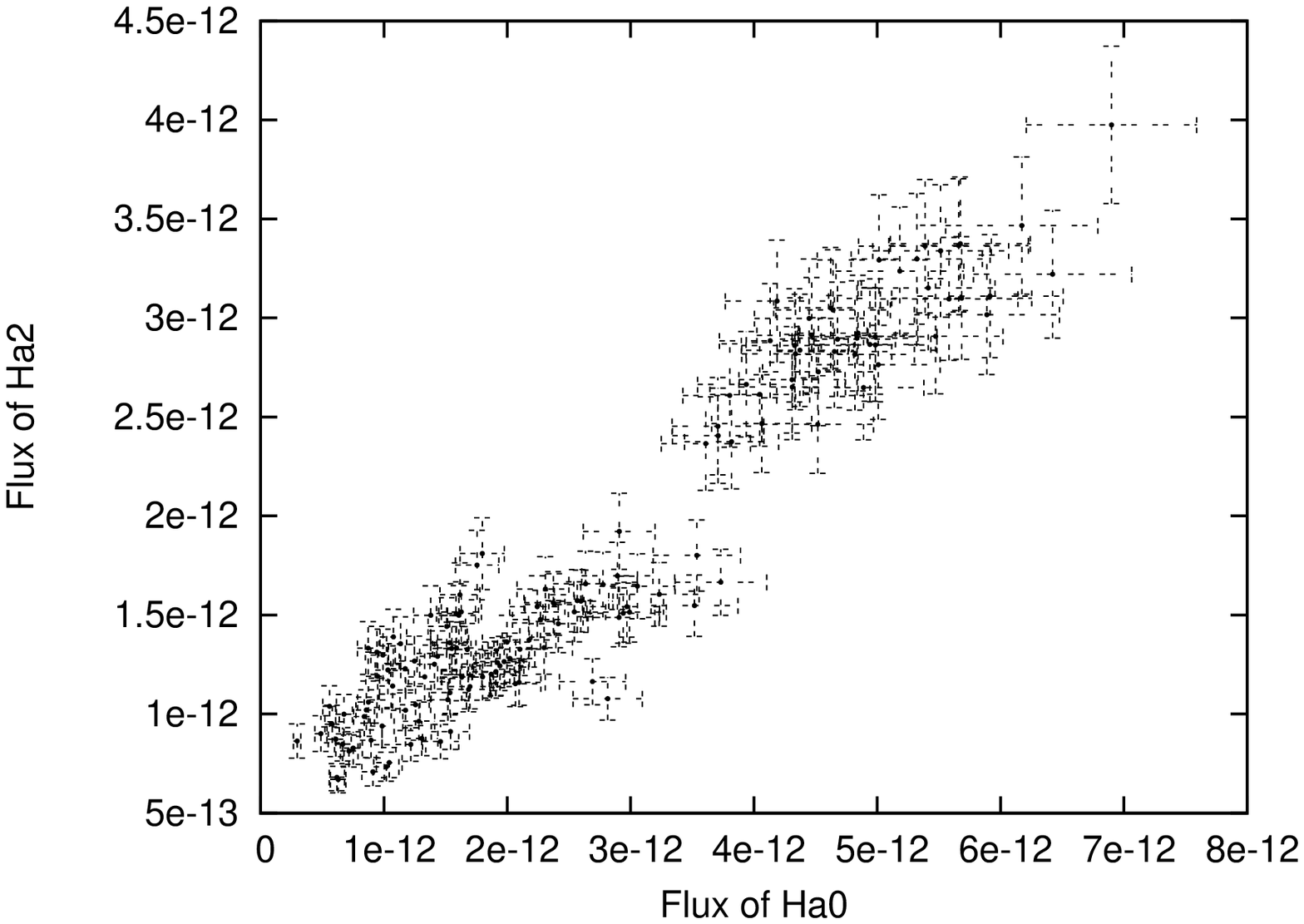}
\includegraphics[width=6cm]{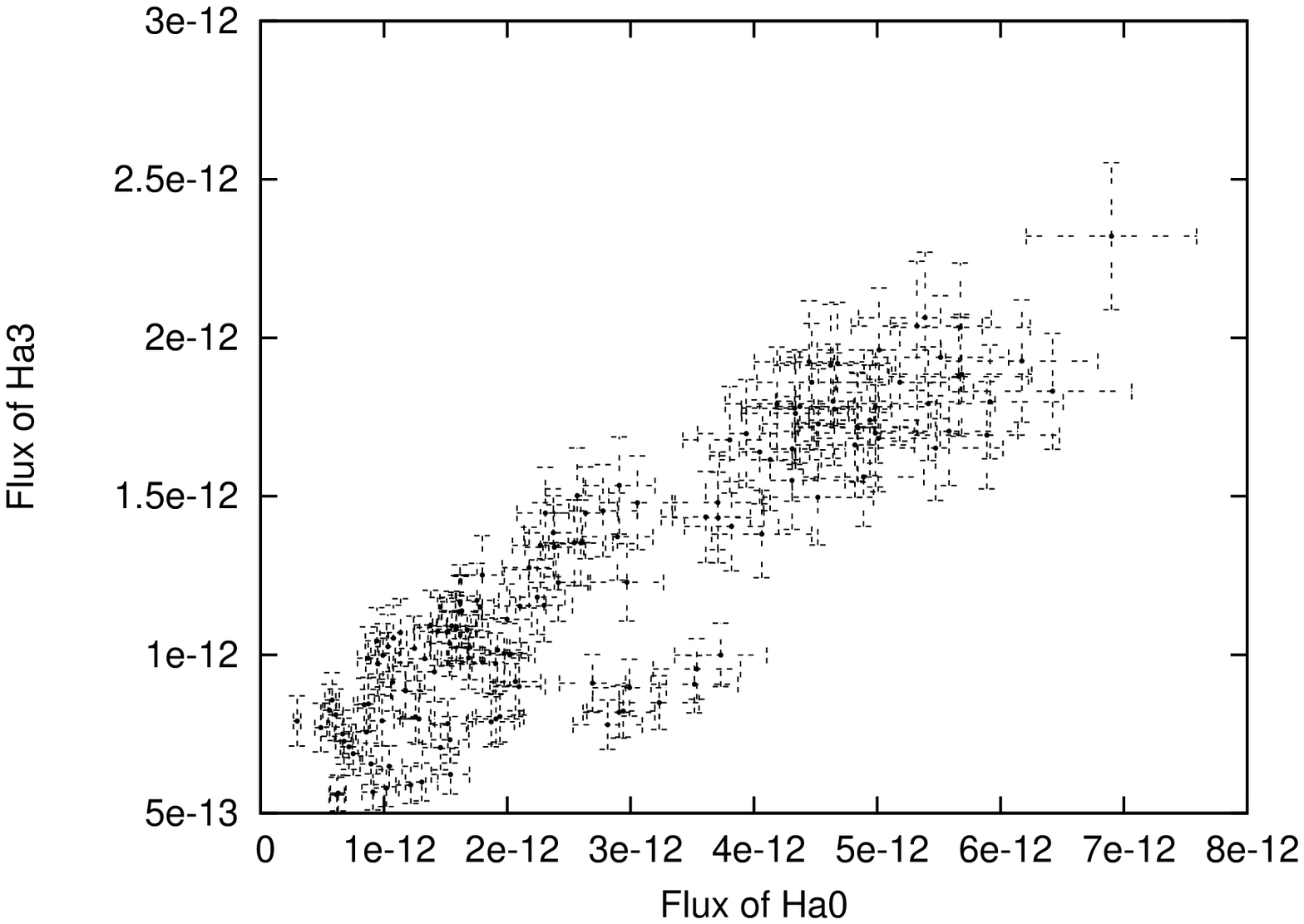}
\includegraphics[width=6cm]{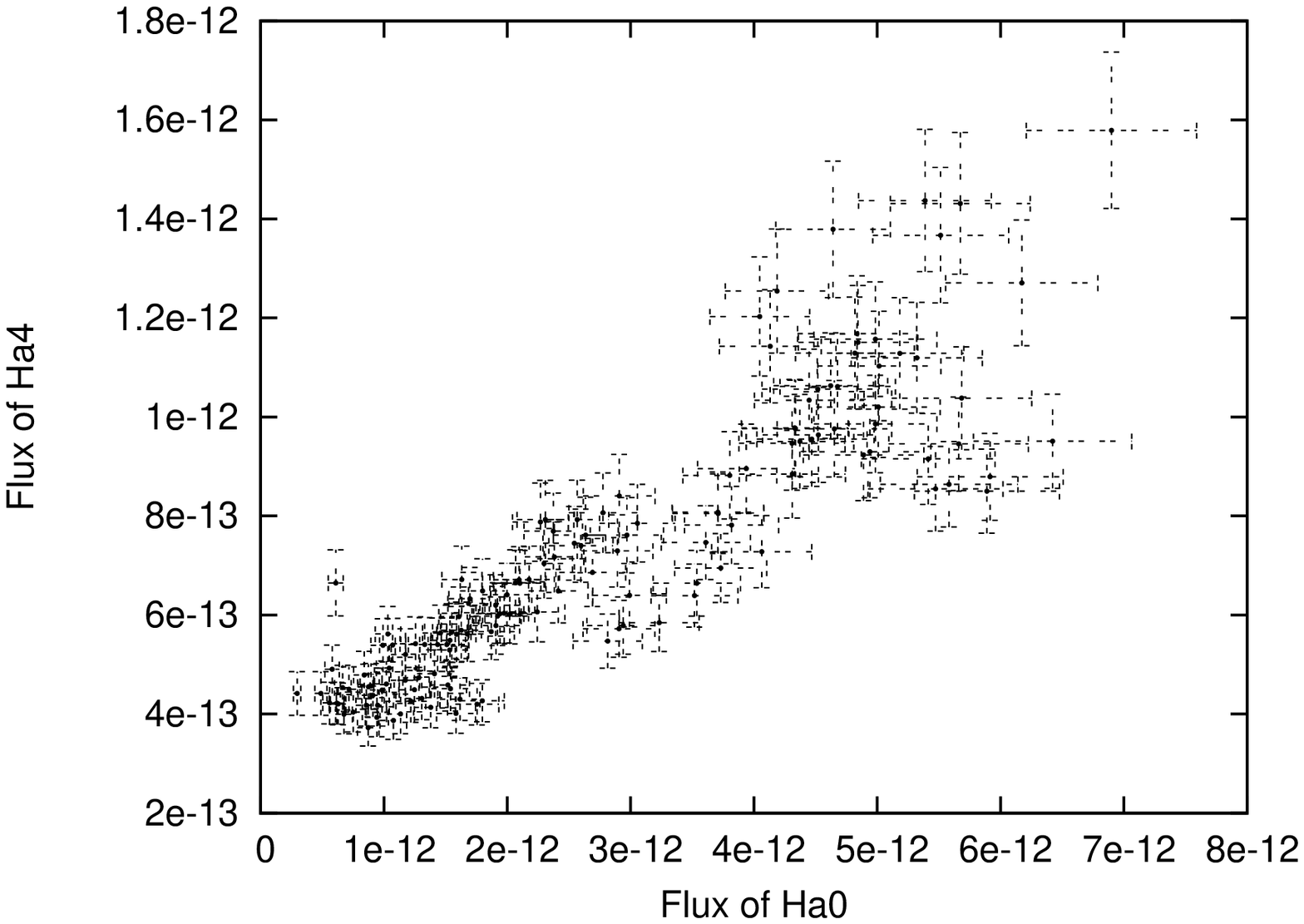}
\includegraphics[width=6cm]{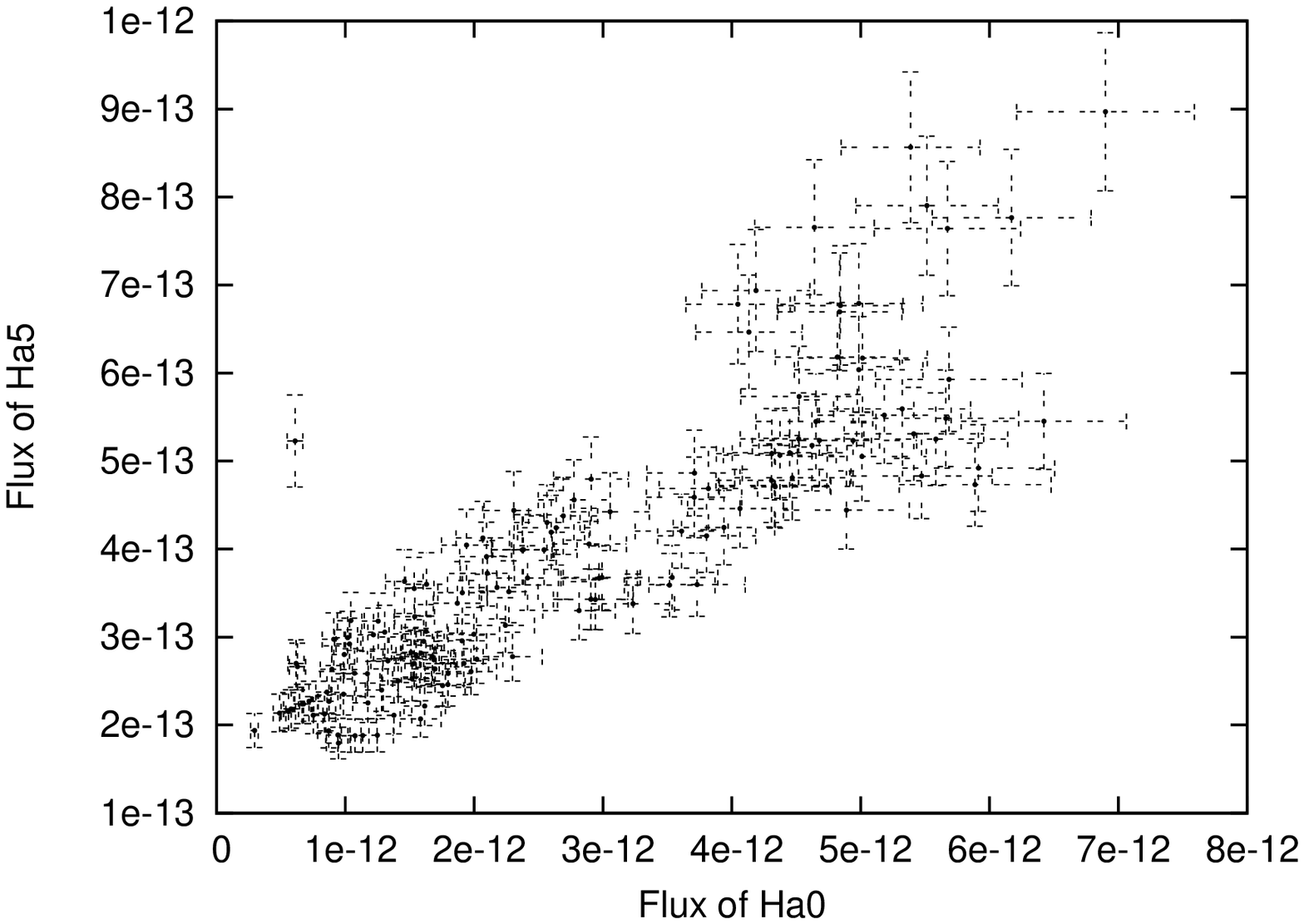}
\caption{The flux of different H$\alpha$ segments as a function of
the flux of the central H$\alpha$ segment. The line flux is given in
$ \rm erg \ cm^{-2}s^{-1}$.} \label{fig13}
 \end{figure*}
}

In Fig. \ref{fig13} (available electronically only) the flux of
different H$\alpha$ segments as a
function of the flux of the central segment (H$\alpha$0) are
presented. It can be seen that there are different relations between
different H$\alpha$ segments and H$\alpha$0: for segments near the
center (H$\alpha$-1, H$\alpha$-2, H$\alpha$1 and H$\alpha$2) the
relation is almost linear, indicating that the core of the line
originates in the same substructures (Fig. \ref{fig13}); segments in
the H$\alpha$ wings (the near blue wing H$\alpha$-3 and the red wing
H$\alpha$3, H$\alpha$4 and H$\alpha$5) also show a linear response
to the central segment (but the scatter of the points is larger than
in previous case) which also indicates that a portion of the emission in the
center and in these segments is coming from the same emission
region. On the contrary, the far blue wing (H$\alpha$-4 and
H$\alpha$-5) responds weakly to the line center, especially
H$\alpha$-5 that shows practically independent variations with
respect to the central segment.

\section{The Balmer decrement}

\subsection{Integral Balmer decrement}

From month-averaged profiles of the broad H$\alpha$ and H$\beta$
lines we determined their integrated fluxes in the range between -
and +5500 ${\rm km \ s^{-1}}$ of radial velocities. We call
"Integral Balmer decrement" (BD) the integrated flux ratio
$F$(H$\alpha$)/$F$(H$\beta$).

\begin{figure}
\centering
\includegraphics[width=8.5cm]{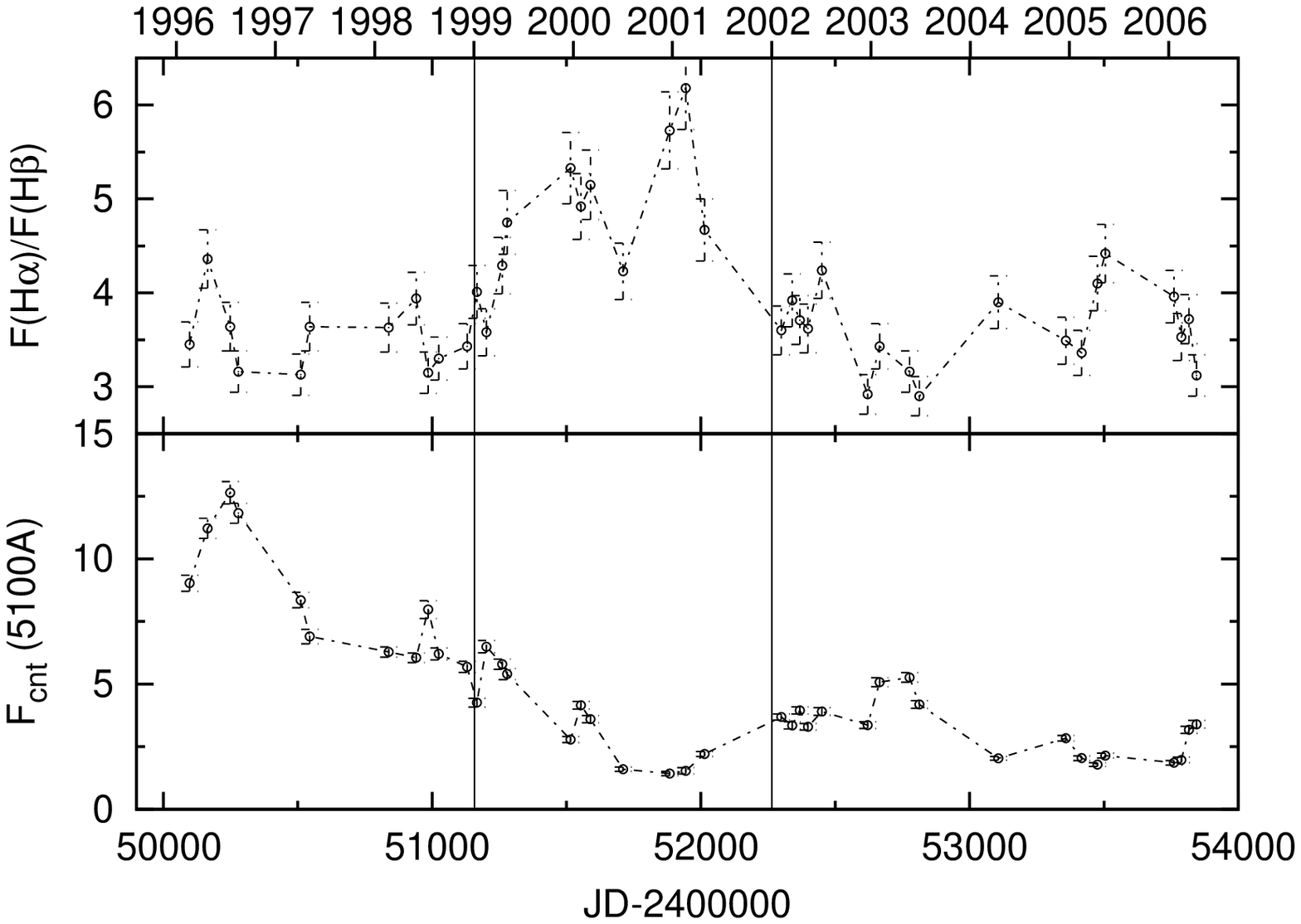}
\caption{Variations of the integrated Balmer decrement
BD=F(H$\alpha$)/F(H$\beta$ (upper panel)
          and of the continuum flux at $\lambda 5100\, \AA$ (bottom panel)
          in 1996--2006.
          The abscissae gives the Julian date (bottom) and the corresponding year (top).
          The continuum flux is in units
          $10^{-14}$\,erg\,cm$^{-2}$\,s$^{-1}$\,\AA$^{-1}$.
          The vertical lines correspond to years 1999 and 2002.}\label{fig14}
\end{figure}

Fig. \ref{fig14} shows the behavior of the integrated BD (upper
panel) and continuum flux at the wavelength 5100\,\AA\, (bottom
panel). In 1999--2006 an anticorrelation between the changes of the
integrated BD and continuum flux was observed. It was especially
noticeable in 1999--2001.

Table \ref{tab5}  presents year-averaged values
of the BD and continuum flux determined from month-averaged profiles
in each year.

{  We found that  in 1996--1998  the continuum flux was rather large
and was changing within the limits
 $F_{\rm c}\sim(6-12) \times 10^{-14} \rm \ erg \ cm^{-2}s^{-1}\AA^{-1}$,
but the BD was practically not changing. In Paper I
we already noted the absence of correlation between the continuum
flux and the integrated flux of the broad lines for the
above-mentioned flux values. Also, we found (see Figure \ref{fig14} that the
Balmer decrement is systematically higher in 1999--2001.}

\onlfig{15}{
\begin{figure*}[]
\includegraphics[width=16cm]{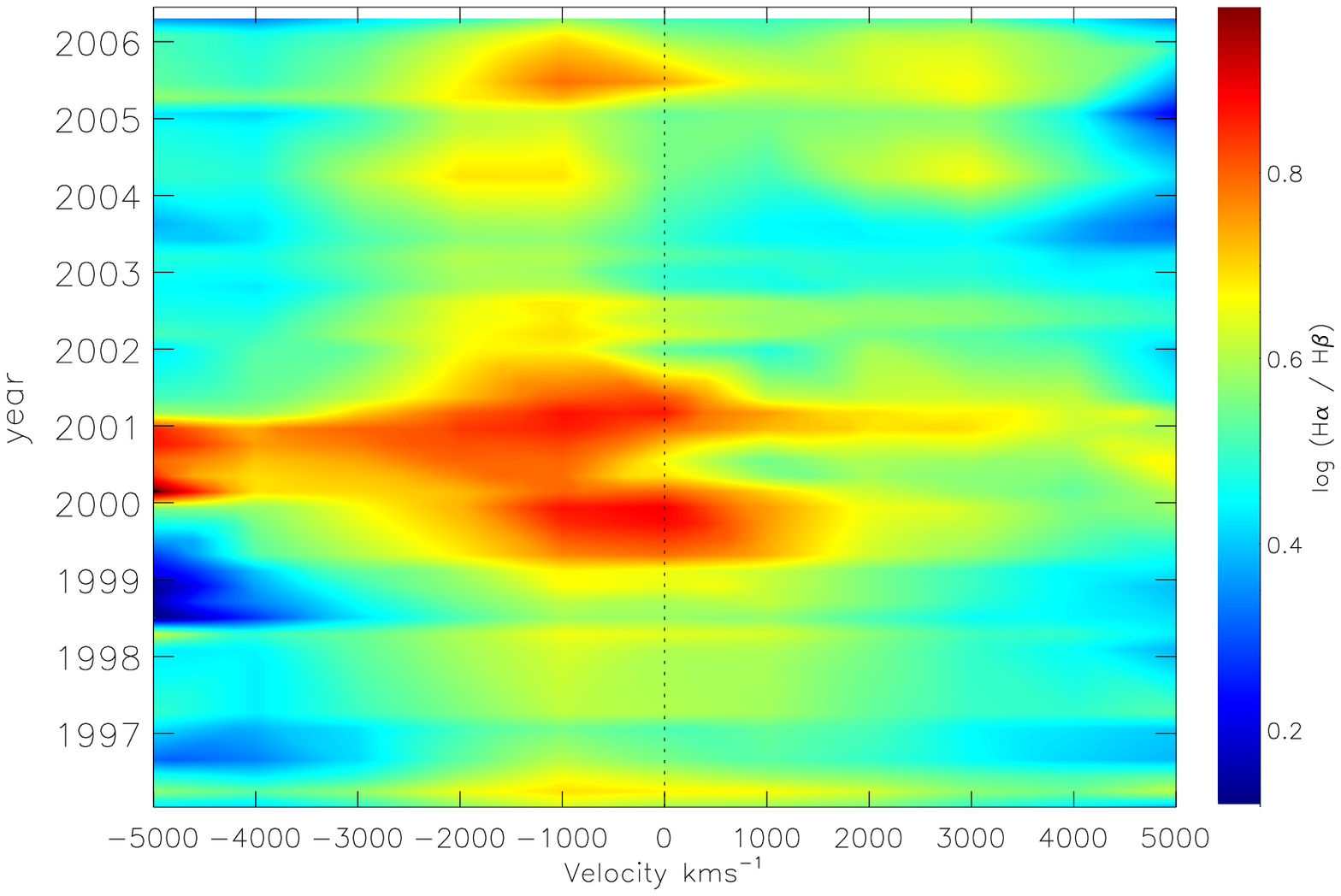}
\caption{The variations of the Balmer decrement for different parts
of the broad emission lines. The flux is given in $\rm erg \ cm^{-2}
\ s^{-1} \AA^{-1}$} \label{fig15}
 \end{figure*}
}

\begin{figure}
\centering
\includegraphics[width=7.5cm]{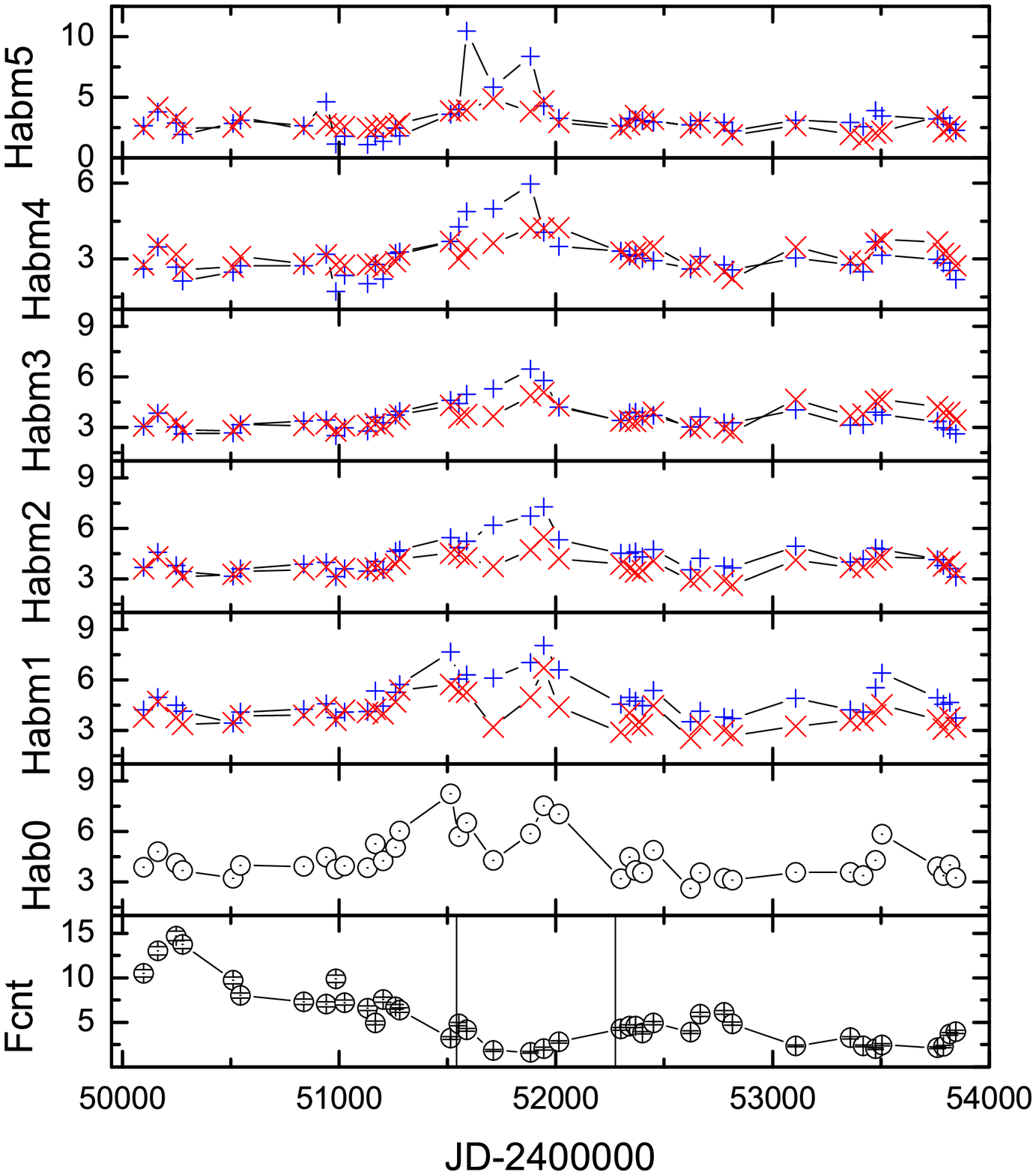}
\caption{ Variations of the BD of different segments of the line
profiles and of the continuum flux (bottom panel) in 1996--2006. The
BD of segments in the blue wing (numbers from -5 to -1 from Table
\ref{tab3}) are denoted with plus ($+$), in red wing (numbers from
+5 to +1 from Table \ref{tab3}) with crosses ($\times$), and of
central part (number 0) with open circles. The vertical lines
correspond to years 1999 and 2002. The abscissae shows the Julian
date (bottom) and the corresponding year (top). The ordinate shows
the BD for different segments of the line profile. The continuum
flux is in units
           $10^{-14}$\,erg\,cm$^{-2}$\,s$^{-1}$\,\AA$^{-1}$.}\label{fig16}
\end{figure}

Table \ref{tab5} also gives average data of the BD and continuum
flux for the periods of 1996--1998, 1999, 2000--2001, 2002--2006. It
is evident that in the period of 1996--1998 the BD did not vary
within the error bars in spite of strong variations of the continuum.
 Radical changes (the increasing of the BD) started in 1999.

The BD reached its maximum in 2000--2001. Then in 2002--2006 average
values of the BD coincide with those of 1996--1998. Thus, it can be
concluded with confidence that from 1999  to 2001 we observed an
obvious increase of the BD.

\subsection{Balmer decrement of different profile segments}

From month-averaged profiles of H$\alpha$ and H$\beta$ we determined
the BD for the segments of Table \ref{tab3}. Fig. \ref{fig15}
(available electronically only) shows the BD variation in a 2D plane
(year--$V_r$), and Fig. \ref{fig16} gives the changes of the BD of
different segments of the profiles during the whole monitoring
period.

It can be seen from Figs. \ref{fig15} and  \ref{fig16} that in
2000--2001 (JD=51200--52300) the values of the BD in different
segments were always on average noticeably larger than in the other
years, and the blue wing had always on average a noticeably larger
BD than the red one. In 1996--1998 BDs of the blue and red wings
practically coincided, and in 2002--2006 they either coincided or
the BD of the blue wing was slightly larger.

During the whole monitoring period the value of the BD along the
line profile in central segments (in segments 0,$\pm$1 corresponding
to the velocity range from -1500 $\ {\rm km \ s^{-1}}$ to 1500 $\
{\rm km \ s^{-1}}$) were considerably larger (by $\sim1.5$) on
average than at the periphery (segments from $\pm$2 to $\pm$5
corresponding to the velocity range from ((-5500)-(-1500)$\ {\rm km
\ s^{-1}}$) to (+5500)-(+1500)$\ {\rm km \ s^{-1}}$)).

\begin{figure*}
\centering
\includegraphics[width=5.7cm]{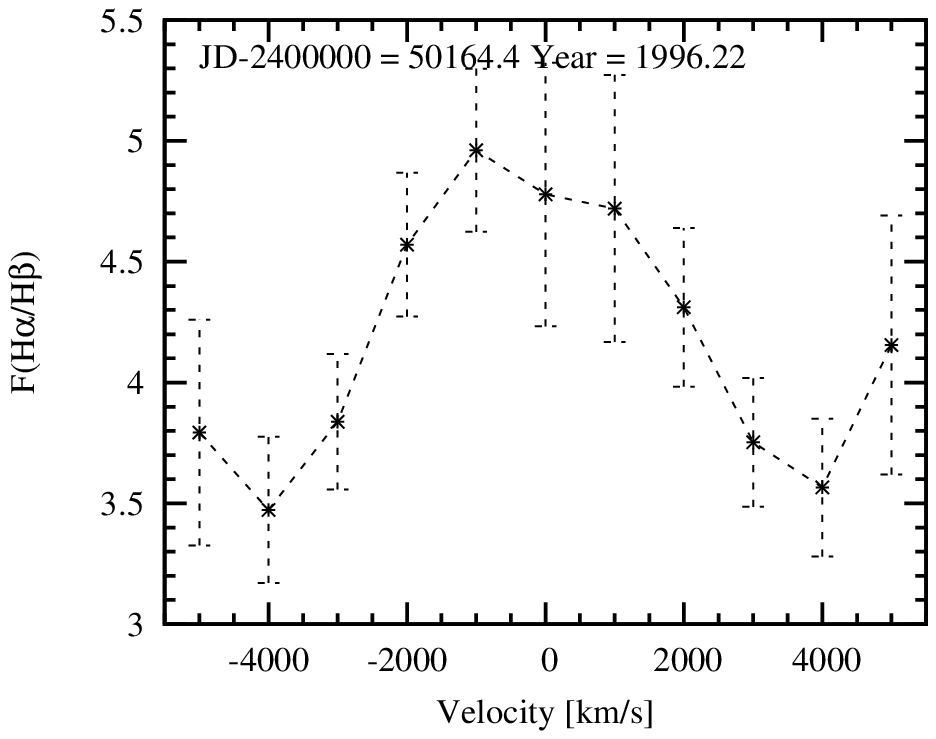}
\includegraphics[width=5.7cm]{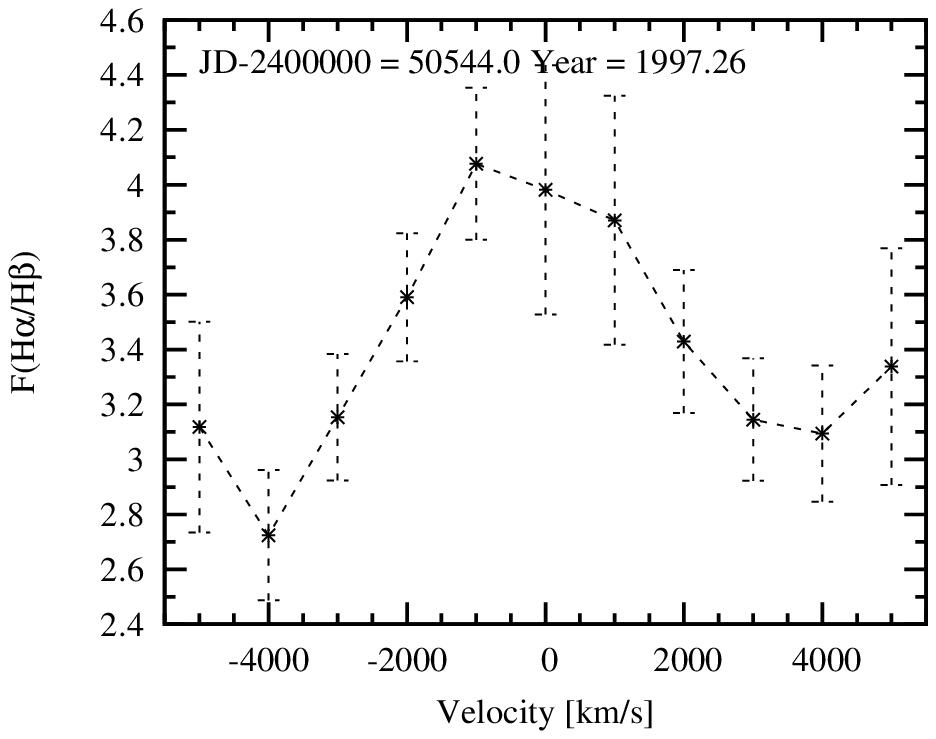}
\includegraphics[width=5.7cm]{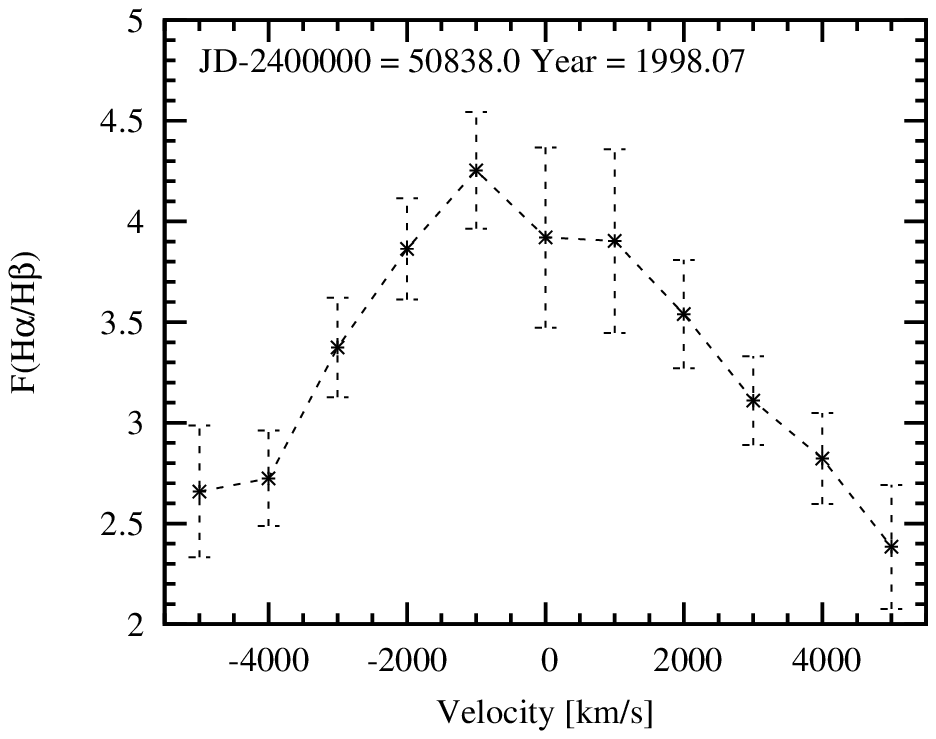}
\includegraphics[width=5.7cm]{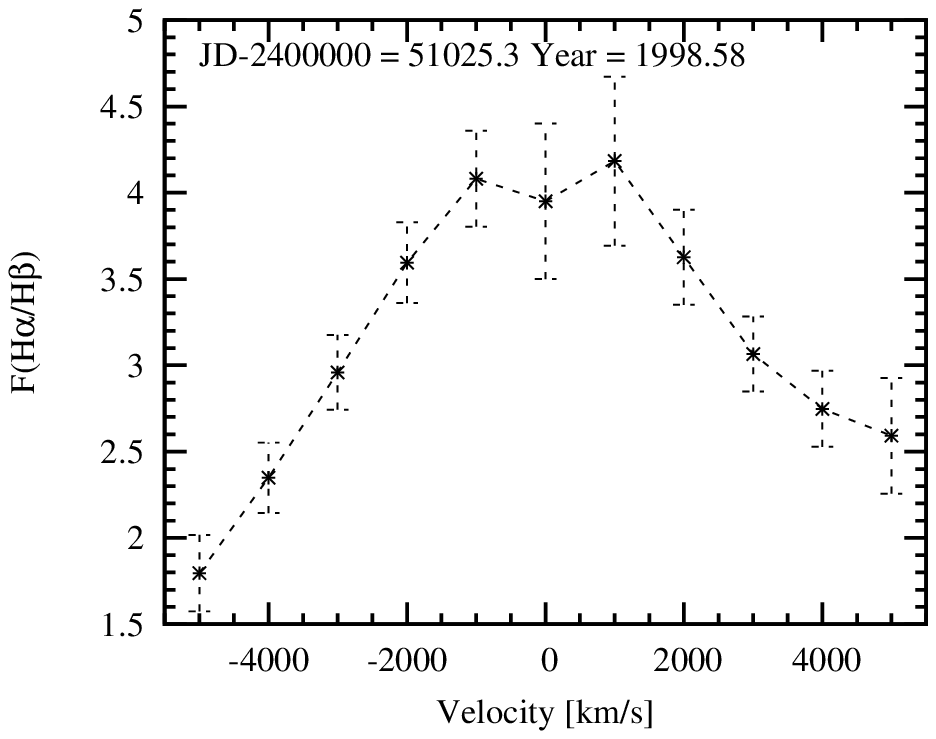}
\includegraphics[width=5.7cm]{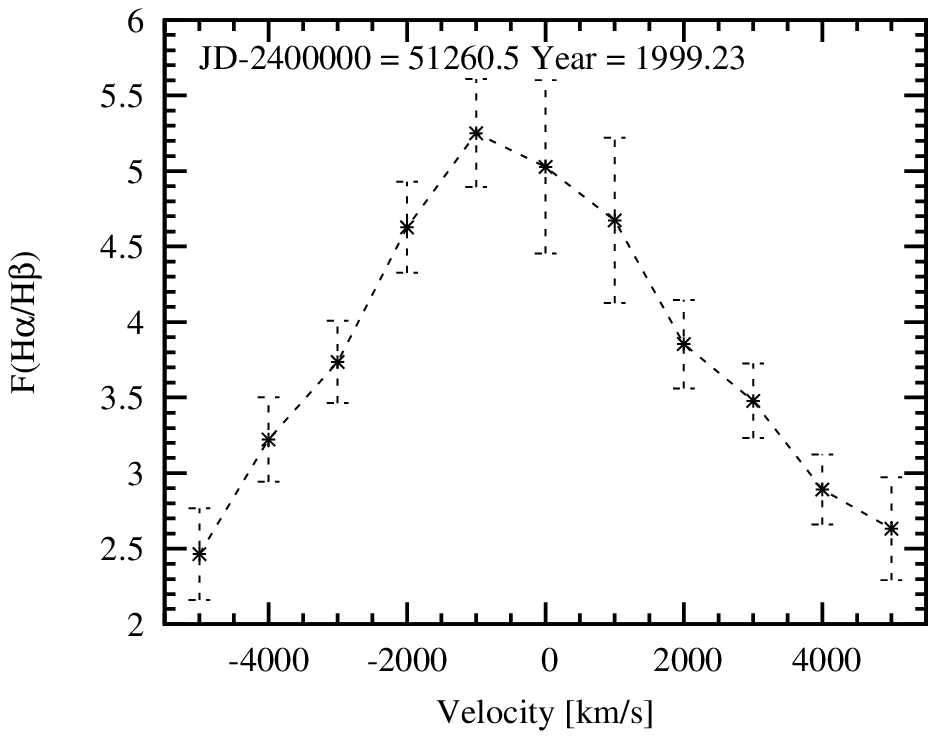}
\includegraphics[width=5.7cm]{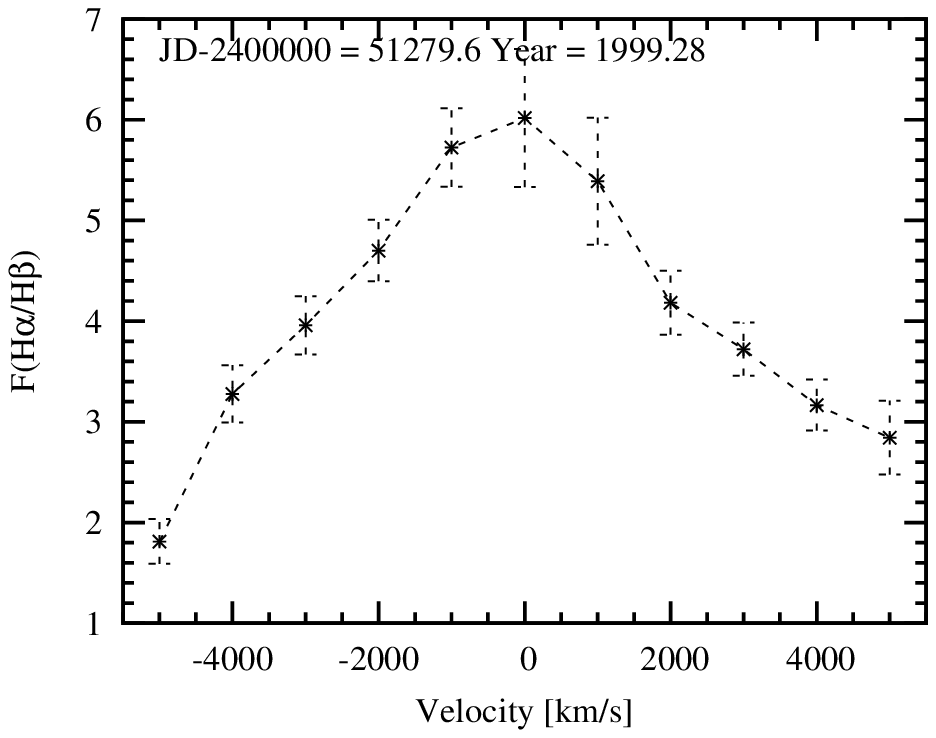}
\includegraphics[width=5.7cm]{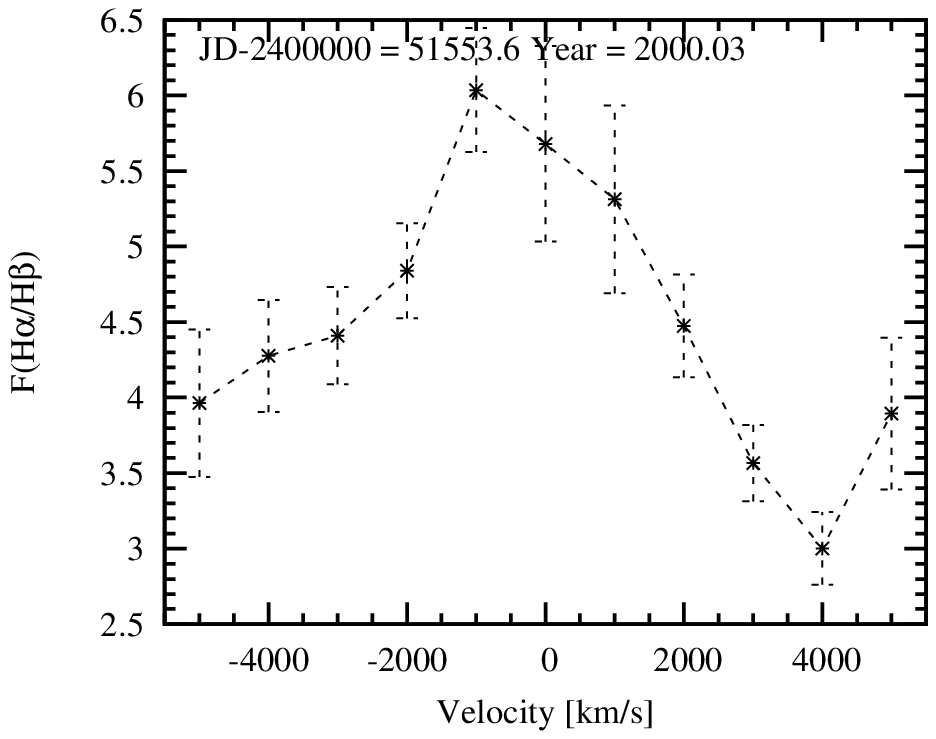}
\includegraphics[width=5.7cm]{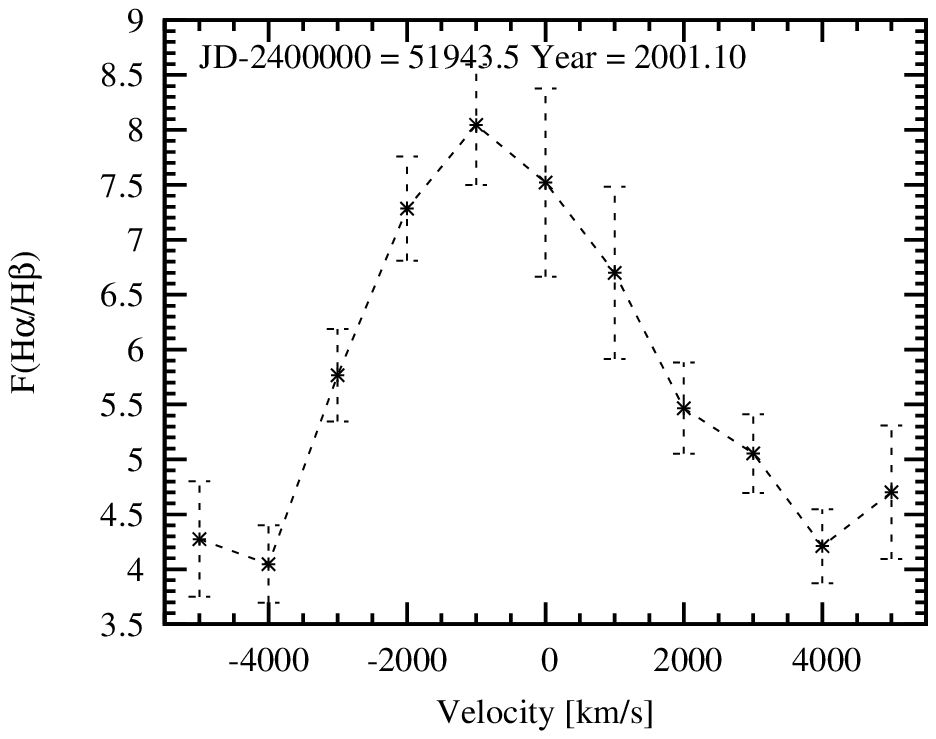}
\includegraphics[width=5.7cm]{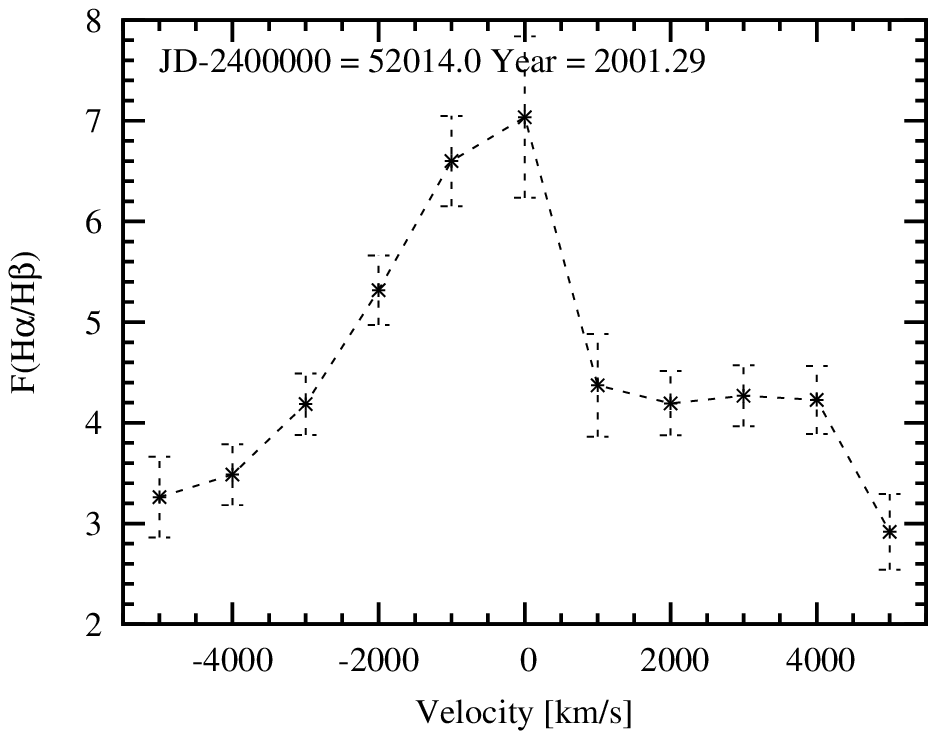}
\includegraphics[width=5.7cm]{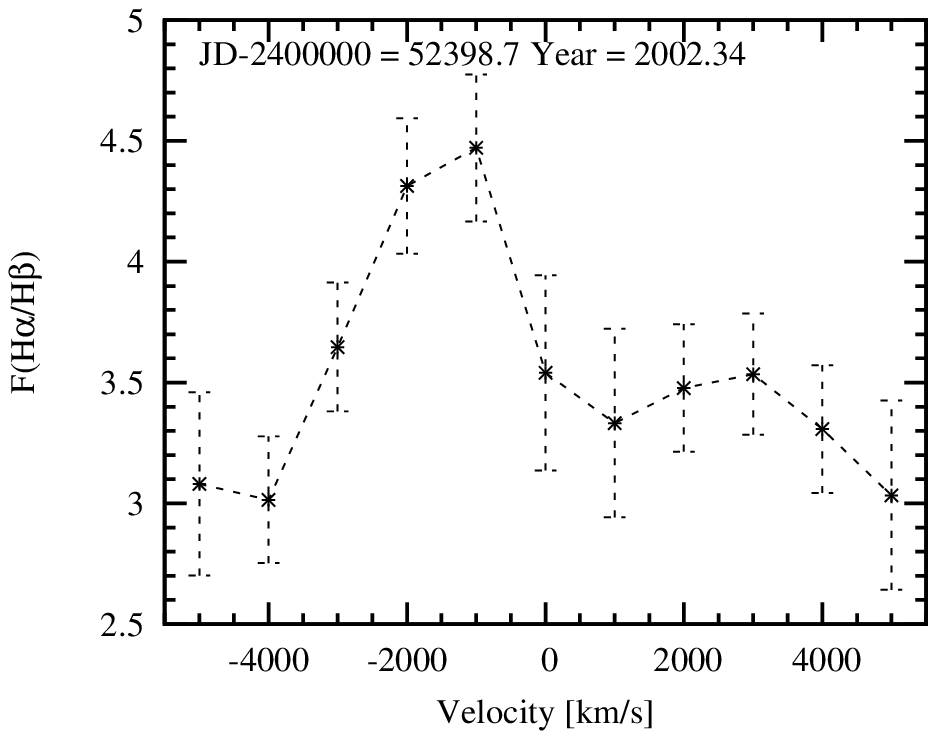}
\includegraphics[width=5.7cm]{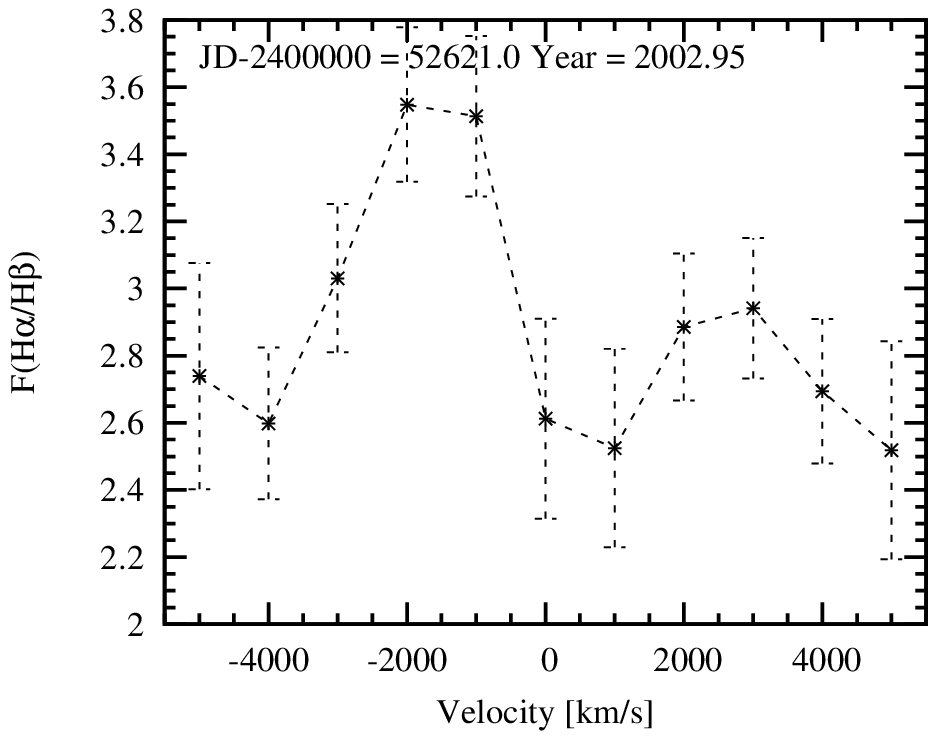}
\includegraphics[width=5.7cm]{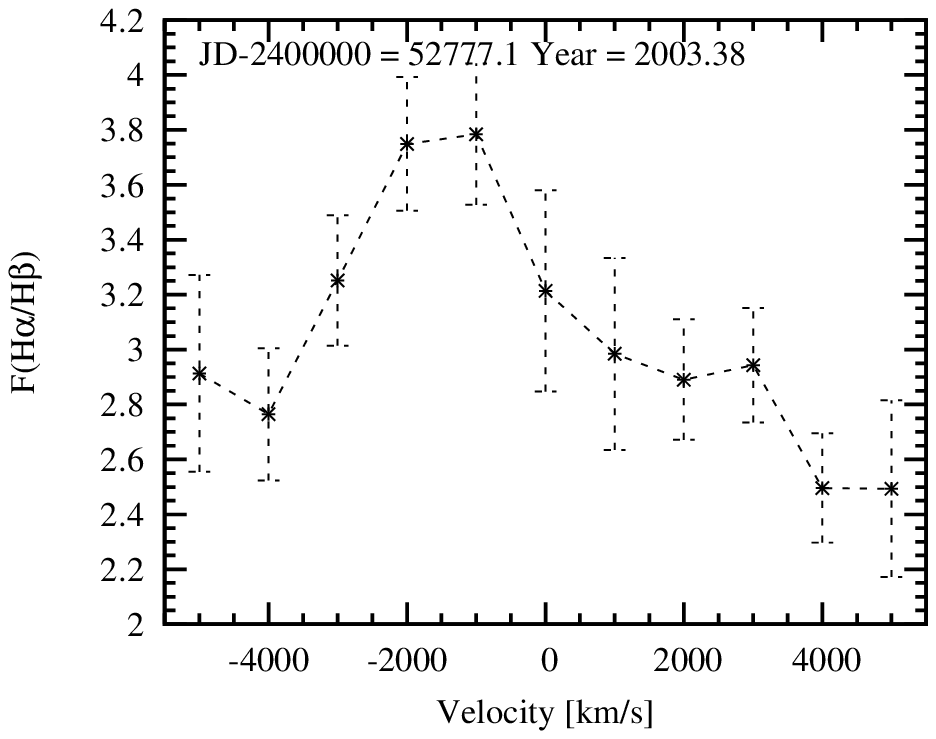}
\includegraphics[width=5.7cm]{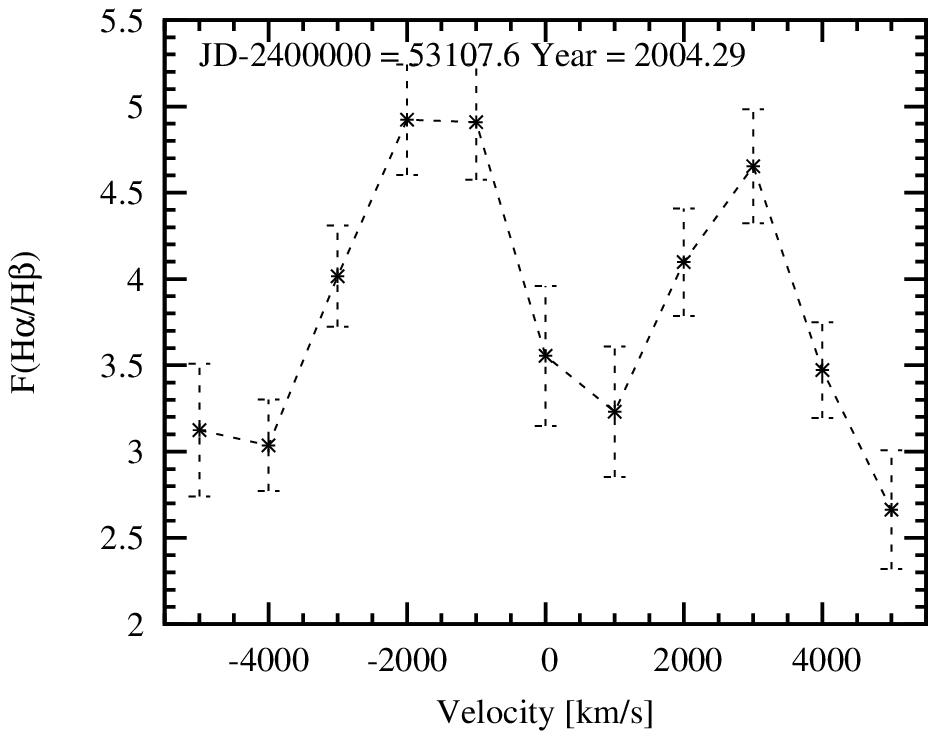}
\includegraphics[width=5.7cm]{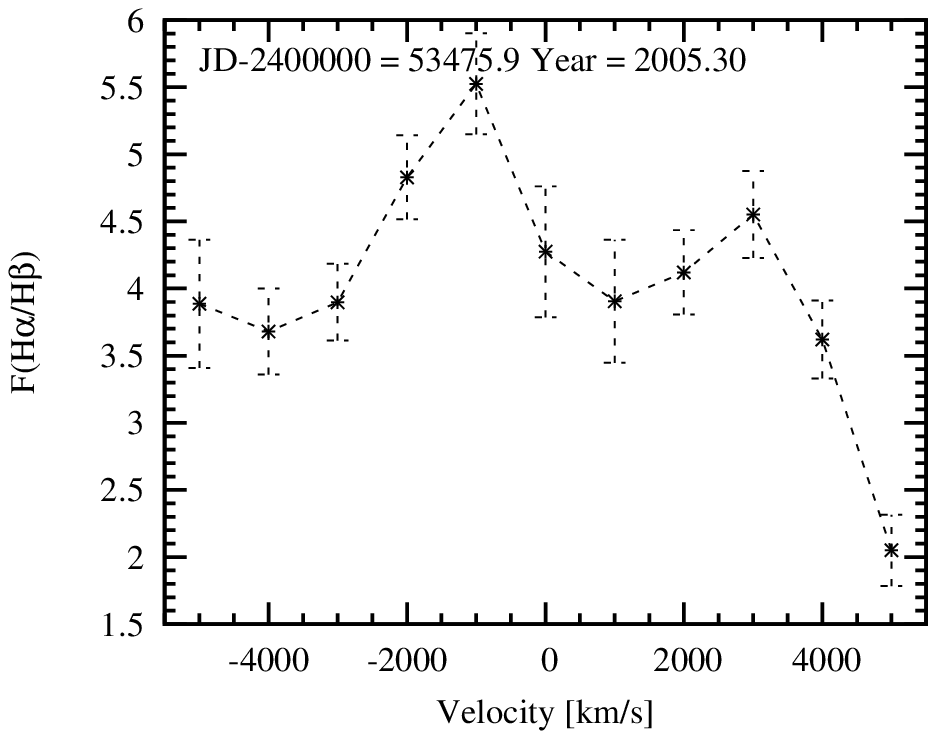}
\includegraphics[width=5.7cm]{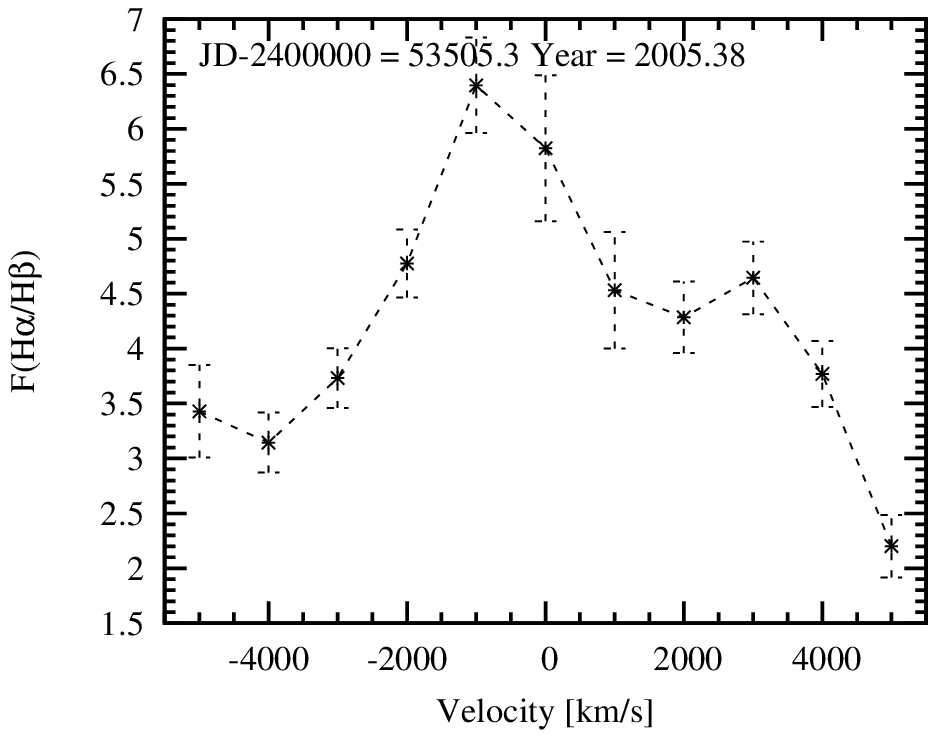}
\includegraphics[width=5.7cm]{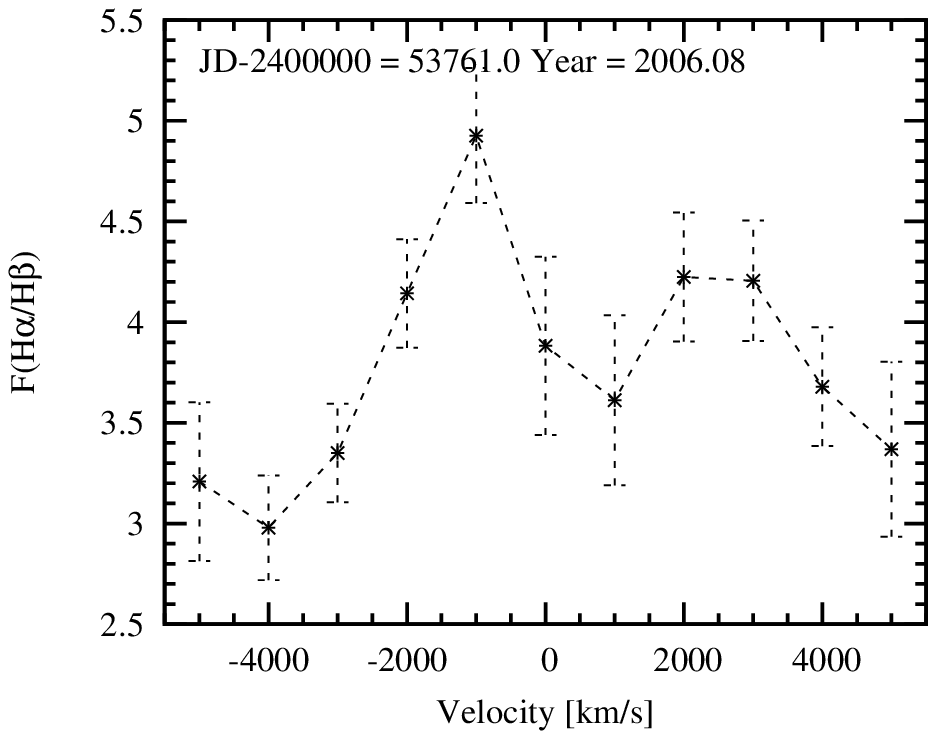}
\caption{Variations of the Balmer decrement
BD=F(H$\alpha$)/F(H$\beta$) as a function of the radial velocity for
{  month-averaged spectra} each year of the monitoring period. The
abscissae gives the radial velocities relative to the narrow
components. The {  month-averaged} Julian date and the corresponding
year are given at the top of each plot. }\label{fig17}
\end{figure*}

\subsubsection{Balmer decrement variations as a function of
the radial velocity}

As an illustration of the variation of the Balmer decrement with the
radial velocity during the monitoring period, we show in Fig.
\ref{fig17} the BD as a function of the radial velocity. Since the
BD vs. velocity remains the same during one year, we give a few
examples just for {  month-averaged spectra} for each year of the
monitoring period.

On the other hand, the behavior of the Balmer decrement as a
function of the radial velocity differs in different years. As a
rule, in 1996--1998 the maximum of the BD was at $\sim-1000\ {\rm km
\ s^{-1}} $, while the BD was slowly decreasing in the velocity
range from 0 to 1000 ${\rm km \ s^{-1}}$, and sharply decreasing in
the region $>\pm1000\ {\rm km \ s^{-1}} $, usually more strongly in
the region of negative velocity.  Recall here the results obtained
from photoionization model by Korista \& Goad (2004) where they
found velocity dependent variations in the Balmer decrement. They
found that the Balmer decrement is steeper in the line core than in
line wings, as we obtained in some cases (see Fig. \ref{fig17}), but
it is interesting that in all periods this peak is offset from the
central part to the blue side and also there are several cases where
the Balmer decrement is steeper in the wings. On the other hand in
the case  if velocity field is dominated by the central massive
object, one can expect symmetrical BD in the blue and red part of
velocity field (Korista \& Goad 2004). In our case we obtained
different asymmetrical shapes of BD vs. velocity field. {  The BD
seems to have  a systematic change in behavior starting around 2002
- i.e. corresponding to period III, when the "red bump" appears,
showing a two maxima in the BD vs. velocity field. This may indicate
that velocity field in the BLR is not dominated by central massive
black hole, i.e. it is in favor of some kind of streams of the
emitting material as e.g. outflow or inflow.}

In 1999--2001 the maximum values of
the BD $\sim(6-8)$ were observed in the velocity region $\pm 1000\
{\rm km \ s^{-1}}$, while a steeper decrease of the BD was more
often observed in the region of the positive velocity. The change of
the BD relative the radial velocity in 2002--2006 differed strongly
from those in 1996--2001. In these years 2 peaks (bumps) were
observed in the BD distribution: 1) at radial velocities from -2000
to -1000 with larger values of the BD; 2) at $+3000\ {\rm km \
s^{-1}}$ with somewhat smaller (by $\sim(0.5-1.0)$) values of the
BD.

\begin{figure*}
\centering
\includegraphics[width=6.cm]{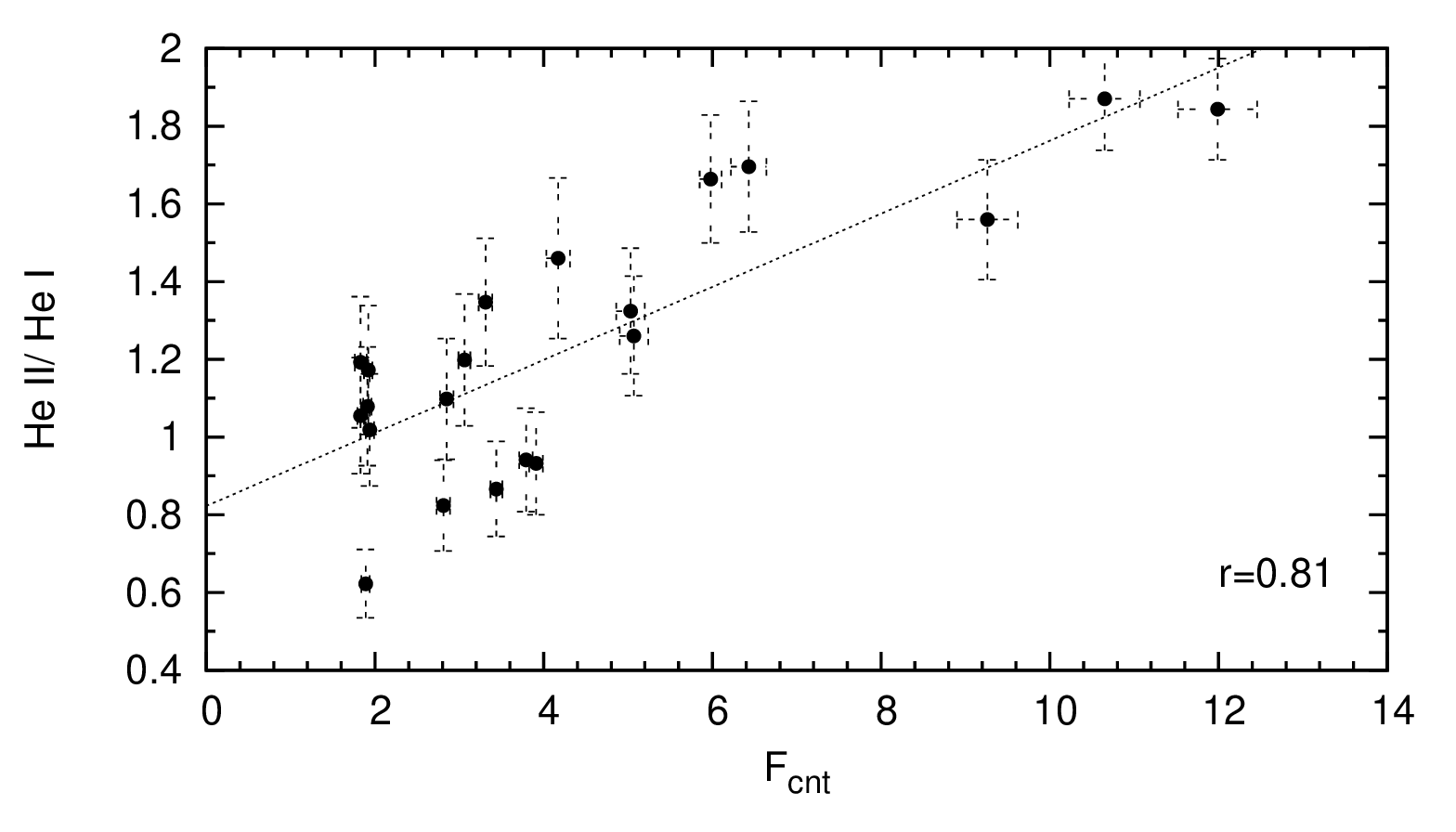}
\includegraphics[width=6.cm]{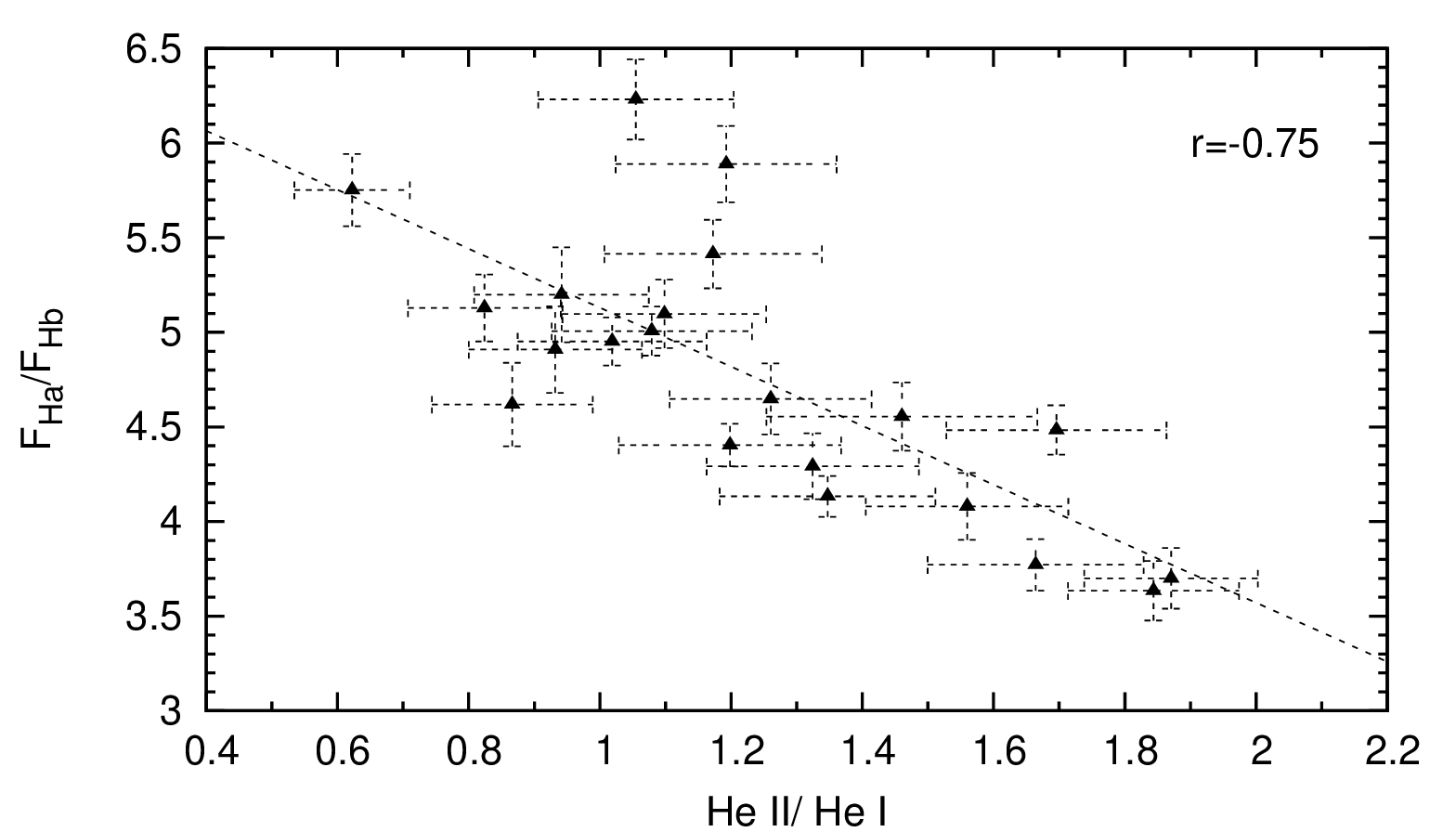}
\includegraphics[width=6.cm]{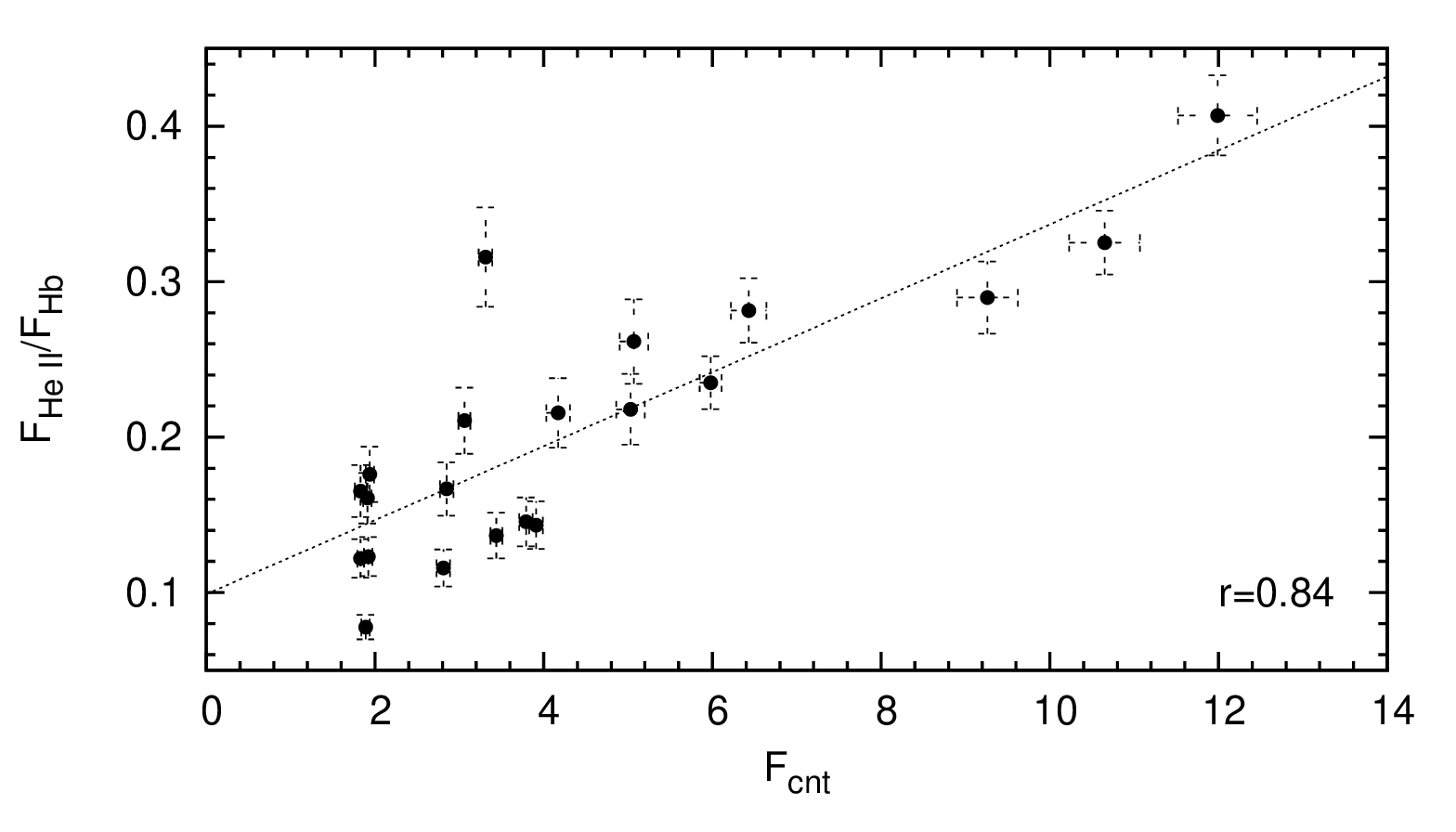}
\caption{Left: The variation of the ratio of the helium He II
$\lambda$4686 and He I $\lambda$5876 lines as a function of the
continuum flux. Middle: The BD as a function of the helium line
ratio. Right: The variation of the ratio of the helium He II
$\lambda$4686 and H$\beta$ lines as a function of the continuum
flux. The continuum flux is in units $10^{-14} \rm erg \, cm^{-2}
sec^{-1} \AA^{-1}$.}\label{fig18}
\end{figure*}

\subsection{Balmer decrement and helium line ratio}

In order to probe the physical conditions in the BLR, we studied the
flux ratio of He II $\lambda$4686 and HeI $\lambda$5876 broad lines.
From all the available spectra, we selected  only 21 spectra where
the broad helium lines could be precisely measured, and where the
two helium lines were observed on the same night. Note that in the
case of the minimum activity, the broad component of the He II
$\lambda$4686 line could not be detected at all.

We use the helium lines He II $\lambda$4686 and He I $\lambda$5876
since these two lines come from two different ionization stages and,
thus, are very sensitive to changes of the electron temperature and
density of the emitting region (Griem 1997).

In Fig. \ref{fig18} we give the HeII/HeI vs. continuum flux (first panel), BD vs.
HeII/HeI (second panel) and He II/H$\beta$ vs. continuum flux (third
panel). As it can be seen from Fig. \ref{fig18} (first panel) there is a good
correlation between the continuum flux and the helium line ratio
(the correlation coefficient r=0.81) which indicates that these
lines are probably coming from the region photoionized by the
continuum source. On the other hand, the Balmer decrement decreases
as the ratio of the two helium lines increases (the anti-correlation
is a bit weaker here, the correlation coefficient is r=0.75),
indicating that as the ionization gets stronger, the BD decreases
(Fig. \ref{fig18}, second panel).

The photoionization model (Korista \& Goad 2004) shows that the
Balmer lines should show less flux variation with continuum state
than He I, that is less varying than He II. In this case one can
expect that He II/H$\beta$ changes much more than the He II/He I.
The observed changes in the He II/H$\beta$ flux ratio (See Fig.
\ref{fig18}, third panel) is probably not due to differences in the
way these two lines respond to the level of the ionizing continuum
flux (see Shapovalova et al. 2008, Fig. 10 in the paper), but rather
to changes in the shape of the ionizing SED. In our case, this ratio
drops about 4 times from the high to low continuum states, while the
He II/He I line ratio changes around 3 times.

{  There is some connection between the helium and Balmer lines
(Fig. \ref{fig18}, left), but the physical properties of the
emitting regions of these lines are probably different. One of the
explanation of this correlation may be that a part of the Balmer
lines is coming from the same region as He I and He II lines. Since
the ratio of He II/ He I is sensitive to the change in temperature
and electron density, in the case of higher He II/ He I ratio the
ionization is higher and population of higher level in Hydrogen has
higher probability, consequently the ratio of H$\alpha$/H$\beta$ is
smaller than in the case of lower He II/ He I ratio.}

\begin{table}
\begin{minipage}[t]{\columnwidth}
\caption[]{Year-averaged and period-averaged variations of the
integral Balmer decrement (BD) and continuum fluxes. Columns: 1-year
or year-intervals; 2- F(H$\alpha$)/F(H$\beta$), H$\alpha$ and
H$\beta$ flux ratio or integrated Balmer Decrement (BD); 3 -
$\sigma$(F(H$\alpha$)/F(H$\beta$)), the estimated
F(H$\alpha$)/F(H$\beta$) error; 4 - F(5100), continuum flux at
$\lambda 5100\, \AA$; 5 - $\sigma$, estimated continuum flux error.
The last four lines give the mean BD, mean continuum flux and their
estimated errors in year-periods.} \label{tab5} \centering
\renewcommand{\footnoterule}{}
\begin{tabular}{lcccc}
\hline
\hline
 year     &     \multicolumn{2}{l}{F(H$\alpha$)/F(H$\beta$)$\pm \sigma$} &
\multicolumn{2}{l}{F(5100)$\pm \sigma$\footnote{in units of $10^{-14} \rm erg \ cm^{-2}s^{-1}\AA^{-1}$ }}   \\

\hline
 1996      &           3.65 & 0.51 &  11.18 & 1.55   \\
 1997      &           3.39 & 0.36 &   7.62 & 1.03   \\
 1998      &           3.58 & 0.35 &   6.08 & 1.19   \\
 1999      &           4.49 & 0.74 &   5.12 & 1.62   \\
 2000      &           5.01 & 0.62 &   2.75 & 1.33   \\
 2001      &           5.43 & 1.07 &   1.87 & 0.47   \\
 2002      &           3.67 & 0.44 &   3.59 & 0.29   \\
 2003      &           3.16 & 0.27 &   4.84 & 0.58   \\
 2004      &           3.70 & 0.29 &   2.44 & 0.57   \\
 2005      &           3.96 & 0.54 &   1.99 & 0.18   \\
 2006      &           3.58 & 0.36 &   2.60 & 0.80   \\
\hline
mean (1996-1998):    & 3.54 & 0.13 &   8.29 & 2.62  \\
\hline
mean 1999(Jan.-Apr.):& 4.21 & 0.59 &   5.89 & 0.56  \\
1999 (Dec.):         & 5.33 &      &   2.78 & 0.12  \\
\hline
mean (2000-2001):    & 5.22 & 0.30 &   2.31 & 0.62  \\
\hline
mean (2002-2006):    & 3.61 & 0.29 &   3.09 & 1.14  \\
\hline

\end{tabular}
\end{minipage}
\end{table}


\section{Summary of the results}

In this paper we investigated different aspects of broad H$\alpha$
and H$\beta$ line variations (shapes, ratios, widths and
asymmetries), so in this section we outline the summaries of our
obtained results.

\subsection{Summary of the line profile variations}

i) The line profile of H$\alpha$ and H$\beta$ were changing during
the monitoring period, showing blue (1996-1999) and red (2002-2006)
asymmetry. We observe bumps (shelves or an additional emission) in
the blue region in 1996--1997 at $V_r\sim-4000$ and $-2000 \ \rm km
\ s^{-1}$, in 1998--99 at $V_r\sim-2000 \ \rm km \ s^{-1}$ , and in
2000--2001 at -2000 and -500 $\rm km \ s^{-1}$ already. However, in
2002--2004 these details disappeared in the blue wing of both lines.
It became steeper than the red one and did not contain any
noticeable features. In 2002 a distinct bump appeared in the red
wing of both lines at $+3100\ \rm km  \ s^{-1}$. The radial velocity
of a bump in the red wing changed from   $+3100\ \rm km  \ s^{-1}$
in 2002 to $\sim2100 \ \rm km\ s^{-1}$ in 2006. Possibly, the
appearance of the blue and red bumps are related with a
jet-component.

ii) In 2005 (May-June), when the nucleus of NGC 4151 was in the
minimum activity state, the line profiles had a double-peaked
structure with two distinct peaks (bumps) at radial velocities of
-2586; +2027 $\ {\rm km \ s^{-1}}$ in H$\beta$ and -1306; +2339 $\
{\rm km \ s^{-1}}$ in H$\alpha$.

iii) In 1996-2001 we observed a broad deep absorption line in
H$\beta$ at a radial velocity  that took values
from $V_r\sim-1000$ to $V_r\sim-400 \ \rm km \ s^{-1}$.

iv) the FWHM is not correlated with the continuum flux, while the
asymmetry tends to anti-correlate with it (coefficient correlation
r$\sim$-0.5). The FWHM and asymmetry in H$\beta$ are larger than in
H$\alpha$. {  It may be explained with the fact that H$\beta$
originates closer to the SMBH, and thus has larger widths.}

v) We divided the line profiles into 11 identical segments, each of
width 1000 $\rm km \ s^{-1}$ and investigated the correlations
between segments vs. continuum flux and segments vs. segments flux.
We found that the far red wings (from 3500 to 5500 $\rm km \
s^{-1}$) and central (at $V_r=\pm3500 \ \rm km \ s^{-1}$) segments
for the continuum flux $F_{\rm c}<7 \times 10^{-14} \rm erg cm^{-2}
s^{-1} A^{-1}$ respond to the continuum almost linearly, and
this segments are probably coming from the part of the BLR
ionized by the AGN source.

vi) The central ($\pm3500 \ \rm km \ s^{-1}$) segments for $F_{\rm
c}>7 \times 10^{-14} \rm erg cm^{-2} s^{-1} A^{-1}$ do not show any
linear relation with the continuum flux. Probably in periods of high
activity this segments of line is partly originating in
substructures which are not photoioninized by the AGN continuum.

vii) The far blue wing (from -5500  to -4500) seems to originate in
a separate region, so it does not respond to the continuum flux
variation as well as to the variation of other line segments, except
in the case of high line flux where it responds to the far red wing
(see Fig. \ref{fig12}). This may indicate that the far blue wing is emitted
by a distinct region which is not photoionized, and also that the
emission is highly blueshifted (as e.g. an outflow with velocities
$>$3500 $\ {\rm km \ s^{-1}}$).

viii) The far red wing is very sensitive to the continuum flux
variation, and is thus probably coming from the part the closest to
the center, i.e. this part of the line seems to be purely
photoionized by the AGN source.

From all the facts mentioned above, one can conclude that the broad
lines of NGC 4151 are produced at least in three kinematically (and
physically) distinct regions.

\subsection{Summary of the BD variations}

The Balmer decrement (BD=F(H$\alpha$)/F(H$\beta$)) varied from 2 to
8 during the monitoring period. It is interesting that BDs are quite
different along line profiles as well. In 1996--1998 there was no
significant correlation between the BD and the continuum flux (the
continuum flux was $F_{\rm c}\sim (6-12) \times 10^{-14} \rm erg \
cm^{-2} s^{-1} A^{-1}$). In 1999--2001 maximal variations of the BD were
observed, especially in the blue part of lines. The maximal value of
the BD along the line profiles strongly differed: in 1996--2001 the
BD was maximum at $V_r\sim \pm1000\ \rm km \ s^{-1} $ and in
2002--2006 the BD had 2 peaks - at velocity from -2000 to -1000 and
at $V_r\sim3000\ \rm km \ s^{-1}$ with somewhat smaller (by
$\sim$(05-1)) values of the BD. In the last case it is possible that
the second bump (the fainter one) is caused by the interaction
between the receding sub-parsec jet and environment.

The different values of the BD observed during the monitoring period
(as well as different values of the BD along the profiles) also
indicate a multicomponent origin of the broad lines. Such different
ratios may be caused by absorption, but also by different physical
conditions in different parts of the BLR.

\section{Discussion}

It is interesting to compare our results with those found in the UV
and X-radiation. Crenshaw \& Kraemer (2007) found a width of 1170 km
s$^{-1}$ (FWHM) for the UV emission lines significantly smaller than
our results for the Balmer lines (FWHM$\sim$ 5000-7000 km s$^{-1}$,
 around 5-6 times smaller). This can be interpreted as the
existence of an intermediate component between the broad and narrow
(emission) line regions (see e.g. Popovi\'c et al. 2009). In the
same paper, they found an evidence that the UV emission lines arise
from the same gas responsible for most of the UV and X-ray
absorption. The absorption can be seen in outflow at a distance of
0.1 pc from the central nucleus. Also, it is interesting that
Crenshaw \& Kraemer (2007) favor a magnetocentrifugal acceleration
(e.g., in an accretion disk wind) over those that rely on radiation
or thermal expansion as the principal driving mechanism for the mass
outflow. Obscuration should play an important role, but we found
that while an absorption can clearly be detected in the H$\beta$
line, the H$\alpha$ line displays a small amount of absorption (see
Figs. \ref{fig2} and \ref{fig3}). Absorption in the H$\beta$ line
was previously reported by Hutchings et al. (2002), therefore
obscuration of the optical continuum should be present in NGC 4151.
The location of the obscuring material was estimated by Puccetti et
al. (2007). They analyzed the X-ray variability of the nucleus. They
found that the location of the obscuring matter is within
3.4$\times$10$^4$ Schwarzschild radii from the central X-ray source
and suggested that absorption variability plays a crucial role in
the observed flux variability of NGC 4151.

Let us discuss some possible scenarios for the BLR. As we mentioned
above, the outflow is probably induced by the magnetocentrifugal
acceleration, and starts very close to the black hole. If we adopt a
black hole mass around 4$\times 10^7\ M_\odot$ (obtained from
stellar dynamical measurements, see Onken et al. 2008, which is in
agreement with Bentz et al. 2006), the acceleration (and line
emission) is starting at $\sim$10$^{-4}$ pc from the black hole and
the outflow is emitting until $\sim 0.01$ pc; taking into account
the absorption velocities (e.g. Hutchings et al. 2002) around -1000
km s$^{-1}$,

{  As it was mentioned above, the broad H$\alpha$ and H$\beta$ lines
show different widths and asymmetries during the monitoring period,
indicating a complex structure of the BLR. Also, we found that the
line profiles of H$\alpha$ and H$\beta$ could be different at the
same epoch. Therefore one should consider a multi-component origin
of these lines.

To propose a BLR model, one should take into account that: a) there
is an absorption component in the blue part of lines indicating some
kind of outflow that may start in the BLR (see \S 3.2). Also
Crenshaw \& Kraemer (2007) reported a mass outflow from the nucleus
of the NGC 4151, but they confirmed the observed outflow in the
Intermediate Line Region (ILR).; b) the flux in the far red wing
correlates well with the continuum flux, indicating that it
originates in an emission region very close to the continuum source,
i.e. to the central black hole; c) the flux of the far blue wing
does not correlate with the continuum, {  that may indicate some
kind of shock wave contribution to the Balmer line}.

All these facts indicates that an outflow probably exists in the
BLR. In this case a complex line profiles (with different features)
can be expected due to changes in the outflowing structure, as it is
often seen in the narrow lines observed in the jet induced shocks
(see e.g. Whittle \& Wilson 2004). Of course we cannot exclude that
there is the contribution of some different regions to the composite
line profile, as e.g. there can be also contribution of the ILR
which may be with an outflow geometry. In forthcoming paper we will
try to adjust the observational results with those predicted by
various models of the kinematics and structure of the BLR (a
bi-conical outflow, an accelerating outflow (wind), a Keplerian
disk, jets, etc).

It is interesting to note that the line profiles are changing during
monitoring period, especially after 2002, when the 'red bump'
appears. After that the asymmetry of both lines (see Table 2) and BD
show different behavior than in the period 1996 - 2002. Moreover, in
the III period, the line profiles of H$\alpha$ and H$\beta$ very
much changed in the shape from double peaked profiles (see Figs. 1
and 2) to a quite asymmetrical profiles (as was observed in 2006).}

The integrated Balmer decrement was maximum in 1999--2001 (see Fig.
\ref{fig14}). The BD changes shape from 2002, showing two peaks in
the BD vs. velocity field profile. Most probably, this is connected
with strong inhomogeneities in distribution of the absorbing
material during different periods of monitoring.  As the line flux
in H$\alpha$ and H$\beta$ at small fluxes ($<7 \times 10^{-14} \rm
erg \ cm^{-2} s^{-1} A^{-1}$) correlates well with that of the
continuum (Paper 1), we infer that the change of the integrated
Balmer decrement in 1999--2001 is also caused, at least partially,
by changes in the continuum flux. Indeed, when the ionization
parameter decreases for a constant density plasma, an increase of
F(H$\alpha$)/F(H$\beta$) intensity ratio is expected (see for
instance Wills et al. 1985). This is due to the decrease of the
excitation state of the ionized gas: the temperature of the ionized
zone being smaller, the population of the upper levels with respect
to the lower ones decreases. In 1996--1998 BDs did not correlate
with continuum, i.e. in this case the main cause of BD variations is
not related to the active nucleus and probably the shock initiated
emission is dominant. We detected that the FWHM of lines does not
correlate with continuum. This confirms our assumption that broad
lines can formed in several (three) different subsystems or that the
emission is affected by an outflow that produces shock initiated
emission.

\section{Conclusions}

This work is subsequent to the Paper I (Shapovalova et al. 2008) and
is dedicated to a detailed analysis of the broad H$\alpha$ and
H$\beta$ line profile variations during the 11-year period of
monitoring (1996--2006). From this study (Section 3) it follows that
the BLR in NGC 4151 is complex, and that broad emission lines are
the sums of components formed in different subsystems:

1) the first component is photoionized by the AGN continuum (far
red line wings, $V_r\sim +4000$ and $+5000 \ \rm km \ s^{-1}$ and
central (at $V_r=\pm3500 \ \rm km \ s^{-1}$) segments for continuum
flux $F_{\rm c}<7 \times 10^{-14} \rm erg cm^{-2} s^{-1} A^{-1}$).
This region is the closest to the SMBH.

2) the second component is independent of changes of the AGN continuum
(far blue line wings, $V_r\sim-5000 \ \rm km \ s^{-1}$; $-4000 \ \rm km \
s^{-1}$). It is possibly generated by shocks initiated by an
outflow.

3) the third component, where the central parts of lines
($V_r<4000 \ \rm km \ s^{-1}$) are formed, in high fluxes $F_{\rm c} > 7 \times
10^{-14} \rm erg \ cm^{-2} s^{-1} A^{-1}$ is also independent of the
AGN continuum (possibly, outflow and jet).

Finally, we can conclude that the BLR of NGC 4151 may not be purely
photoionized, i.e. besides photoionization, there could be some
internal shocks contributing to the broad lines  see also
discussion in Paper I). At least there are three different
subregions with different velocity field and probably different
physical conditions, that produce complex variability in the broad
lines of NGC 4151 and changes in the line profile (very often
temporarily bumps).  Our investigations indicate
that the reverberation might not be valid as a tool to determine the
BLR size in this AGN and that this AGN is not perfect for this
method. Consequently, the results for the BLR size of NGC 4151
(given in the introduction) should be taken with caution.

\section*{Acknowledgments}

 Authors would like to thank Suzy Collin for her suggestions how
to improve this paper. Also, we thank the anonymous referee for very
useful comments. This work was supported by INTAS (grant N96-0328),
RFBR (grants N97-02-17625 N00-02-16272, N03-02-17123 and
06-02-16843, 09-02-01136), State program 'Astronomy' (Russia),
CONACYT research grant 39560-F and 54480 (M\'exico) and the Ministry
of Science and Technological Development of Republic of Serbia
through the project Astrophysical Spectroscopy of Extragalactic
Objects (146002).

\Online

\end{document}